\documentclass[aps,prb,twocolumn,superscriptaddress]{revtex4-1}
\usepackage{graphicx,psfrag,color}
\usepackage{mathbbol,amsmath,amsfonts,amssymb,bm,dsfont,enumerate}
\usepackage{wasysym}
\usepackage{hyperref}
\hypersetup{
    colorlinks,
    citecolor=blue,
    filecolor=blue,
    linkcolor=blue,
    urlcolor=blue
}

\def\a{{\mathbf{a}}}
\def\j{{\mathbf{j}}}

\def\k{{\mathbf{k}}}

\def\r{{\mathbf{r}}}
\def\B{{\mathbf{B}}}
\def\b{{\mathbf{b}}}

\def\T{{\bm{T}}}

\begin{document}

\title{A generalization of Bloch's theorem for arbitrary boundary conditions: 
Theory}

\begin{abstract}

We present a generalization of Bloch's theorem 
to finite-range lattice systems of independent fermions, in which translation symmetry is 
broken solely due to {\it arbitrary} boundary conditions, by providing exact, 
analytic expressions for {\it all} energy eigenvalues and eigenstates.
Starting with a re-ordering of the fermionic basis that transforms the single-particle 
Hamiltonian into a {\it corner-modified banded block-Toeplitz} matrix, a key step is a 
Hamiltonian-dependent bipartition of the lattice, which splits the eigenvalue problem 
into a system of bulk and boundary equations. The eigensystem inherits most of its 
solutions from an auxiliary, infinite translation-invariant Hamiltonian
that allows for non-unitary representations of translation -- hence complex values 
of crystal momenta with specific localization properties. 
A reformulation of the boundary equation in terms 
of a {\em boundary matrix} ensures compatibility with the boundary conditions, and determines
the allowed energy eigenstates in the form of {\em generalized Bloch states}. We show how the boundary 
matrix quantitatively captures the interplay between bulk and boundary properties, leading to the 
construction of efficient indicators of bulk-boundary correspondence. 
Remarkable consequences of our generalized Bloch theorem are the engineering of Hamiltonians  
that host perfectly localized, robust zero-energy edge modes, and the predicted emergence, 
for instance in Kitaev's Majorana chain, of localized excitations whose amplitudes decay in 
space exponentially with a {\it power-law prefactor}. We further show how the  theorem 
may be used to construct numerical and algebraic diagonalization algorithms for the class of 
Hamiltonians under consideration, and use the proposed bulk-boundary indicator to 
characterize the topological response of a multi-band time-reversal invariant $s$-wave topological 
superconductor under twisted boundary conditions, 
showing how a fractional Josephson effect can occur {\it without} entailing  
a fermionic parity switch.  Finally, we establish connections to the transfer matrix method and 
demonstrate, using the paradigmatic Kitaev's chain example, that a defective (non-diagonalizable) 
transfer matrix signals the presence of solutions with a power-law prefactor.

\end{abstract}

\author{Abhijeet Alase}
\affiliation{\mbox{Department of Physics and Astronomy, Dartmouth
College, 6127 Wilder Laboratory, Hanover, New Hampshire 03755, USA}}

\author{Emilio Cobanera}
\affiliation
{\mbox{Department of Physics and Astronomy, Dartmouth
College, 6127 Wilder Laboratory, Hanover, New Hampshire 03755, USA}}

\author{Gerardo Ortiz}
\affiliation{\mbox{Department of Physics, Indiana University,
Bloomington, Indiana 47405, USA}}
\affiliation{\mbox{Department of Physics, University of Illinois, 1110 W Green Street, 
Urbana, Illinois 61801, USA}}

\author{Lorenza Viola}
\affiliation{\mbox{Department of Physics and Astronomy, Dartmouth
College, 6127 Wilder Laboratory, Hanover, New Hampshire 03755, USA}}

\date{\today}
\maketitle


\section{Introduction}

Modern electronic transport theory in crystalline solids relies on two 
fundamental tenets. On the one hand, because of the Pauli exclusion principle, electrons  
satisfy Fermi-Dirac statistics; on the other,  Bloch's theorem allows 
labeling of the one-electron wave-functions in terms of their crystal momenta. The set 
of allowed momenta, defining the so-called Brillouin zone,  is determined 
by symmetry and the fact that Born-von-Karman 
(periodic) boundary conditions (BCs) are enforced on the system\cite{ashcroftmermin}. It is 
the organization of electrons within the Brillouin zone that is key to defining its conduction properties.
While the assumption of a perfect crystal with a unit cell that is periodically repeated emphasizes 
the (discrete) symmetry of translation, the torus topological constraint imposed by 
the Born-von-Karman condition further eliminates the potential emergence of edge or boundary 
electronic states in a real, finite crystal. Although much of the transport properties are determined 
by bulk electrons, technologically relevant processes on the surface of solids are known to 
lead to intriguing phenomena, such as surface superconductivity \cite{gor'kov} or 
Kondo screening of magnetic impurities resulting in exotic surface spin textures \cite{kondoleonid}. 
Early theoretical investigations by Tamm and Shockley \cite{Tamm, Shockley} initiated the 
systematic study of surface state physics, that witnessed a landmark achievement with the 
discovery of the quantum Hall effect \cite{hall}, and that today finds its most striking applications 
in topological insulating and superconducting materials \cite{bernevig}.  

The organization of bulk electrons   
comes with a twist. The quantum electronic 
states labeled by crystal momenta organize in ways subject to classification 
according to integer values of topological invariants defined over the entire Brillouin zone
\cite{zak,chiu16}. 
The first Chern number, determined in terms of the Berry connection, is one of those 
topological invariants, defining a topologically non-trivial electronic phase whenever 
its value differs from zero \cite{bernevig}. For instance, the transverse 
conductivity of a quantum Hall fluid is proportional to such a Chern number. 
Perhaps surprisingly, there appears to be a connection between 
a non-vanishing value of the topological invariant, a bulk property, and the emergence 
of ``robust'' boundary states, an attribute of the surface. This principle is known as the {\it bulk-boundary 
correspondence} \cite{bernevig,Graf13,chiu16}.
At first, this relation seems odd, since surface properties  are 
totally independent from those of the bulk; for example, one can deposit impurities, 
generate strain and reconstruction, or add externally applied electric fields only on the surface. 
Nonetheless, it seems reasonable to assume that as long as the symmetry protecting the surface 
states is not broken by external means, a bulk-boundary correspondence will still hold, 
although the quantum surface state will, in general, get transformed 
\cite{isaev11}. In other words, 
although the mere existence of a boundary mode may be robust, only 
{\em classical} information may be protected in general \cite{isaev11,ips}.

It is apparent that Bloch's theorem and its consequences pertain to the realm of bulk physics. 
A crystal {\it without boundaries} is required to establish it. But, can one generalize Bloch's 
theorem for independent electrons to arbitrary BCs, so that bulk and surface states 
can be handled on an equal footing, and physical insight about the interplay between 
bulk and boundary may be gained? 
In light of our previous discussion, it is clear that to accomplish 
such a task one needs to give up on some concepts, such as the notion of a Brillouin zone. 
If possible, such a generalization would allow us to formulate a bulk-boundary correspondence 
principle that makes use of {\em both} bulk and boundary information. It is tempting to argue 
that the relative importance of BCs diminishes as the size of the crystal grows. Notwithstanding,  
for example, recent work shows that BCs impact the quasi-conserved local
charges of one-dimensional systems, with important consequences
for bulk quench dynamics \cite{fagotti,gluza16}. More generally, the statistical 
mechanics of topologically nontrivial systems begs some answers directly relevant to 
the questions above \cite{quelle15,quelle16}. 

In this paper, we generalize Bloch's theorem to systems of independent electrons 
subject to arbitrary BCs. Intuitively speaking, one may expect such a result on the 
basis that translation symmetry is only mildly broken by BCs -- namely, clean (disorder-free) 
systems are translationally-invariant {\it away} from the boundary. Our generalized Bloch 
theorem makes this idea precise, by providing an exact  
(often in fully closed-form)
description of the eigenstates of the system's Hamiltonian in terms of generalized eigenstates of 
{\it non-unitary} representations of translation symmetry in infinite space, that is, 
with boundaries at infinity and no torus topology \cite{commentAM}. As a 
result, both exponentially decaying edge modes and more exotic modes with {\em power-law} 
prefactors can emerge, provided the BCs allow them.
Our generalized Bloch theorem leverages the bulk-boundary separation
of the Schr\"odinger equation we introduced in Ref.\,[\onlinecite{abc}] and the
full solution of the bulk equation rigorously established in Ref.\,[\onlinecite{JPA}].
It extends the diagonalization procedure described in Ref.\,[\onlinecite{abc}],
and recently used in Ref.\,[\onlinecite{KatsuraTwisted}], to a more general class of 
Hamiltonians and BCs, which in particular allows for different modifications to be 
imposed on different boundaries.
A unifying theme behind these results is an effective analytic continuation to the complex 
plane of the standard Bloch's Hamiltonian off the Brillouin zone. This analytic
continuation is remarkably useful because the original problem reduces to 
a matrix  polynomial function \cite{JPA}.  Interestingly, a recent
study made use of similar polynomial structures
for the purpose of topological classification \cite{read16}.

The outline of this paper is as follows. 
In Sec.\,\ref{sec:setting} we discuss a re-arrangement of the fermionic basis that allows  
us to reduce the diagonalization of the original many-electron finite-range quadratic Hamiltonian
in second quantization, subject to specified BCs, to the one of a single-particle Bogoliubov-de 
Gennes Hamiltonian that has the structure of a {\it corner-modified block-Toeplitz matrix}, as
introduced in Ref.\,[\onlinecite{JPA}]. 
Section\,\ref{sec:theory}  develops 
a structural characterization of the energy eigenstates for the many-electron 
systems under consideration, culminating into our generalization of 
Bloch's theorem. Like the usual Bloch's theorem, such a generalization is first 
and foremost a practical tool for calculations, granting direct access to exact 
energy eigenvalues and eigenstates. 
In Sec.\,\ref{sec:algo}, we provide two new procedures -- one numerical 
and another algebraic -- for carrying out the exact diagonalization of the single-particle Hamiltonian, 
based on the generalized Bloch theorem. The algebraic procedure, which may provide {\em closed-form} 
solutions to the problem, is explicitly illustrated through a number of examples in 
Sec.\,\ref{sec:examples}. While, in order to illustrate our methodology, 
we focus largely on one-dimensional systems here, 
we anticipate that additional applications to higher-dimensional problems 
will be addressed in a companion paper \cite{PRB2}.
Remarkably, while mid-gap modes with power-law prefactors
have been predicted for systems with long-range couplings, 
we show analytically that they can also 
prominently manifest in short-range tight-binding models of topological insulators and 
superconductors \cite{pientka13,degottardi13,vodola14,ortiz14,yazdani14}.

Crucially,
our generalized Bloch theorem also allows derivation of a boundary indicator for the
bulk-boundary correspondence, which contains information from both the bulk {\it and} the BCs and, as remarked in Ref.\,[\onlinecite{abc}], is computationally more efficient than other indicators
also applicable in the absence of translational symmetry \cite{kitaev01}. 
This is the subject of Sec.\,\ref{sec:indicator}. In the same section, we expand on the analysis 
of the two-band time-reversal invariant $s$-wave topological superconducting 
wire we introduced previously \cite{swavePRL, swavePRB}, by employing 
our newly defined indicator of bulk-boundary correspondence -- constructed by 
using the generalized Bloch theorem, as opposed to the simplified Ansatz
we presented in Ref.\,[\onlinecite{abc}].
Specifically, this indicator is employed in the analysis of the 
Josephson response of the $s$-wave 
superconductor in a bridge configuration, sharply diagnosing the occurrence 
of a fractional $4\pi$-periodic Josephson effect. Remarkably, we find that 
this is possible {\em without} a conventional fermionic parity switch, which 
we explain based on a suitable transformation into two
decoupled systems, each undergoing a parity switch.
Section \,\ref{sec:TM}
establishes some important connections between our generalized Bloch theorem
and the widely employed {\em transfer matrix} approach \cite{Lee}. Interestingly, from the standpoint 
of computing energy levels, our bulk-boundary separation is in many ways complementary to 
the transfer matrix method.  While the latter can 
handle bulk disorder (at a computational cost), it does not, {\em a priori}, lend itself to investigating the 
space of arbitrary BCs in a transparent way. 
On the contrary, our generalized Bloch theorem can handle arbitrary BCs 
efficiently, as long as the bulk respects translational invariance -- 
with arbitrary (finite-range) disorder on the boundary being permitted.
Looking afresh at the transfer matrix approach from the 
generalized Bloch theorem's perspective yields a remarkable result: the generalized eigenvectors 
of the transfer matrix, whose role has been appreciated only recently \cite{DwivediBB},  
describe energy eigenstates with power-law corrections to an otherwise exponential behavior. 
Our generalized Bloch theorem further suggests a way to extend the transfer matrix approach 
to a disordered bulk and arbitrary BCs.
A discussion of the main implications of our work, along with outstanding research questions, 
concludes in Sec. \ref{sec:end}, whereas additional technical material is included in separate appendixes.

\section{From independent fermions to Toeplitz matrices}
\label{sec:setting}

We begin by describing the class of model Hamiltonians 
investigated in this and the companion paper \cite{PRB2}. 
The upshot of this section will be a \emph{non-conventional re-ordering} of the 
physical subsystems' labels that allows recasting the single-particle 
(Bogoliubov-de Gennes (BdG))  
Hamiltonians in Toeplitz form,  essential for 
the exact diagonalization procedure we will describe. 

Consider a \(D\)-dimensional, translation-invariant {\em infinite} system of independent fermions. 
Such a system is described in full generality by a quadratic, not necessarily 
particle-number-conserving, Hamiltonian in Fock space. 
In a lattice approximation, 
the vector position of a given fermion in the regular crystal lattice can be written 
as the sum of a Bravais lattice vector and a basis vector \cite{ashcroftmermin}.
We will include these basis vectors as part of the internal 
labels, and denote Bravais lattice vectors as 
\(\j \equiv \sum_{\mu=1}^D j_\mu\a_\mu\), 
with \(\a_1,\dots,\a_D\) primitive vectors and each 
\(j_\mu \in \mathbb{Z}\). An orthonormal 
basis of the Hilbert space of single-particle 
states is thus labeled by Bravais lattice vectors \(\j\),  
and a finite number of internal labels \(m=1,\dots,d_{\rm int}\). 
We denote by $c^{\;}_{\j m}$ ($c^\dagger_{\j m}$)
the fermionic annihilation (creation) operator 
corresponding to lattice vector $\j$ and internal state $m$.
The Hamiltonian of a translation-invariant 
system can then be written as
\begin{eqnarray}
\label{Hamtransinv}
\widehat{\bm{H}} = \sum_{\r}\sum_{\j}\big[\hat{\Phi}^\dagger_{\j}K_{\r}\hat{\Phi}^{\;}_{\j+\r}
+\frac{1}{2}(\hat{\Phi}^\dagger_{\j}\Delta_{\r}\hat{\Phi}_{\j+\r}^\dagger +\text{h.c.})\big],\quad
\end{eqnarray}
with $\hat{\Phi}^\dagger_{\j} \equiv 
\begin{bmatrix}c^\dagger_{\j 1} & \cdots & 
c^\dagger_{\j d_{\rm int}}\end{bmatrix}$, 
$\r$ a Bravais lattice vector, and the $d_{\rm int}\times d_{\rm int}$ hopping and pairing 
matrices $K_{\r}$, $\Delta_{\r}$ satisfying
\(
K_{-\r}=K_{\r}^\dagger$, $\Delta_{-\r}=-\Delta_{\r}^{\rm T},
\)
where the superscript ${\rm T}$ denotes the transpose operation. For  arrays, such
as $\hat{\Phi}^{\dagger}_{\j}$ and $\hat{\Phi}_{\j}^{\;}$, we stick to the convention that
those appearing on the left (right) of a matrix are row (column) arrays.

Since the infinite system is translation-invariant in all $D$ directions, it is customary to
introduce the volume containing the electrons by imposing 
Born-von Karman (periodic) BCs over a macroscopic volume 
commensurate with the primitive cell of the underlying Bravais lattice. If the allowed 
$\j$'s in the macroscopic volume correspond to $j_\mu=1,\dots,N_\mu$, 
then, 
\[
\hat{\Phi}_{\k}^\dagger\equiv  
\sum_{\j} \frac{e^{i\k\cdot \j}}{\sqrt{M}}\hat{\Phi}_{\j}^\dagger
\]
defines the Fourier-transformed array of creation operators of {\em real} Bloch wavevector (or crystal momentum),  
$\k \equiv \sum_{\mu=1}^{D} \frac{k_\mu}{N_\mu} \b_\mu$, with $k_\mu$ integers such that 
$\k$ lies inside the Brillouin zone. The  total number of primitive cells 
is given by $M=N_1 N_2 \dots N_D$, and $\b_\mu$ defines the reciprocal lattice vectors 
satisfying $\a_\mu \cdot \b_{\nu} = 2\pi\delta_{\mu\nu}$, with $\delta_{\mu\nu}$
representing Kronecker's delta  \cite{ashcroftmermin}. Finally, by letting $*$ denote complex conjugation, 
one can express the Hamiltonian of Eq.\,\eqref{Hamtransinv} in momentum space as
\begin{multline*}
\widehat{\bm{H}} = \frac{1}{2}\sum_\k [\hat{\Phi}_{\k}^\dagger K_{\k}\hat{\Phi}^{\;}_{\k}
+ \hat{\Phi}_{-\k}^\dagger K_{-\k}^{*}\hat{\Phi}^{\;}_{-\k}\\
+ \hat{\Phi}_{\k}^\dagger \Delta_{\k}\hat{\Phi}_{-\k}^\dagger
+ \hat{\Phi}_{\k} \Delta_{-\k}^{*}\hat{\Phi}_{-\k}],
\end{multline*}
which has a block structure in terms of the matrices
\[
K_{\k} \equiv \sum_\r e^{i\k\cdot\r}K_{\r},\quad \Delta_{\k} \equiv \sum_\r e^{i\k\cdot\r}\Delta_{\r}. 
\]

Now let us turn our attention to systems that are periodic along $(D-1)$ directions and 
terminated by two parallel hyperplanes perpendicular to the direction $\a_1$. 
We then write the allowed values of $\j$ as 
\begin{eqnarray*}
\j = j \a_1+\j_\perp,\quad j=1,\dots,N=N_1, \quad 
\j_\perp = \sum_{\mu=2}^{D}j_\mu \a_\mu .
\end{eqnarray*}
In this scenario,
each Bloch wavevector $\k$ is no longer  a good quantum number. However,
we can still block-diagonalize the Hamiltonian
in the partial basis 
\begin{equation}
\label{phikperp}
\hat{\Phi}_{\k_\perp}^\dagger = \sqrt{N}
\sum_{\j_\perp} \frac{e^{i\k_\perp\cdot \j_\perp}}{\sqrt{M}}\hat{\Phi}_{\j_\perp}^\dagger, \ \ \ 
\k_\perp =\sum_{\mu=2}^{D} \frac{k_\mu}{N_\mu} \b_\mu ,
\end{equation}
where 
$\hat{\Phi}_{\j_\perp}^\dagger$ is defined to be the array
\[
\hat{\Phi}_{\j_\perp}^\dagger \equiv 
\begin{bmatrix}
\hat{\Phi}_{\a_1+\j_\perp}^\dagger & \hat{\Phi}_{2\a_1+\j_\perp}^\dagger &
\hat{\Phi}_{N\a_1+\j_\perp}^\dagger
\end{bmatrix}.
\] 
A system with sudden termination at hyperplanes correspondig to $j=1$ and $j=N$
is modeled by open (or hardwall) BCs, in which case the Hamiltonian can be expressed 
as $\widehat{H}_{N} \equiv  \sum_{\k_\perp}\widehat{H}_{N,\k_\perp}$,
\begin{multline}
\label{HamOBC}
\widehat{H}_{N,\k_\perp} = \frac{1}{2}(\hat{\Phi}_{\k_\perp}^\dagger K_{\k_\perp}\hat{\Phi}^{\;}_{\k_\perp}
+ \hat{\Phi}_{-\k_\perp}^\dagger K_{-\k_\perp}^{*}\hat{\Phi}^{\;}_{-\k_\perp}\\
+ \hat{\Phi}_{\k_\perp}^\dagger \Delta_{\k_\perp}\hat{\Phi}_{-\k_\perp}^\dagger
+ \hat{\Phi}_{\k_\perp} \Delta_{-\k_\perp}^{*}\hat{\Phi}_{-\k_\perp}),
\end{multline}
in terms of $Nd_{\rm int} \times Nd_{\rm int}$ matrices
\begin{eqnarray*}
[K_{\k_\perp}]_{jj'} = K_{j'-j,\k_\perp} \equiv 
\sum_{\r_\perp} e^{i\k\cdot \r}K_{\r}, \quad \r = (j'-j)\a_1+\r_\perp ,
\end{eqnarray*}
and analogously defined matrices $\Delta_{\k_\perp}$. We will henceforth 
assume that the {\em range} $R$ of hopping and pairing along the $\a_1$ direction is finite.
This means that 
\begin{equation}
\label{Range}
K_{r,\k_\perp} = \Delta_{r,\k_\perp}=0, \quad \forall\ \k_\perp \ \ \text{if}\quad  |r|>R.
\end{equation}

In this paper, we are interested in BCs more general than open BCs. They are
modeled by a Hermitian many-body operator $\widehat{W}$ on Fock space which 
satisfies the following restrictions (see also Appendix \ref{finitediff}):
\begin{itemize}
\item $\widehat{W}$ has no effect beyond the ``boundary slab'', containing basis vectors
\[
\quad \j = b\a_1 +\j_\perp,\quad b=1,\dots,R,\ N-R+1,\dots,N; 
\]
\item $\widehat{W}$ is periodic along the $D-1$ directions $\a_2,\dots,\a_D$,
and has a decomposition analogous to that of $\widehat{H}_N$.
\end{itemize}
Because of the latter restriction, $\widehat{W} \equiv \sum_{\k_\perp} \widehat{W}_{\k_\perp}$ 
with
\begin{multline*} \hspace*{-0.5cm}
[\widehat{W}_{\k_\perp }]_{b b'} =
\frac{1}{2}(\hat{\Phi}_{b,\k_\perp}^\dagger W^{(K)}_{\k_\perp}\hat{\Phi}^{\;}_{b',\k_\perp}
+ \hat{\Phi}_{b,-\k_\perp}^\dagger (W^{(K)}_{-\k_\perp})^{*}\hat{\Phi}^{\;}_{b',-\k_\perp}\\
+ \hat{\Phi}_{b,\k_\perp}^\dagger W^{(\Delta)}_{\k_\perp}\hat{\Phi}_{b',-\k_\perp}^\dagger
+ \hat{\Phi}_{b,\k_\perp} (W^{(\Delta)}_{-\k_\perp})^{*}\hat{\Phi}_{b',-\k_\perp}),
\end{multline*}
where  $b,b' \in \{1,\dots,R,\ N-R+1,\dots,N\}$, 
$W_{\k_\perp}^{(K)}$ is Hermitian and $W_{\k_\perp}^{(\Delta)}$ is 
antisymmetric for each $\k_\perp$.
Then, the model Hamiltonian, with arbitrary BCs, becomes 
$$ \widehat{H}=\widehat{H}_N+\widehat{W}=\sum_{\k_\perp} \widehat{H}_{\k_\perp}, \quad 
\widehat{H}_{\k_\perp} = \widehat{H}_{N,\k_\perp} + \widehat{W}_{\k_\perp}.$$
From now on, we will focus on diagonalizing one such block 
$\widehat{H}_{\k_\perp}$, for a fixed value of $\k_\perp$.
We will investigate the interplay between \(\k_\perp\) and our diagonalization algorithm,
(and, more generally, disordered BCs), in Ref.\,[\onlinecite{PRB2}]. 

The next step consists of deriving the BdG Hamiltonian 
for this block. The conventional way \cite{blaizot} is to use the (Nambu) basis
$\hat{\Psi}_{\k_\perp}^\dagger \equiv 
\begin{bmatrix} \hat{\Phi}_{\k_\perp}^\dagger & \hat{\Phi}_{-\k_\perp} \end{bmatrix}$, 
with $\hat{\Phi}^\dagger_{\k_\perp}$ defined in Eq.\,\eqref{phikperp},
so that  $\widehat{H}_{\k_\perp}$  can be expressed in the form,  
\[
\widehat{H}_{\k_\perp} = 
\frac{1}{2}\hat{\Psi}_{\k_\perp}^\dagger\widetilde{H}_{\k_\perp}\hat{\Psi}_{\k_\perp} 
+ \frac{1}{2}\text{tr}(K_{\k_\perp}+W_{\k_\perp}^{(K)})
\]
in terms of a Hermitian matrix $\widetilde{H}_{\k_\perp}$ (note that 
the matrix $W_{\k_\perp}^{(K)}$ has entries $[W_{\k_\perp}^{(K)}]_{jj'}=0$ if any of $j,j'$
take values from the set $\{R+1,\dots,N-R\}$).
This relation leads us to a BdG Hamiltonian $\widetilde{H}_{\k_\perp} \equiv 
\widetilde{H}_{N,{\k_\perp}}+\widetilde{W}_{\k_\perp}$ with
\begin{eqnarray*}
&&\widetilde{H}_{N,{\k_\perp}} = \begin{bmatrix}K_{\k_\perp} & \Delta_{\k_\perp} \\ 
-\Delta_{-\k_\perp}^* & -K_{-\k_\perp}^*\end{bmatrix},\\ 
&&\widetilde{W}_{\k_\perp} = \begin{bmatrix}W_{\k_\perp}^{(K)} & W_{\k_\perp}^{(\Delta)} \\ 
{-W_{-\k_\perp}^{(\Delta)}}^* & {-W_{-\k_\perp}^{(K)}}^*\end{bmatrix}.
\end{eqnarray*}
The diagonalization of the BdG Hamiltonian 
\(\widetilde{H}_{\k_\perp}\) implies that of \(\widehat{H}_{\k_\perp}\), as detailed for 
example in Ref.\,[\onlinecite{blaizot}].

The $2\times2$ block-structure of $\widetilde{H}_{\k_\perp}$ emphasizes the intrinsic 
charge-conjugation symmetry  under the anti-unitary operator
$\mathcal{C} \equiv  (\mathds{1}_{Nd_{\sf int}} \tau_x)\,{\mathcal C}_{\text{cc}}$, 
i.e., \(\mathcal{C} \widetilde{H}_{\k_\perp} \mathcal{C}^{-1} = -{\widetilde{H}_{-\k_\perp}}, \)
where  $\tau_x$ is the Pauli $\sigma_x$-matrix in the Nambu basis, 
and ${\mathcal C}_{\text{cc}}$ denotes complex conjugation.
Such a block-structure, however, does {\em not} explicitly highlight the role of 
translation invariance.
For this reason, we reorder the (Nambu) basis according to \cite{abc} 
\begin{eqnarray*}
\hat{\Psi}_{\k_\perp}^\dagger \equiv \begin{bmatrix}
\hat{\Psi}_{1,\k_\perp}^\dagger  & \cdots & \hat{\Psi}_{N,\k_\perp}^\dagger
\end{bmatrix},\quad
\hat{\Psi}_{j,\k_\perp}^\dagger \equiv  \begin{bmatrix}\hat{\Phi}_{j,\k_\perp}^\dagger &
\hat{\Phi}^{\;}_{j,-\k_\perp}\end{bmatrix} , 
\end{eqnarray*}      
so that  the BdG Hamiltonian transforms to 
\[ \widetilde{H}_{\k_\perp} \mapsto H_{\k_\perp} \equiv H_{N,{\k_\perp}}+W_{\k_\perp},
\]
in terms of a {\em banded block-Toeplitz matrix} $H_{N,{\k_\perp}}= H_N$, 
with entries $[H_{N}]_{jj'}=h_{j'-j}$ along the diagonals,
and a block matrix $W_{\k_\perp}= W$, where
\[
h_r = \begin{bmatrix} 
K_{r,\k_\perp} & \Delta_{r,\k_\perp} \\ -\Delta_{r,-\k_\perp}^* & -K_{r,-\k_\perp}^*
\end{bmatrix},
\vspace{-0.5cm}\]
\[
[W]_{bb'} = \begin{bmatrix} 
W^{(K)}_{bb',\k_\perp} & W^{(\Delta)}_{bb',\k_\perp} \\ 
-(W_{bb',-\k_\perp}^{(\Delta)})^* & -(W_{bb',-\k_\perp}^{(K)})^*
\end{bmatrix}.
\]
Explicitly, in array form, we have:
\begin{eqnarray*}
\label{bandblocktoep}
H_N=
\begin{bmatrix}
h_{0}       & \dots  & h_R       & &    &0              &\cdots   & 0       \\
\vdots                &\ddots  & \      & \ddots    & &     & \ddots  & \vdots  \\ 
h_R^\dagger &       &\ \ddots  &          &\;\ \ddots & &  & 0      \\
& \ddots\\
& & & & & &  \ddots &\\
0         & &        &\;\ \ddots  &\       &\quad\ddots         &         &\;\  h_R    \\
\vdots                &\ddots  & & &\;\ \ddots  &\              &\quad\ddots   & \;\ \vdots  \\    
0                     &\cdots  &0   & &    & h_R^\dagger   &\cdots   & \;\ h_0
\end{bmatrix},
\end{eqnarray*}
\begin{eqnarray*}
W = \ \ 
\begin{bmatrix}
w^{(l)}_{11} & \dots  & w^{(l)}_{1R} &\quad  & 0 & \quad  &w_{11}& \dots  & w_{1R}\\
\vdots      &\ddots  & \vdots  &  \quad &\vdots  &\quad &\vdots         & \ddots  & \vdots  \\ 
w^{(l)}_{R1} &\dots       &w^{(l)}_{RR} & \quad& \vdots &\quad & w_{R1} &\dots   &w_{RR}      \\
 & & & \quad&\quad & & & \\
 & & & \quad&\quad & & & \\
0 & \cdots & \cdots& \quad&0  &\quad & \cdots & \cdots & 0\\
 & & & \quad &\quad & & & \\
  & & & \quad&\quad & & & \\
w^\dagger_{11}& \dots  & w^\dagger_{1R}  &\quad&\vdots & \quad &w^{(r)}_{11} & \dots  & w^{(r)}_{1R} \\
\vdots       &\ddots  &\vdots &\quad &\vdots & \quad&\vdots        &\ddots   & \vdots  \\    
w^\dagger_{R1} &\dots   &w^\dagger_{RR}  & \quad&0  &\quad  & w^{(r)}_{R1} &\dots    &w^{(r)}_{RR}  
\end{bmatrix},
\end{eqnarray*}
where we have used the notation
\begin{eqnarray}
\label{Wmatrix}
&w^{(l)}_{bb'} \equiv  W_{bb'} ,\quad
w^{(r)}_{bb'} \equiv W_{N-b+1,N-b'+1}, \\ \nonumber
& 
w_{bb'} \equiv W_{b,N-b'+1} .
\end{eqnarray}
Here, the superscript $(l)$ [or $(r)$] indicates the entries that allow 
hoppings only near the left [or right] boundary, whereas the ones 
without superscript allow hoppings from the left to the right boundary
slabs. The matrix $H=H_N+W$ is a  
{\em corner-modified}
banded block-Toeplitz matrix as defined in Ref.\,[\onlinecite{JPA}],
and is amenable to the exact solution approach described therein \cite{RemarkSymm}.

This transformed BdG Hamiltonian allows us to write the second-quantized Hamiltonian 
$\widehat{H}_{\k_\perp}$ in the form
\begin{multline*}
\widehat{H} = \frac{1}{2}\sum_{j=1}^{N}\hat{\Psi}^\dagger_{j}
h_{0}\hat{\Psi}^{\;}_{j} + \frac{1}{2}\sum_{r=1}^{R}\Big(\sum_{j=1}^{N-r}\hat{\Psi}^\dagger_{j}
h_{r}\hat{\Psi}^{\;}_{j+r} +\text{h.c.}\Big)\\
+\frac{1}{2}\sum_{b,b'}\hat{\Psi}^\dagger_b W_{bb'} \hat{\Psi}^{\;}_{b'} +
\frac{1}{2}\text{tr}(K+W^{(K)}),
\end{multline*}
where we have dropped the label $\k_\perp$ everywhere.
In particular, for one-dimensional systems ($D$=1), we recover (up to a constant) the 
class of Hamiltonians considered in Ref.\,[\onlinecite{abc}], provided that $W$ is expressible 
as 
\[
\widehat{W} = \frac{1}{2}\sum_{r=1}^{R}\sum_{b=N-R+1}^{N}\Big(\hat{\Psi}^\dagger_b \, g_r \, \hat{\Psi}^{\;}_{b+r-N} + \text{h.c.}\Big),
\]
for some $2d_{\rm int}\times 2d_{\rm int}$ matrices $g_r$.

Notice that for particle number-conserving systems ($\Delta=0=W^{(\Delta)}$), 
the single-particle Hamiltonian is just $H = K+W^{(K)}$, which is already
a corner-modified, banded block-Toeplitz matrix. In such cases, the re-ordering of
the basis is not required, and one may directly apply the diagonalization procedure
described in the following sections to $H$, with internal
blocks of dimension $d_{\rm int}$.
In order to have a uniform notation, we shall use
\begin{eqnarray*}
d \equiv \left\{
\begin{array}{lcl}
d_{\rm int}& \mbox{if} &\Delta=0 \;\,\text{(number-conserving)}\\
2d_{\rm int}& \mbox{if} &\Delta\neq 0  \;\,\text{(number-non-conserving)}
\end{array}\right. .
\end{eqnarray*}

\section{Algebraic characterization of energy eigenstates}
\label{sec:theory}

A main goal of this work is to diagonalize the single-particle Hamiltonian $H=H_N+W$,
which is a corner-modified, banded block-Toeplitz matrix. 
In this section, we investigate the structure of its energy eigenstates, 
which will culminate in a generalization of Bloch's theorem to systems described by 
such model Hamiltonians.
Our analysis will illustrate, in particular, that for non-generic parameter values, 
Hamiltonians may display a finite number of exceptional (singular) energies corresponding to 
dispersionless, {\em flat bands}. The latter represent a macroscopic number of energy
eigenstates that are localized in the bulk and, thus, are completely insensitive 
to BCs. It is remarkable that the analytic continuation of the Bloch Hamiltonian can 
still encompass this situation. We will show how to use it to construct the localized flat band energy 
eigenstates {\it directly in real space}. 

\subsection{An impurity problem as a motivating example}
\label{motivating}

Consider the simple tight-binding Hamiltonian
\begin{eqnarray*}
\widehat{H}_N = -t\sum_{j=1}^{N-1}(c^\dagger_j c_{j+1}+ c^\dagger_{j+1} c_j),
\end{eqnarray*}
defined on an open chain of $N$ (even) lattice sites with nearest-neighbor hopping strength $t$,
and lattice constant $a=1$. The corresponding single-particle Hamiltonian is
\begin{eqnarray*}
H_N = -t\sum_{j=1}^{N-1}\big(|j\rangle\langle j+1|+|j+1\rangle\langle j|\big),
\end{eqnarray*}
and breaks translation-invariance due to the presence of the boundary,
so that the crystal momentum is not a good quantum number. In fact,
for any $k \in (0,2\pi]$, the state 
$|k\rangle = \frac{1}{\sqrt{N}}\sum_{j=1}^{N}e^{ikj}|j\rangle$ (labeled by $k$) obeys 
\begin{eqnarray}
\label{metalobc}
H_N|k\rangle	= -2t\cos k |k\rangle 
+\frac{t}{\sqrt{N}}\Big(|1\rangle + e^{ik(N+1)}|N\rangle\Big),		
\end{eqnarray}
with a similar relation holding for $-k$
\begin{equation}
\label{metalobc2}
H_N|-k\rangle	= -2t\cos k |-k\rangle +
\frac{t}{\sqrt{N}}\Big(|1\rangle + e^{-ik(N+1)}|N\rangle\Big).
\end{equation}
The first term on the right-hand side of Eqs.\,\eqref{metalobc}-\eqref{metalobc2} indicates that 
$|k\rangle$ and $|-k\rangle$ ``almost'' (for large $N$) satisfy the eigenvalue relation with energy $-2t\cos k$, 
while the two terms in the brackets show that the eigenvalue relation is violated 
near the two edges of the chain. Under {\it periodic} BCs,
$-2t\cos k$ is the actual energy eigenvalue of the eigenstate $|k\rangle$ (and $|-k\rangle$), 
and $k$ is the crystal momentum, given by $k=2\pi q/N,\ q=1,\dots,N \in(0,2\pi]$\cite{ashcroftmermin}.  

Because of the identical first term $-2t\cos k$ in Eqs.\,\eqref{metalobc} 
and \eqref{metalobc2}, the states $|k\rangle$ and $|-k\rangle$ can be 
linearly combined in order to cancel off the similar-looking boundary 
contributions. For $\alpha,\beta\in\mathds{C}$, the eigenvalue relation 
\begin{eqnarray*}
H_N\Big(\alpha|k\rangle + \beta|-k\rangle\Big)=-2t\cos k \Big(\alpha|k\rangle + \beta|-k\rangle\Big),
\end{eqnarray*}
is recovered provided that the constraint
\begin{eqnarray*}
\frac{t}{\sqrt{N}}(\alpha +\beta )|1\rangle + 
\frac{t}{\sqrt{N}}(\alpha e^{ik(N+1)} + \beta e^{-ik(N+1)})|N\rangle=0
\end{eqnarray*}
is satisfied. For this to hold, the coefficients of both $|1\rangle$ and 
$|N\rangle$ must vanish, which leads to the kernel equation
\begin{eqnarray}
\label{metalbm}
t
\begin{bmatrix} 
1 & 1 \\ 
e^{ik(N+1)} & e^{-ik(N+1)}\end{bmatrix}
\begin{bmatrix} \alpha \\ \beta \end{bmatrix}\equiv
B\begin{bmatrix} \alpha \\ \beta \end{bmatrix}=0.
\end{eqnarray}
The determinant of the above ``boundary matrix'' $B$ must vanish, which happens 
if the condition $e^{i 2k(N+1)}=1$ is satisfied, that is, when 
$ k={\pi q}/{(N+1)},$ $q=1,\dots,N$.
For each of these values of \(k\), $\alpha=-\beta=1/\sqrt{2}$ provides the 
required kernel vector of the boundary matrix, with the resulting $N$ eigenvectors
\begin{eqnarray*}
|\epsilon_{k}\rangle \equiv \frac{|k\rangle - |-k\rangle}{\sqrt{2}} = 
i \sqrt{\frac{2}{N}}\sum_{j=1}^{N}\sin (k j)  |j\rangle,
\end{eqnarray*}
of  energy $\epsilon_{k} = -2t\cos k$. Notice that the allowed values of $k$ differ from the 
case of periodic BCs \cite{Mikeska}. 

Encouraged by these results, let us change the Hamiltonian
by adding an on-site potential at the edges, 
\begin{eqnarray*}
W= w(|1\rangle\langle 1|+|N\rangle\langle N|),\quad w \in {\mathbb R}, 
\end{eqnarray*}
so that the 
total single-particle Hamiltonian becomes $H=H_N+W$.
The boundary 
matrix $B$ changes to
\begin{eqnarray*}
B \equiv
\begin{bmatrix} 
t+w e^{ik} & t+w e^{-ik}\\ 
te^{ik(N+1)}+w e^{ikN} & te^{-ik(N+1)}+w e^{-ikN}\end{bmatrix}.
\end{eqnarray*}
While it is harder to predict analytically the values of $k$ 
for which it has a non-trivial kernel, 
it is interesting to examine the limit \(w\gg t\). 
Then, we can approximate the relevant kernel condition as 
\begin{eqnarray*}
B \begin{bmatrix} \alpha \\ \beta \end{bmatrix} \approx 
w
\begin{bmatrix}e^{ik} & e^{-ik}\\ 
e^{ikN} & e^{-ikN}\end{bmatrix}
\begin{bmatrix} \alpha \\ \beta \end{bmatrix}=0,
\end{eqnarray*}
showing nontrivial solutions if \(e^{i 2k(N-1)}=1\). 
There are now $(N-2)$ $k$-values yielding stationary eigenstates as before. 
The two missing eigenstates are localized at the edges, and can be taken to be 
\(|1\rangle\) and \(|N\rangle\), to leading order in \(t/w \ll 1\). These localized states 
are reminiscent of Tamm-Shockley modes \cite{Tamm, Shockley}.

In hindsight, it is natural to ask whether this approach to
diagonalization may be improved and extended to more general
Hamiltonians. The answer is Yes, and this paper provides the appropriate tools.

\subsection{The bulk-boundary system of equations}
\label{sec:BBseparation}

\begin{figure}
\includegraphics[width=8cm]{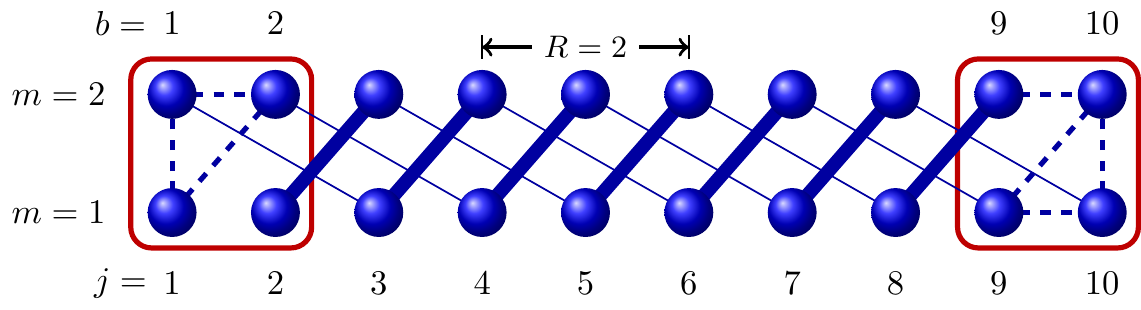}
\caption{(Color online) 
Bulk-boundary separation for a system with two fermionic modes 
per unit cell, $d=2$, and next-nearest-neighbor hopping, $R=2$. 
Each (blue) circle stands for a fermionic mode.
Thick and thin solid lines indicate two different hopping strengths
in the bulk.
Since the size of the boundary depends on the range $R$, 
the boundary comprises the first and last two 
unit cells of the chain. Dotted lines stand for arbitrary
hopping strengths at the boundary.}
\label{BBseparation}
\end{figure}

The above motivating example suggests that it may be possible to 
isolate the extent to which boundary effects prevent bulk 
eigenstates from becoming eigenstates of the actual Hamiltonian. 
Consider Eqs.\,\eqref{metalobc} and \eqref{metalobc2} in particular.
We may condense them into a single {\em relative eigenvalue 
equation,
\(
P_BH_N|\pm k\rangle= (- 2t\cos k) P_B|\pm k\rangle, \)}
in terms of the projector 
\( P_B \equiv \sum_{j=2}^{N-1}|j\rangle \langle j|.\)
The extension of this observation to the general class of 
Hamiltonians \(H=H_N+W\)  
requires only knowledge of the range \(R\) in Eq.\,\eqref{Range}. 
The block-structure of $H_N$ defines a subsystem decomposition of the 
single-particle state space \cite{abc},
\[
\mathcal{H} \cong \mathds{C}^N\otimes\mathds{C}^{d}
\equiv \mathcal{H}_L \otimes \mathcal{H}_I,
\]
where $\mathcal{H}_L$ and $\mathcal{H}_I$ are lattice and internal state spaces 
of dimensions $N$ and $d$, respectively. 
Let $\{|j\rangle,\ j=1,\dots,N\}$ and $\{|m\rangle,\ m=1,\dots,d\}$
be their respective orthonormal bases.  Define {\it bulk and boundary projectors}, 
\begin{eqnarray*}
P_B \equiv \sum_{j=R+1}^{N-R}|j\rangle\langle j|\otimes \mathds{1}_{d},\quad 
P_\partial \equiv \mathds{1}-P_B,
\end{eqnarray*}
with $\mathds{1} \equiv \mathds{1}_N\otimes \mathds{1}_{d}$ 
the identity matrix on $\mathcal{H}$, and \(\mathds{1}_{N}\), \(\mathds{1}_{d}\) 
the identity matrices on $\mathcal{H}_L$ and $\mathcal{H}_I,$
respectively (see Fig.\,\ref{BBseparation}). 
The defining property of the bulk projector is that it annihilates any 
boundary contribution $W$, that is, $P_BW=0$.
Because \(P_B+P_\partial=\mathds{1}\), the {\it bulk-boundary
system of equations},
\begin{align}
\label{bbsystem}
\left\{\begin{array}{r}
P_B H_N|\epsilon\rangle=\epsilon P_B|\epsilon\rangle, \\
(P_\partial H_N+W)|\epsilon\rangle=\epsilon P_\partial|\epsilon\rangle ,
\end{array}\right.
\end{align}
may be seen to be completely equivalent to the standard eigenvalue
equation, \(H|\epsilon\rangle=\epsilon|\epsilon\rangle\) \cite{JPA}.

This bulk-boundary separation of the eigensystem problem is advantageous 
because the bulk equation is, in a well-defined sense, {\em translation-invariant}. 
Let us define a {\it left-shift} operator  
\( T \equiv \sum_{j=1}^{N-1}|j\rangle\langle j+1| \) on the lattice space ${\cal H}_L$
(see Appendix \ref{app:opalg}). Then, one may verify that 
\begin{eqnarray}
\label{HBBT}
\hspace*{-4mm}
H_N=
\mathds{1}_{N}\otimes h_0+\sum_{r=1}^R (T^r \otimes h_r+{T^\dagger}^r \otimes h_r^\dagger) .
\end{eqnarray}
By extending $T$ infinitely on both directions, we obtain a translation-invariant {\it auxiliary Hamiltonian},  
\begin{eqnarray}
\label{HBBL}
\bm{H}  \equiv 
\mathbf{1}\otimes h_0+\sum_{r=1}^R (\bm{T}^r\otimes h_r+\bm{T}^{-r}\otimes h_r^\dagger),
\end{eqnarray}
where \(\bm{T} \equiv \sum_{j\in \mathbb{Z}}|j\rangle\langle j+1|\)
now denotes the generator of discrete translations on the (infinite-dimensional) 
vector space spanned by $\{|j\rangle\}_{j\in\mathds{Z}}$, and 
$\mathbf{1}$ the corresponding identity operator.
The subtle difference between Hamiltonians $H_N$ and $\bm{H}$
is that while $T$ is not invertible, $\bm{T}$ {\em is}, 
and in fact $\bm{T}^{-1}=\bm{T}^\dagger$. This difference is decisive in
solving the corresponding eigenvalue problems.
On the one hand, the eigenvalue equation
\(
\bm{H}|\Psi_\epsilon\rangle=\epsilon|\Psi_\epsilon\rangle
\)
is equivalent to the infinite system of linear equations
\begin{equation}
\label{goahead}
h_0|\psi_{j}\rangle + \sum_{r=1}^{R}\big(h_{r}|\psi_{j+r}\rangle+h^\dagger_{r}|\psi_{j-r}\rangle\big) = 
\epsilon|\psi_j\rangle,\quad  j\in\mathds{Z},
\end{equation}
where 
\( |\Psi_\epsilon\rangle \equiv \sum_{j\in \mathbb{Z}}|j\rangle \otimes |\psi_j\rangle.
\) 
On the other, the bulk equation \(P_BH_N|\epsilon\rangle=\epsilon P_B|\epsilon\rangle\),  
with $|\epsilon\rangle \equiv \sum_{j=1}^N|j\rangle \otimes |\psi_j\rangle$
is equivalent to Eq.\,\eqref{goahead} but restricted to the  finite domain $R < j\le N-R$.
Hence, the bulk equation is underdetermined (there are \(2R\) more 
vector variables than constraints). 
In particular, if \(|\Psi_\epsilon\rangle\) 
is an eigenstate of the infinite Hamiltonian as above, then 
\begin{eqnarray*}
\label{restriction}
|\epsilon\rangle\equiv \sum_{j=1}^N|j\rangle\langle j|\Psi_\epsilon\rangle=\bm{P}_{1,N}|\Psi_\epsilon \rangle 
\end{eqnarray*}
is a solution of the bulk equation. It is in this sense of shared 
solutions with \(\bm{H}\) that the bulk equation is, as anticipated, translation-invariant.

\subsection{Exact solution of the bulk equation}
\label{sec:bulkeq}

Let us revisit the energy eigenvalue equation, Eq.\,\eqref{goahead}. 
If the goal were to diagonalize the infinite-system Hamiltonian
\(\bm{H}\), then one should focus on finding energy eigenvectors 
associated to normalized states in Hilbert space. 
However, our model systems are of {\em finite} extent,
and we are only interested in using \(\bm{H}\) as an auxiliary operator
for finding the translation-invariant solutions of the bulk equation.
Hence, we will allow \(\bm{H}\) to act on {\em arbitrary} vector sequences 
of the form $ \Psi = \sum_{j\in\mathds{Z}}|j\rangle|\psi_j\rangle,$
possibly ``well outside" the Hilbert state space, and so we will drop 
Dirac's ket notation. From the standpoint of 
solving the bulk equation, {\em every} sequence that satisfies
\( \bm{H}\Psi=\epsilon\Psi  \)
is acceptable, so one must find them all. 
In the space of all sequences, the translation symmetry \(\bm{T}\) remains 
invertible but is no longer unitary, because the notion of adjoint operator 
is not defined. 
This is important, 
because it means that translations need not have their eigenvalues on 
the unit circle, or be diagonalizable. Nonetheless, \([\bm{T},\bm{H}]=0\),
and so both features have interesting physical consequences {\it for
finite systems}. 

We will refer to the space of solutions of the bulk equation as the 
{\it bulk solution space} and denote it by
\begin{eqnarray*}
\mathcal{M}_{1,N}(\epsilon) \equiv {\rm Ker}\, P_B(H_N-\epsilon\mathds{1}_d),
\end{eqnarray*}
for any {\it fixed} energy $\epsilon$. Let
\(
\mathcal{M}_{-\infty,\infty}(\epsilon) \equiv {\rm Ker}\ (\bm{H}-\epsilon {\bf 1})
\)
denote the space of eigenvectors of \(\bm{H}\) of energy 
\(\epsilon\) within the space of all sequences. In terms
of these spaces, our arguments in Sec.\,\ref{sec:BBseparation}
establish the relation 
\begin{eqnarray}
\label{softy}
\bm{P}_{1,N}\mathcal{M}_{-\infty,\infty}\subseteq \mathcal{M}_{1,N},
\end{eqnarray}
where we dropped the argument $\epsilon$.
Translation invariance is equivalent to the properties 
\(\bm{T}\mathcal{M}_{-\infty,\infty}\subseteq\mathcal{M}_{-\infty,\infty}\)
and \(\bm{T}^{-1}\mathcal{M}_{-\infty,\infty}\subseteq\mathcal{M}_{-\infty,\infty}\)
\cite{footnote2}.
If the matrix $h_R$ is invertible, Eq.\,\eqref{softy} becomes 
\(
\bm{P}_{1,N}\mathcal{M}_{-\infty,\infty}=\mathcal{M}_{1,N}
\) \cite{JPA}.

Since $\bm{T}$ commutes 
with $\bm{T}^{-1}$, the generator of translations to the right, these two
symmetries share eigenvectors of the form 
\(
\Phi_{z,1}|u\rangle\equiv \sum_{j\in\mathds{Z}}z^j|j\rangle|u\rangle, 
\)
with \(z\) an arbitrary non-zero complex number and $|u\rangle$ 
any internal state: there are \(d\) linearly independent eigenvectors of 
translations for each \(z\neq 0\). 
As a simple but important consequence of the
identities  
\begin{align*}
\label{phiz}
\bm{T}\Phi_{z,1}|u\rangle = z\Phi_{z,1}|u\rangle,\quad 
\bm{T}^{-1}\Phi_{z,1}|u\rangle = z^{-1}\Phi_{z,1}|u\rangle,
\end{align*} 
one finds that   
\begin{eqnarray}
\label{redbulkHamaction}
\bm{H}\Phi_{z,1} \, |u\rangle = 
\Phi_{z,1} H(z)|u\rangle,
\end{eqnarray}
where the linear operator
\[
H(z)=h_0+\sum_{r=1}^{R}(z^r  h_r+z^{-r}h_r^\dagger) ,
\]
acts on the internal space ${\cal H}_I$ only. This \(H(z)\) 
is precisely the {\it reduced bulk Hamiltonian} $h_B(z)$ of Ref.\,[\onlinecite{abc}],
obtained here by way of a slightly different argument.
Since \(H_k=H(z=e^{ik})\) 
is the usual Bloch Hamiltonian of a one-dimensional system with 
Born-von-Karman BCs, $H(z)$ is the analytic 
continuation of \(H_k\) off the Brillouin zone.

One can similarly continue the energy dispersion relation off the 
Brillouin zone, by relating \(\epsilon\) to \(z\) via
\begin{eqnarray}
\label{chareq}
\det (H(z)-\epsilon\mathds{1}_{d})=0.
\end{eqnarray}
In practice, it is advantageous to use the polynomial
\begin{eqnarray}
\label{charz}
P(\epsilon,z) \equiv z^{dR}\det (H(z)-\epsilon\mathds{1}_{d}).
\end{eqnarray}
We will say that \(\epsilon\) is {\em regular} if \(P(\epsilon,z)\) is
not the zero polynomial, and {\em singular} otherwise. That is, 
\(P(\epsilon,z)=0\) identically for all $z$ if \(\epsilon\) is singular. 
Such a (slight) abuse of language\cite{JPA} is permitted since we are 
interested in varying $\epsilon$ for a fixed Hamiltonian.
For any given Hamiltonian of  {\it finite range} $R$, there 
are at most a finite number of singular energies. Physically, singular 
energies correspond to flat bands, as one can see by restriction to the 
Brillouin zone. We can now state a first useful result, whose formal 
proof follows from the general arguments in Ref.\,[\onlinecite{JPA}]:

\medskip

{\bf Theorem 1.} {\em 
If $\epsilon$ is regular, the number of independent solutions of the bulk equation 
is \(\dim \mathcal{M}_{1,N} (\epsilon) =2Rd\), for any system size \(N>2R\).}

\medskip

This result ties well with the  physical meaning of the number 
\(2Rd = \text{dim(Range} \,P_\partial)\) as counting 
the total number of degrees of freedom on the boundary, which is equal 
to the dimension of the boundary subspace. 
The condition \(N>2R\) implies that the system is big enough to contain
at least one site in the bulk. 

\subsubsection{Extended-support bulk solutions at regular energies}
\label{sec:extended}

The solutions of the bulk equation that are inherited from \(\bm{H}\) 
have non-vanishing support on the full lattice space ${\cal H}_L$,
and are labeled by the eigenvalues of 
\(\T\),  possibly together with a second ``quantum number''
that appears because \(\T\) is not unitary on the space of all sequences. 
For any $z \ne 0$, if $|u\rangle$ satisfies the eigenvalue 
equation 
\(
H(z)|u\rangle = \epsilon|u\rangle, \) 
then Eq.\,\eqref{redbulkHamaction} implies that $\Phi_{z,1}|u\rangle$
is an eigenvector of $\bm{H}$ with eigenvalue $\epsilon$. 
In order to be more systematic, let $\{z_\ell\}_{\ell=1}^n$ denote the $n$
{\it distinct} non-zero 
roots of Eq.\,\eqref{charz}, and $\{s_\ell\}_{\ell=1}^n$
their respective multiplicities.
For generic values of $\epsilon$, $H(z_\ell)$ has exactly
$s_\ell$ eigenvectors 
$\{|u_{\ell s}\rangle\}_{s=1}^{s_\ell}$ in ${\cal H}_I$, satisfying 
\begin{eqnarray*}
H(z_\ell)|u_{\ell s}\rangle = \epsilon|u_{\ell s}\rangle,\quad s=1,\dots,s_\ell.
\end{eqnarray*}
Since 
\( \bm{H}\Phi_{z_\ell,1}|u_{\ell s}\rangle=\epsilon\Phi_{z_\ell,1}|u_{\ell s}\rangle,
\) the states
\begin{eqnarray}
\bm{P}_{1,N}\Phi_{z_\ell,1}|u_{\ell s}\rangle = \sum_{j=1}^Nz_\ell^j|j\rangle|u_{\ell s}\rangle 
\equiv |z_\ell,1\rangle |u_{\ell s}\rangle 
\end{eqnarray}
\noindent
are solutions of the bulk equation. Intuitively, these 
states are ``eigenstates of the Hamiltonian up to BCs." 

For a few isolated values of $\epsilon$, 
$H(z_\ell)$ can have less than $s_\ell$ eigenvectors.
However, the number of eigenvectors of $\bm{H}$ is still $s_\ell$ \cite{JPA}, as we 
illustrate here by example. Suppose for concreteness that
\[ \bm{H}-\epsilon\bm{1} = -\frac{t}{2}(\T+\T^{-1})-\epsilon\bm{1}=
-\frac{t}{2}\T^{-1}\prod_{\ell=1}^2(\T-z_\ell). 
\]
Since \(R=1\) and \(d=1\), we expect two eigenvectors
for each value of \(\epsilon\). One concludes that the 
eigenspace of energy \(\epsilon\) is spanned by the 
sequences $\Phi_{z_\ell,1},\ \ell=1,2,$ if \(z_1\neq z_2\).
But, if \(\epsilon=\pm t\), then \(z_1=z_2=\mp 1\), and  
\[
\bm{H}\mp t \bm{1} = -\frac{t}{2}\, \T^{-1}(\T-z_1)^2.
\] 
How can one get two independent solutions in this case? The
answer is that, in addition to \(\Phi_{z_1,1}\), the factor
$(\bm{T}-z_1)^2$ contributes another sequence to the kernel 
of \(\bm{H}-\epsilon\bm{1}\), namely, 
\(
\Phi_{z_1,2} = \sum_{j\in\mathds{Z}}jz_1^{j-1}|j\rangle. \)
There are two eigenvectors in total, even though there is only
one root. 

Returning to the general case, the sequences \cite{JPA, normalization}
\begin{align}
\label{geneig}
\Phi_{z,v} &= 
\frac{1}{(v-1)!}\partial_z^{v-1}\Phi_{z,1} = 
\sum_{j\in\mathds{Z}}\frac{j^{(v-1)}}{(v-1)!}z^{j-v+1}|j\rangle,\\
j^{(v)} &\equiv j(j-1)\dots(j-v+1),\quad j^{(0)}\equiv 1,\nonumber
\end{align}
span the kernel of $(\bm{T}-z)^s$ for \(v=1,\dots,s\). In other
words, $\Phi_{z,v}$ is a {\em generalized eigenvector} 
of the translational symmetry $\bm{T}$ of rank $v$ with eigenvalue $z$. 
We refer to eigenvectors with $v>1$ as the {\it power-law solutions}
of the bulk equation (solutions with a power-law prefactor). They exist 
because translations are not diagonalizable in the full space of sequences 
(as opposed to the Hilbert space of square-summable sequences), leading to the new
quantum number \(v\). 

The power-law solutions of the bulk equation may be found from the 
action of $\bm{H}$ on the generalized eigenvectors of $\bm{T}$. 
For arbitrary internal state \(|u_x\rangle\), we have:
\begin{equation}
\label{whyhb}
\bm{H}\Phi_{z,x}|u_x\rangle=
\frac{1}{(x-1)!}\partial_z^{x-1}\Phi_{z,1}H(z)|u_x\rangle .
\end{equation}
Then one can show from Eqs.\,\eqref{geneig} and \eqref{whyhb} 
that the action of $\bm{H}$ on the vector sequence 
$\Psi = \sum_{x=1}^{v}\Phi_{z,x}|u_x\rangle$, 
where $\{|u_x\rangle\}$ are arbitrary internal states, is given by
\begin{eqnarray}
\label{genhbaction}
\bm{H}\Psi & = & \sum_{x=1}^{v} \sum_{x'=1}^{v} \Phi_{z,x}[H_v(z)]_{xx'}|u_{x'}\rangle .
\end{eqnarray}
Here, $H_v(z)$ is an {\it upper triangular block-Toeplitz} matrix with 
non-trivial blocks 
\begin{equation}
\label{genhb}
[H_v(z)]_{xx'} \equiv 
\frac{1}{(x'-x)!}\partial_z^{x'-x}H(z),\quad 1\le x\le x'\le v.
\end{equation}
In matrix form, by letting $H^{(x)}\equiv \partial_z^{x}H(z)$, we have 
\[
H_v(z)=
\begin{bmatrix}
\ \ H^{(0)} & H^{(1)}     & \frac{1}{2} H^{(2)}  & \cdots       &  \frac{1}{(v-1)!} H^{(v-1)}          \\
 0          &\! \ddots    &\! \! \ddots          &\! \! \ddots  & \vdots             \\
\vdots      &\! \ddots    &\! \ddots             &\! \! \ddots  & \frac{1}{2}H^{(2)} \\
\vdots      &             &\! \ddots             &\! \ddots     & H^{(1)}            \\
0           &  \cdots     & \cdots              &0             & H^{(0)}            
\end{bmatrix}.
\]
We refer to $H_v(z)$ as the {\it generalized 
reduced bulk Hamiltonian of order $v$}. Notice that $H_1(z)=H(z)$.
In the partial basis 
\begin{eqnarray}
\label{basisphi}
\Phi_{z} = 
\begin{bmatrix}
\Phi_{z,1} & \dots & \Phi_{z,v}
\end{bmatrix},
\end{eqnarray}
organized as a row vector, the entries of 
$ |u\rangle = \begin{bmatrix}|u_1\rangle & \dots & |u_v\rangle\end{bmatrix}^{\rm T}
$ are the vector-valued coordinates of $\Psi$, 
\( \Psi=\Phi_{z}|u\rangle =\sum_{x=1}^{v}\Phi_{z,x}|u_x\rangle. \)
Then, Eq.\,\eqref{genhbaction} can be rewritten as
\begin{eqnarray*}
\bm{H} \Phi_{z}|u\rangle = 
\Phi_{z} H_v(z)|u\rangle.
\end{eqnarray*}
Now it becomes clear that for $\Psi$ to be an eigenvector of $\bm{H}$, 
the required condition is $H_v(z)|u\rangle = \epsilon|u\rangle$, 
which is analogous to the condition derived for the generic case $v=1$.
If a root $z_\ell$ of Eq.\,\eqref{chareq} has multiplicity $s_\ell$, then 
$\bm{H}$ has precisely $s_\ell$
linearly independent eigenvectors corresponding to $z_\ell$.
This provides a characterization of the eigenstates of $\bm{H}$,
which may be regarded as extending Bloch's theorem to 
\(\bm{H}\) viewed as a linear transformation on the space of all 
vector-valued sequences, and whose rigorous justification follows 
from Ref.\,[\onlinecite{JPA}]:

\medskip

\noindent{\bf Theorem 2.} {\em 
For fixed, regular \(\epsilon\), let $\{z_\ell\}_{\ell=1}^n$
denote the distinct non-zero roots of Eq.\,\eqref{chareq},  
with respective multiplicities $\{s_\ell\}_{\ell=1}^{n}$. Then, the 
eigenspace of $\bm{H}$ of energy \(\epsilon\) is a direct sum of 
$n$ vector spaces spanned by generalized eigenstates of $\bm{T}$ 
of the form 
\begin{eqnarray*}
\label{megusta}
\Psi_{\ell s}=
\Phi_{z_\ell}|u_{\ell s}\rangle
=\sum_{v=1}^{s_\ell}\Phi_{z_\ell, v}|u_{\ell s v}\rangle,\quad s=1,\dots,s_\ell,
\end{eqnarray*}
where the linearly independent vectors 
$\{|u_{\ell s}\rangle\}_{s=1}^{s_\ell}$ 
are chosen in such a way that 
\( H_{s_\ell}(z_\ell)|u_{\ell s}\rangle = \epsilon|u_{\ell s}\rangle \), and 
$|u_{\ell s}\rangle = \begin{bmatrix}|u_{\ell s 1}\rangle & \dots & |u_{\ell s s_\ell}\rangle\end{bmatrix}^{\rm T}$.}

\medskip

Once the eigenvectors of $\bm{H}$ are calculated, the bulk 
solutions of extended support are readily obtained by projection. 
Let, for $v \geq 1$,  
\[ |z,v\rangle \equiv  \bm{P}_{1,N}\Phi_{z,v}=\sum_{j=1}^N \frac{ j^{(v-1)} }{(v-1)!} 
z^{j-v+1}|j\rangle
\]
be the projections of generalized eigenvectors of $\bm{T}$. Then
\begin{eqnarray*}
\mathcal{B}_{\rm ext} \equiv \{|\psi_{\ell s}\rangle,\ s=1,\dots,s_\ell, \ \ell=1,\dots,n\}
\end{eqnarray*} 
describes a basis of the translation-invariant solutions of the bulk 
equation, where 
\begin{eqnarray}
\label{psils}
|\psi_{\ell s}\rangle = \sum_{v=1}^{s_\ell}|z_{\ell},v\rangle|u_{\ell s v}\rangle\quad \forall \ell,s.
\end{eqnarray}

{\em Remark.---} The bulk equation bears power-law solutions only at a 
few isolated values of $\epsilon$ \cite{Trench85}. However,  {\em linear 
combinations} of $v=1$ solutions show power-law-like behavior, as soon as 
two or more of the roots of 
Eq.\,\eqref{chareq} are sufficiently close to each other. 
Suppose, for instance, that for some value 
of energy $\epsilon$, two of the roots of Eq.\,\eqref{chareq} 
coincide at $z_*$. For energy differing from $\epsilon$  
by a small amount $\delta \epsilon$, the double root $z_*$
bifurcates into two roots slightly away from each other, with values 
$z_*\pm \delta z$. The relevant bulk solution space is spanned by 
\begin{eqnarray*}
|z_*+\delta z,1\rangle + |z_*+\delta z,1\rangle &\approx& 2|z_*,1\rangle, \nonumber\\
|z_*+\delta z,1\rangle - |z_*+\delta z,1\rangle &\approx& 2(\delta z/z_*)|z_*,2\rangle, 
\end{eqnarray*}
showing that the second vector has indeed a close resemblance to the power-law solution 
$|z_*,2\rangle$.
Similar considerations apply if $d>1$, as it is typically  
the case in physical applications. 
Assuming that the relevant bulk solutions
at energy $\epsilon+\delta\epsilon$ are described by analytic vector functions $|\psi(z_*+\delta z)\rangle$
and $|\psi(z_*-\delta z)\rangle$, then, from the above analysis, it is clear that for energy $\epsilon$,
the power-law bulk solution will be proportional to
\begin{eqnarray}
\label{shortcut}
\lim_{\delta z\rightarrow 0}(|\psi(z_*+\delta z)\rangle-|\psi(z_*-\delta z)\rangle) \propto\partial_z |\psi(z_*)\rangle .
\;\;\;
\end{eqnarray}
We will make use of this observation for the calculation of power-law solutions
in Sec.\,\ref{topocomb}.

\subsubsection{Emergent solutions at regular energies}
\label{sec:emergent}

While the extended solutions of the bulk equation correspond 
to the nonzero roots of Eq.\,(\ref{chareq}), the 
polynomial $P(\epsilon,z)$ defined in Eq.\,\eqref{charz} may also 
include $z_0=0$ as a root of multiplicity $s_0$, that is, we 
may generally write
\begin{eqnarray*} 
P(\epsilon,z)=z^{dR}\det(H(z)-\epsilon\mathds{1}_d) 
\equiv c\prod_{\ell=0}^n(z-z_\ell)^{s_{\ell}}, \quad c\ne 0.
\end{eqnarray*}
However, \(|z=0\rangle|u\rangle=0\)
does not describe any state of the system.  This observation 
suggests that the extended solutions of the bulk equation may 
fail to account for all \(2Rd\) solutions we expect for regular 
$\epsilon$. That this is indeed the case follows from a known 
result in the theory of matrix polynomials \cite{teran15},  
implying that  \( 2Rd=2s_0+\sum_{\ell=1}^{n} s_\ell \) for 
matrix polynomials associated to Hermitian Toeplitz matrices \cite{JPA}. 
Hence, the number of solutions of the bulk equation of the 
form given in Eq.\,\eqref{psils} is 
\begin{equation} 
\sum_{\ell=1}^ns_\ell = 2Rd- 2s_0.
\label{s0}
\end{equation}
We call the missing \(2s_0\) solutions of the bulk equation 
{\it emergent}, because they are no longer controlled by \(\bm{H}\) 
and (nonunitary) translation symmetry, but rather they appear 
only because of the truncation of the infinite lattice down to 
a finite one, and only if \(\det h_R=0\) \cite{JPA}. Emergent 
solutions are a direct, albeit non-generic, manifestation 
of translation-symmetry-breaking; nonetheless, remarkably, 
they can also be determined by the analytic continuation of 
the Bloch Hamiltonian, in a precise sense. 

While full technical detail is provided in Appendix  \ref{app:ext2eme}, 
the key to computing the emergent solutions is to relate the 
problem of solving the bulk equation to a {\em half-infinite} 
Hamiltonian, rather than the doubly-infinite \(\bm{H}\) 
we have exploited thus far. 
Let us define the {\em 
unilateral shifts} 
\begin{eqnarray*}
\bm{T}_{{-}} =\sum_{j=1}^{\infty}|j\rangle\langle j+1|,\quad
\bm{T}_{{-}}^\star =\sum_{j=1}^{\infty}|j+1\rangle\langle j|.
\end{eqnarray*}
The Hamiltonian
\begin{eqnarray}
\label{halfinfham}
\bm{H}_- \equiv \bm{1}_-\otimes h_0+
\sum_{r=1}^R(\bm{T}_-^r\otimes h_r+\bm{T}_-^{\star\, r}\otimes h_r^\dagger)
\end{eqnarray}
is then the half-infinite counterpart of \(\bm{H}\). The corresponding 
half-infinite bulk projector is 
\begin{eqnarray*}
\bm{P}_B^- \equiv \sum_{j=R+1}^\infty|j\rangle\langle j|=\bm{T}_-^{\star\, R}\bm{T}_-^R.
\end{eqnarray*}
Suppose there is a state \(\Upsilon^{-}\), that solves the equation
$\bm{P}^-_B(\bm{H}_--\epsilon\bm{1}_-)\Upsilon^{-}=0.$
Then one can check that \(|\psi\rangle=\bm{P}_{1,N}\Upsilon^{-}\)
is a solution of the bulk equation, Eq.\,\eqref{bbsystem}. 
Clearly, some of the bulk solutions we arrive at in this way using 
\(\bm{H}_-\) will coincide with those obtained from \(\bm{H}\).
These are precisely the extended 
solutions we already computed in Sec. \ref{sec:extended}.
In contrast, the emergent solutions are obtained {\em only} from \(\bm{H}_-\).

Since \(\bm{T}_-\bm{T}_-^\star=\bm{1}_-\), we may write 
\( \bm{P}^-_B(\bm{H}_--\epsilon\bm{1}_-)=\bm{T}_-^{\star R}K^-(\epsilon,\bm{T}_-), 
\)
in terms of the matrix polynomial 
\begin{align}
&K^-(\epsilon,z) \equiv z^R (H(z)-\epsilon\mathds{1}_d).
\quad 
\label{kminus}
\end{align}
Half of the emergent solutions, namely, the ones localized on the left
edge, are determined by the kernel of \(K_{s_0}^-(\epsilon,z_0=0) \equiv K^- (\epsilon)\),
with \([K^-_v(\epsilon,z)]_{xx'}\) constructed as in Eq.\,\eqref{genhb}.  
Explicitly, such a matrix,  which was obtained by different means 
in Ref.\,[\onlinecite{JPA}], takes the form 
\begin{eqnarray} 
\label{kmatrix}
K^{-}(\epsilon)\! \equiv\! \! & \\
&\hspace*{-1cm}    
\begin{bmatrix}
h^\dagger_R & \cdots & h_0-\epsilon\mathds{1}_d & \cdots & h_R & & 0 &\cdots& 0 \\
	 &	 \hskip -0.5em\ddots & & \hskip -1em\ddots & & \ddots  & & \ddots &\vdots  \\
	 &&&&&&\ddots&& 0\\
   & &  &\hskip -4em\ddots  & & \ \ \ddots  &&\ddots &  \\
   &&&&&&&&\ h_R\\
   & & & &\ddots  & && \ddots& \ \vdots \\
   & & & & & && &h_0-\epsilon\mathds{1}_d \\
0    & & & & & &\hskip -1em\ddots & & \ \vdots\\
\vdots    &\ \ddots & &  &  & && &\\
0  &\cdots &\hskip -2em 0 &  &  & && &\ \ h^\dagger_R 
\end{bmatrix} \nonumber,
\end{eqnarray}  
for systems with fairly large \(s_0>2R+1\). Let \(\{ |u^-_s\rangle \}_{s=1}^{s_0}\) 
denote a basis of the kernel of \(K^-(\epsilon)\), with
\[
|u^-_s\rangle = \begin{bmatrix}
|u^-_{s1}\rangle & |u^-_{s2}\rangle & \dots & |u^-_{ss_{0}}\rangle
\end{bmatrix}^{\rm T}.
\]
Then,
\begin{equation}
|\psi_{s}^{-}\rangle=\sum_{j=1}^{s_0}|j\rangle|u^-_{sj}\rangle ,
\quad s=1,\dots,s_0, 
\label{basisj-}
\end{equation} 
are the emergent solutions with support on the first $s_0$ lattice 
sites, with $s_0$ obeying Eq.\,\eqref{s0}.

We are still missing \(s_0\) emergent solutions for the right edge.
They may be  constructed from the kernel of the lower-triangular block  
matrix
\( K^+ (\epsilon) \equiv [K^{-} (\epsilon)]^\dagger =[K^-_{s_0}(\epsilon, z_0=0)]^\dagger.\)
Let \( \{ |u^+_s\rangle\}_{s=1}^{s_0}\) denote a basis of the kernel
of \(K^+(\epsilon)\), with
\[
|u^+_s\rangle = \begin{bmatrix}
|u^+_{s1}\rangle & |u^+_{s2}\rangle & \dots & |u^+_{ss_{0}}\rangle
\end{bmatrix}^{\rm T}.
\]
Then,
\begin{eqnarray}
|\psi^{+}_s\rangle = \sum_{j=1}^{s_0}|N-s_0+j\rangle|u^{+}_{sj}\rangle 
\quad s=1,\dots,s_0, 
\label{basisj+}
\end{eqnarray} 
are the emergent bulk solutions associated to the right edge, 
supported on the lattice sites $N-s_0+1, \ldots, N$. 
Again, for
mathematical justifications, see Appendix\, \ref{app:ext2eme}.

In what follows, we shall denote the spaces spanned by left- and right- localized emergent 
bulk solutions by $\mathcal{F}_1^{-}$ and $\mathcal{F}_N^{+}$, 
and their bases by 
$\mathcal{B}^{-}\equiv \{|\psi_s^{-}\rangle\}_{s=1}^{s_0} $  
and $\mathcal{B}^{+}\equiv \{|\psi_s^{+}\rangle\}_{s=1}^{s_0}$, respectively.

\subsubsection{Bulk-localized states at singular energies}
\label{flatbandid}

If \(h_R\) is {\em not} invertible, there can be at most a {\em finite} 
number of singular energy values (usually referred to as
 flat bands), leading to bulk-localized solutions:
these solutions are finitely-supported and appear everywhere in the bulk. Hence, 
a singular energy {\em cannot} be excluded from the physical spectrum of 
a finite system by way of BCs. In contrast, emergent solutions are also 
finitely-supported but necessarily ``anchored'' to the edges 
(and only appearing for regular values of \(\epsilon\)).

Recall that if \(\epsilon\) is singular, then  
\(\det(H(z)-\epsilon\mathds{1})=0\) for any $z$. Thus,  
there exists an analytic vector function,
\begin{equation}
|v(z)\rangle \equiv 
\sum_{\delta=0}^{\delta_0}z^{-\delta}|v_\delta\rangle,\quad \delta_0=(d-1)2Rd, 
\label{delta0}
\end{equation}
satisfying $H(z)|v(z)\rangle = \epsilon|v(z)\rangle$ for all \(z\).
To obtain $|v(z)\rangle$, one can construct the {\em adjugate matrix}
of $(H(z)-\epsilon\mathds{1}_{d})$. (Recall that the adjugate matrix
\(\text{adj}(M)\) associated to a square matrix \(M\) is constructed 
out of the signed minors of \(M\) and satisfies \(\text{adj}(M)M=
\det(M)\mathds{1}\).) Hence,
\[
(H(z)-\epsilon\mathds{1}_{d})\text{adj}(H(z)-\epsilon\mathds{1}_{d})=
\det(H(z)-\epsilon\mathds{1}_{d})\mathds{1}_{d}=0,
\]
and so one can use {\em any} of the non-zero
columns of $\text{adj}(H(z)-\epsilon\mathds{1}_{d})$,  
suitably pre-multiplied by a power of $z$, for the vector 
polynomial $|v(z)\rangle$. By matching powers of $z$, this 
equation becomes 
\begin{eqnarray}
\label{localfoundation}
\begin{bmatrix}
h_{R} & & 0 & \cdots & 0\\
h_{R-1} & h_{R} & & \ddots& \vdots\\
\vdots & \ddots & \ddots & & 0\\
\vdots & \ddots & \ddots & \ddots&\\
h_{R}^\dagger & \ddots& \ddots & \ddots & h_{R}\\
& \ddots & \ddots & \ddots  &\vdots\\
0 & & \ddots & \ddots & \vdots\\
\vdots & \ddots & \phantom{\Big(}& \ddots & h_{R-1}^\dagger\\
0 & \cdots & 0 & & h_{R}^\dagger
\end{bmatrix}
\begin{bmatrix}
|v_{0}\rangle\\
|v_{1}\rangle\\
\vdots\\
|v_{\delta_0}\rangle
\end{bmatrix}=0.
\end{eqnarray}

The idea now is to use the linearly independent solutions of 
Eq.\,\eqref{localfoundation} to construct finite-support 
solutions of the bulk equation. Let us denote such solutions by
\(
|v_\mu\rangle \equiv \begin{bmatrix}
|v_{\mu 0}\rangle&
|v_{\mu 1}\rangle&
\dots & 
|v_{\mu \delta_0}\rangle
\end{bmatrix}^{\rm T} \),  
for \(\mu=1,\dots, \mu_0\).  
One can check directly that the finitely-supported sequences 
\[
\Psi_{j \mu} \equiv 
\sum_{\delta=0}^{\delta_0}|j+\delta\rangle|v_{\mu \delta}\rangle,\quad j\in\mathds{Z},
\quad \mu=1,\dots,\mu_0,
\]
all satisfy \((\bm{H}-\epsilon\bm{1})\Psi_{js}=0\) 
because \(|v_\mu\rangle\) obeys Eq.\,\eqref{localfoundation}.
Hence, the states $\bm{P}_{1,N}\Psi_{j\mu}$ provide finitely-supported 
solutions of the bulk equation. In addition, as long as $2R<j<N-2R-\delta_0$, 
the boundary equation is also satisfied trivially, and so {\em all}
such states become eigenvectors of $H_N+W$ with the
singular energy \(\epsilon\). This is why singular energies,
if present for the infinite system, are necessarily also part of the spectrum of the 
finite system and display macroscopic degeneracy of order $\mathcal{O}(N)$.

Let us further remark that the sequences \(\Psi_{j\mu}\)
and associated solutions of the bulk equation need {\em not} be 
linearly independent. To obtain a complete (rather than
overcomplete), set of solutions for flat bands, one would require 
a technical tool, the {\em Smith normal form} \cite{Gohberg}, 
which is beyond the scope of this paper. See Ref.\,[\onlinecite{JPA}] 
for details.

\subsection{The boundary matrix}
\label{sec:boundeq}

For regular energies, the bulk solutions determine a subspace of 
the full Hilbert space [Theorem 1], whose dimension \(2Rd\ll dN\) for typical 
applications. While not all bulk solutions are eigenstates of the Hamiltonian 
$H=H_N+W$, the actual eigenstates must necessarily appear as bulk solutions. 
Hence, the bulk-boundary separation in Eqs.\,\eqref{bbsystem}, and, in particular, 
the bulk equation, identifies by way of a translational symmetry 
analysis a {\em small} search subspace. In order to find the energy eigenstates 
efficiently, one must solve the boundary equation on this search subspace.
Since the boundary equation is linear, its restriction to the
space of bulk solutions can be represented by a matrix, the {\it boundary matrix} 
\cite{abc}. The latter is a square matrix that combines our basis of bulk solutions 
with the relevant BCs. 

Let ${\cal B}\equiv {\cal B}_{\text{ext}} \cup \mathcal{B}^{-} \cup \mathcal{B}^{+}$ 
be a basis for $\mathcal{M}_{1,N}$. Then, building on the previous section, the Ansatz state
\begin{widetext}
\begin{align}
\label{ansatz}
|\epsilon,{\bm \alpha}\rangle  \equiv |\Psi_{\mathcal{B}}\rangle\bm{\alpha}
=\sum_{\ell=1}^{n}\sum_{s=1}^{s_\ell}\alpha_{\ell s}|\psi_{\ell s}\rangle
+\sum_{s=1}^{s_0}\alpha^{+}_{s}|\psi^{+}_{s}\rangle  
+\sum_{s=1}^{s_0}\alpha^{-}_{s}|\psi^{-}_{s}\rangle\ , 
\end{align}
represents the solutions of the bulk equation parametrized by the 
\(2Rd\) amplitudes \(\bm{\alpha}\), where 
\noindent
\begin{eqnarray}
\label{bulkbasis}
\label{boldalpha}
\hspace*{-6mm}
\begin{array}{l}
\ \ \bm{\alpha} \equiv 
\begin{bmatrix} 
\alpha_{11} & \cdots & \alpha_{ns_n} & \alpha_1^{+} & \cdots & \alpha_{s_0}^{+} & 
\alpha_1^{-} & \cdots & \alpha_{s_0}^{-}\end{bmatrix}^{\rm T}, \\ 
|\Psi_{\mathcal{B}}\rangle \equiv 
\begin{bmatrix} 
|\psi_{11}\rangle & 
\cdots & 
|\psi_{ns_n}\rangle & 
|\psi_1^{+}\rangle & 
\cdots & 
|\psi_{s_0}^{+}\rangle & 
|\psi_1^{-}\rangle & 
\cdots & 
|\psi_{s_0}^{-}\rangle 
\end{bmatrix}\!.  
\end{array}
\end{eqnarray}
Moreover, let as before \(b=1,\dots,R,N-R+1,\dots,N\) label the 
boundary sites. Then,
\begin{equation}
\label{BBfinal}
P_B(H-\epsilon\mathds{1})|\epsilon,\bm{\alpha}\rangle=0 \quad \text{and} \quad
P_\partial(H-\epsilon\mathds{1})|\epsilon,\bm{\alpha}\rangle=
\sum_b|b\rangle\langle b|(H_N+W-\epsilon\mathds{1})|\Psi_{\mathcal{B}}\rangle\bm{\alpha}.
\end{equation}
In particular, the boundary equation is equivalent to the requirement that 
\(
\langle b|(H_N+W-\epsilon\mathds{1})|\Psi_{\mathcal{B}}\rangle\bm{\alpha}=0
\)
for all boundary sites. Since  
\(\langle b|(H_N+W-\epsilon\mathds{1})|\Psi_{\mathcal{B}}\rangle \equiv 
\langle b| H_\epsilon |\Psi_{\mathcal{B}}\rangle \) 
denotes a row array of internal states, it is possible to organize 
these arrays into the boundary matrix  
\begin{eqnarray}
\label{Boundary matrix}
B (\epsilon) & \equiv 
\begin{bmatrix}
\ \ \ \ \ \ \ \ \quad \langle 1|H_\epsilon|\psi_{1 1}\rangle &
\cdots& 
\ \ \ \ \ \ \ \ \quad\langle 1|H_\epsilon|\psi_{n s_n}\rangle& 
\ \ \ \ \ \ \ \ \ \quad\langle 1|H_\epsilon|\psi_{1}^+\rangle&
\cdots&
\ \ \ \ \ \ \ \ \quad\langle 1|H_\epsilon|\psi_{s_0}^{-}\rangle \\
\ \ \ \ \ \ \ \ \quad \vdots&  
 &
\ \ \ \   \quad \vdots &
\ \ \ \ \ \ \ \ \quad\vdots &  
 &  
\ \ \ \ \ \  \quad\vdots\\
\ \ \ \ \ \ \ \ \quad \langle R|H_\epsilon|\psi_{1 1}\rangle &
\cdots& 
\ \ \ \ \ \ \ \ \quad\langle R|H_\epsilon|\psi_{n s_n}\rangle& 
\ \ \ \ \ \ \ \ \ \quad\langle R|H_\epsilon|\psi_{1}^+\rangle&
\cdots&
\ \ \ \ \ \ \ \ \quad\langle R|H_\epsilon|\psi_{s_0}^{-}\rangle \\
\langle N-R+1|H_\epsilon|\psi_{1 1}\rangle&
\cdots&
\langle N-R+1|H_\epsilon|\psi_{n s_n}\rangle&
\langle N-R+1|H_\epsilon|\psi_{1}^+\rangle&
\cdots&
\langle N-R+1|H_\epsilon|\psi_{s_0}^{-}\rangle\\
\ \ \ \ \ \ \ \ \quad \vdots&   
 &
\ \ \ \   \quad\vdots &
\ \ \ \ \ \ \ \ \quad\vdots &  
 &  
\ \ \ \ \ \ \ \ \quad\vdots\\
\ \ \ \ \ \ \ \ \ \quad \langle N|H_\epsilon|\psi_{1 1}\rangle&
\cdots&
\ \ \ \ \ \ \ \ \quad\langle N|H_\epsilon|\psi_{n s_n}\rangle&
\ \ \ \ \ \ \ \ \ \quad\langle N|H_\epsilon|\psi_{1}^+\rangle&
\cdots&
\ \ \ \ \ \ \ \ \quad\quad \langle N|H_\epsilon|\psi_{s_0}^{-}\rangle
\end{bmatrix}\!\!.
\end{eqnarray}
\end{widetext}
By construction, 
the boundary matrix $B$ is a block matrix of block-size \(d\times 1\).
In terms of this matrix, Eq.\,\eqref{BBfinal} provides the useful identity
\begin{eqnarray}
\label{bethelike}
H|\epsilon,\bm{\alpha}\rangle = \epsilon|\epsilon,\bm{\alpha}\rangle + 
\sum_{b,s}|b\rangle B_{bs}(\epsilon)\bm{\alpha}_s,\quad \epsilon\in\mathds{R}.\ \ \ \  
\end{eqnarray}
One may write an analogous equation in Fock space by defining an array
\[ \eta_{\epsilon,\bm{\alpha}}^\dagger \equiv 
\sum_{j=1}^{N}\langle j|\epsilon,\bm{\alpha}\rangle\hat{\Psi}_j^\dagger.
\]
Then Eq.\,\eqref{bethelike} translates into
\begin{equation}
\label{bethemb}
[\widehat{H},\eta_{\epsilon,\bm{\alpha}}^\dagger] = 
\epsilon\,\eta_{\epsilon,\bm{\alpha}}^\dagger + 
\sum_{b,s} \hat{\Psi}_b^\dagger B_{bs}(\epsilon)\bm{\alpha}_s .
\end{equation}
It is interesting to notice that this (many-body) relation remains true even 
if \(\epsilon\) is allowed to be a complex number.

\subsection{The generalized Bloch theorem} 

The bulk-boundary separation of the energy eigenvalue equation shows
that actual energy eigenstates are necessarily linear combinations
of solutions of the bulk equation. This observation leads to a generalization
of Bloch's theorem for independent fermions under arbitrary BCs:

\medskip

{\bf Theorem 3 ({\bf Generalized Bloch theorem}).} {\em 
Let $H=H_N+W$ denote the single-particle Hamiltonian of a clean system subject to 
BCs described by \(W=P_\partial W\). If \(\epsilon\) is a regular energy eigenvalue 
of $H$ of degeneracy \(\cal{K}\), 
the associated eigenstates can be taken to be of the form
\[
|\epsilon,\bm{\alpha}_{\kappa}\rangle  = 
|\Psi_{\mathcal{B}}\rangle\bm{\alpha}_{\kappa},
\quad \kappa=1,\dots,\cal{K},
\]
where \(\{\bm{\alpha}_\kappa,\ \kappa=1,\dots,\cal{K}\}\) is a basis of the 
kernel of the boundary matrix $B(\epsilon)$ at energy $\epsilon$.}

\medskip 

In short, \( (H_N+W)|\epsilon,\bm{\alpha}\rangle=\epsilon|\epsilon,\bm{\alpha}\rangle \)
if and only if \( B\bm{\alpha}=0\), in which case it also follows from Eq.\,\eqref{bethemb}
that $\eta_{\epsilon,\bm{\alpha}}^\dagger$ is a normal fermionic mode of 
the many-body Hamiltonian $\widehat{H}$. From now on, we will refer to energy 
eigenstates of the form \(|\Psi_{\mathcal{B}}\rangle\bm{\alpha}_\kappa\) as 
{\em generalized Bloch states}. Recall that $H$ acts on ${\cal H} = {\mathbb C}^N 
\otimes {\mathbb C}^d$, with couplings of finite range $R$.  A lower bound on 
$N$ should be obeyed, in order for the above theorem to apply.   
If \(\det h_R \neq 0\), since 
there are no emergent solutions nor flat bands, generalized Bloch states 
describe the allowed energy eigenstates as soon as \(N>2R\), independently
of \(d\).  If \(h_R\) fails to be invertible, we should require
that \(N>2\,{\rm max}(s_0,R)\) to ensure that emergent solutions 
on opposite edges do not overlap, and are thus independent. Since 
\(s_0\leq Rd\), this condition is satisfied for any \(N>2Rd\).
In general, $N>2R(d+1)$ always suffices for generalized Bloch states 
to describe {\em generic} energy eigenstates \cite{JPA}. 

We further note that if \(\epsilon\) is {\em not} an energy eigenvalue, 
the kernel of $B(\epsilon)$ is trivial. Thus, the 
degeneracy of a single-particle energy level coincides with the 
dimension of the kernel of $B(\epsilon)$.
Let \(\rho(\omega)\) denote 
the single-particle density of states. Combining its definition with 
the generalized Bloch theorem, we then see that
\[
\rho(\omega)=\sum_{\det B(\epsilon)=0} [\dim {\rm Ker\,} B(\epsilon)] \, \delta(\hbar\omega-\epsilon),
\]
an alternative formula to the usual
\[ \rho(\omega)=-\frac{1}{\pi}{\rm Im\,Tr}\,(H_N+W-\hbar\omega+i 0^+)^{-1}, \]
from the theory of Green's functions \cite{ballentine}. Another interesting and
closely related formula is 
\[ \mathcal{Z}_W={\rm Tr}\, e^{-\beta(H_N+W)}=
\sum_{\det B(\epsilon)=0} \dim {\rm Ker\,} B(\epsilon) \ e^{-\beta\epsilon},  \]
for the partition function of the single-particle Hamiltonian, with the dependence 
on BCs highlighted \cite{quelle15} .

We conclude this section by showing how, for periodic BCs, 
one consistently recovers the conventional Bloch's theorem. In this case, 
the appropriate matrix $W$ reads
\begin{eqnarray*}
W \equiv W_p=\sum_{r=1}^R ( T^{N-r}\otimes h_r^\dagger+{\rm h.c.}),
\end{eqnarray*}
since then one can check that
\[
H_p=H_N+W_p=\mathds{1}_N\otimes h_0+\sum_{r=1}^R (V^r\otimes h_r+{\rm h.c.}),
\]
in terms of the fundamental circulant matrix
\begin{eqnarray*}
V \equiv T+(T^\dagger)^{N-1}=\sum_{j=1}^{N-1}|j\rangle\langle j+1|+|N\rangle\langle 1| .
\end{eqnarray*}
Physically, \(V\) is the generator of translations (to the left)
for a system displaying ring (1-torus) topology. 

The Bloch states are the states that diagonalize \(H_p\) and \(V\)
simultaneously. Theorem 3 guarantees that we
can choose the eigenstates of \(H_p\) to be linear combinations of 
translation-invariant and emergent solutions. Thus, we only need to
check if these linear combinations include 
eigenstates of \(V\). There is no hope of retaining the emergent solutions, because
they are localized and too few in number (at most \(2Rd\)) to be rearranged 
into eigenstates of \(V\). The same holds for translation-invariant
solutions with a power-law prefactor. Hence, the search subspace that is 
compatible with the translational symmetry \(V\) is described by the 
simplified Ansatz \cite{abc}
\[ |\epsilon,\bm{\alpha}\rangle =\sum_{\ell=1}^{n}\alpha_{\ell 1}|\psi_{\ell 1}\rangle .
\]
Now,
\( V|\psi_{\ell 1}\rangle=z_{\ell}|\psi_{\ell 1}\rangle-
z_\ell(1-z_\ell^N)|N\rangle|u_{\ell s_\ell 1}\rangle, \)
and so the generalized Bloch states can only be eigenstates of \(V\) 
if \(e^{ik_\ell N}=1\) with $z_\ell=e^{ik_\ell}$, and all but one entry in \(\bm{\alpha}\) vanish.
That is, $|\epsilon,\bm{\alpha}\rangle 
\equiv  |\epsilon,k_\ell\rangle=|z_\ell, 1 \rangle|u_{\ell 1, 1}\rangle.$ 
As one may verify, 
$H_p|\epsilon,k_\ell\rangle=|z_\ell, 1 \rangle H(z_\ell) |u_{\ell 1, 1}\rangle=
\epsilon |\epsilon,k_\ell\rangle ,$
showing that \(|\epsilon,k_\ell\rangle\) is indeed compatible with the 
boundary matrix. Manifestly, \(|\epsilon,k_\ell\rangle\) is an eingenstate of 
\(H_p\) in the standard Bloch form -- thereby recovering 
the conventional Bloch's theorem for periodic BCs, as desired.

\section{The bulk-boundary algorithms}
\label{sec:algo}

\begin{figure*}
\begin{center}
\includegraphics[width = 12cm]{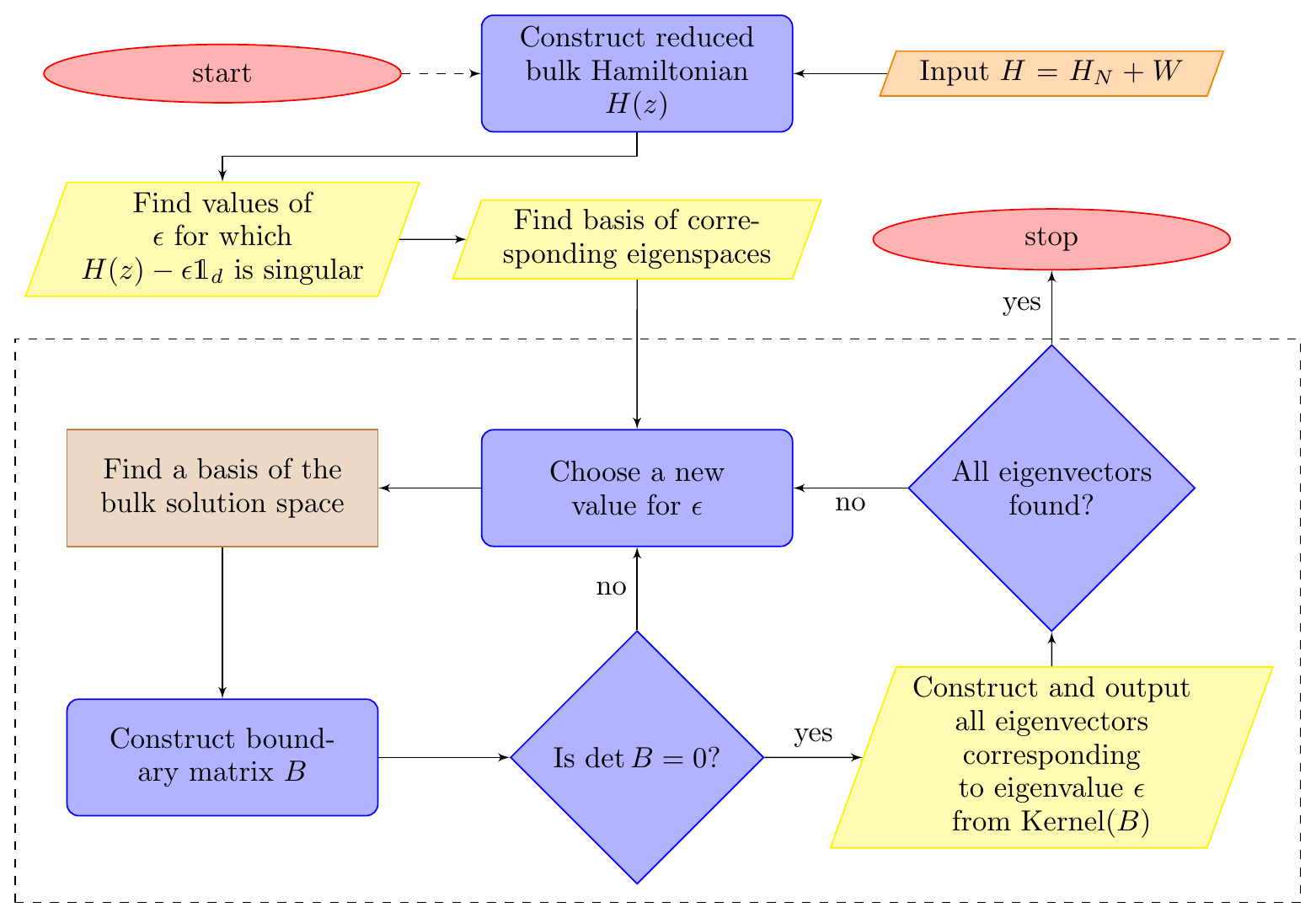}
\caption{(Color online) 
Flowchart of the numerical diagonalization algorithm. The steps inside the dashed rectangle form 
the loop for scanning over $\epsilon$. The crucial step is solving the bulk equation,
which encompasses steps\,(\ref{roots})-(\ref{Kmatrix}) as described in the text.
\label{figflowmain}
}
\end{center}
\end{figure*}

The results of Sec.\,\ref{sec:theory} can be used to develop  
diagonalization algorithms for the relevant class of single-particle Hamiltonians.
We will describe two such algorithms. The 
first treats \(\epsilon\) as a parameter for numerical search.
The second is inspired by the algebraic Bethe Ansatz, as suggested by comparing our 
Eq.\,\eqref{bethelike} to Eq.\,(28) of Ref.\,[\onlinecite{ortiz05}].

\subsection{Numerical ``scan-in-energy'' diagonalization}
\label{sec:scan}

The procedure described in this section is a special instance of 
the Eigensystem Algorithm described in Ref.\,[\onlinecite{JPA}], specialized 
to Hermitian matrices. It employs a search for energy eigenvalues along the real line, 
and takes advantage of the results of Sec.\,\ref{sec:theory} to determine whether 
a given number is an eigenvalue. The overall procedure is schematically depicted
in Fig.\,\ref{figflowmain}.

The first part of the algorithm finds all eigenvectors of $H$ that correspond to
the flat (dispersionless) energy band, if any exists. Two steps are entailed:
\begin{enumerate}
\item \label{1.1} Find all real values of $\epsilon$ for which 
$\det (H(z)-\epsilon \mathds{1}_{d})$ vanishes for any $z$. 
Output these as singular eigenvalues of $H$.

\item \label{1.2}For each of the eigenvalues found in step\,(\ref{1.1}), find and output a basis 
of the corresponding eigenspace of $H$ using any conventional algorithm.

\end{enumerate}
In implementing step (\ref{1.2}) above, one can leverage the analysis of Sec.\,\ref{flatbandid}.
The following part of the algorithm, which repeats until all eigenvectors of $H$ are found, 
proceeds according to the following steps:.

\begin{enumerate}
\setcounter{enumi}{2}
\item \label{seed} Choose a seed value of $\epsilon$, different from those
eigenvalues found already.

\item \label {roots} Find all $n$ distinct non-zero roots of the equation
$\det(H(z)-\epsilon\mathds{1}_{d})=0$. Let these roots 
be $\{z_\ell,\ \ell = 1,\dots,n\}$, 
and their respective multiplicities $\{s_\ell,\ \ell=1,\dots,n\}$.
 
\item For each such roots, construct the generalized reduced bulk 
Hamiltonian $H_{s_\ell}(z_\ell)$ [Eq.\,\eqref{genhb}].

\item Find a basis of the eigenspace of $H_{s_\ell}(z_\ell)$ with eigenvalue $\epsilon$. 
Let the basis vectors be $\{|u_{\ell s}\rangle,\ s=1,\dots,s_\ell\}$. The  
bulk solution corresponding to $(\ell,s)$ is 
$|\psi_{\ell s}\rangle = |z_\ell,1\rangle|u_{\ell s}\rangle$, with 
$\Phi_{z_\ell}$ defined in Eq.\,\eqref{basisphi}.

\item If $h_R$ is non-invertible, find $s_0 = Rd - \sum_{\ell=1}^{n}s_\ell/2$. 
Construct matrices $K^{-}(\epsilon)$ as described in Eq.\,\eqref{kmatrix}, 
and $K^{+}(\epsilon)=[{K^{-}}(\epsilon)]^\dagger$.

\item \label{Kmatrix} Find bases of the kernels of 
$K^{-}(\epsilon)$ and $K^{+}(\epsilon)$. Let the basis vectors
be $\{|u_s^{-}\rangle,\ s=1,\dots,s_0\}$ and $\{|u_s^{+}\rangle,\ s=1,\dots,s_0\}$, respectively.
The emergent bulk solutions corresponding to each $s$ are 
follow from Eqs.\,\eqref{basisj-} and \eqref{basisj+}.

\item Construct the boundary matrix $B(\epsilon)$ [Eq.\,\eqref{Boundary matrix}].

\item \label{last} If $\det B(\epsilon)=0$, output $\epsilon$ as an eigenvalue. 
Find a basis $\{\bm{\alpha}_\kappa,\ \kappa=1,\dots,{\cal K}\}$ of the kernel of $B(\epsilon)$.
Then a basis of the eigenspace of $H$ corresponding to energy 
$\epsilon$ is $\{|\epsilon_\kappa\rangle = 
|\Psi_{\mathcal{B}}\rangle\bm{\alpha}_\kappa,\ \kappa=1,\dots,{\cal K}\}$,
with $|\Psi_{\mathcal{B}}\rangle$ being defined in Eqs.\,\eqref{bulkbasis}.
If all $2dN$ eigenvectors are not yet found, then go back to step\,(\ref{seed}).

\item \label{root-finding} If $\det B(\epsilon)\ne 0$, choose a new value of $\epsilon$ 
as dictated by the  relevant root-finding algorithm \cite{NoteRoot}. 
Go back to step\,(\ref{roots}).
\end{enumerate}

Some considerations are in order, in regard to the fact that 
the determinant of $B(\epsilon)$ plotted as a function of energy $\epsilon$
may display finite-precision inaccuracies, that appear as fictitious roots. 
Such issues arise at those $\epsilon$ where two (or more) of the roots of Eq.\,\eqref{chareq}
cross as a function of $\epsilon$, due to the non-orthogonality of the basis $\mathcal{B}$ that 
results from the procedure described in Sec.\,\ref{sec:bulkeq}.
Let $\epsilon_*$ be a value of energy for which this happens, 
so that the bulk equation bears a power-law solution. For $\epsilon \approx \epsilon_*$
(except $\epsilon_*$ itself), Eq.\,\eqref{chareq} has two roots that are very close in value,
so that the corresponding bulk solutions overlap almost completely. 
This results in a boundary matrix having two nearly identical columns, with determinant vanishing 
in the limit $\epsilon \rightarrow \epsilon_*$, {\em irrespective} of  $\epsilon_*$ being an eigenvalue of $H$ 
 (hence, a physical solution). 
However, if we calculate $B(\epsilon)$ {\em exactly at} $\epsilon_*$, then the basis $\mathcal{B}$
contains power-law solutions, and accurately indicates whether $\epsilon_*$ is an eigenvalue. 
This also means that the function $\det B(\epsilon)$ 
has a discontinuity at $\epsilon=\epsilon_*$.

A simple way to identify those fictitious roots is as follows. 
Rewrite the polynomial in Eq.\,\eqref{charz} as 
\begin{eqnarray}
\label{charpoly}
P(\epsilon,z) = \sum_{r=s_0}^{2Rd-s_0}p_{r}(\epsilon)z^{r},
\end{eqnarray}
which is treated as a polynomial in $z$ with coefficients depending on $\epsilon$
(if $s_0$ changes with $\epsilon$, we use the 
smallest possible value of $s_0$ in Eq.\,(\ref{charpoly})). 
$P(\epsilon,z)$ has double roots at $\epsilon_*$ 
if and only if the {\em discriminant} $D(P(\epsilon_*,z)) =0$ \cite{Gelfand08}.
The latter gives a polynomial expression in $\epsilon$,
of degree $\mathcal{O}(dR)$. By finding the roots of this
equation, one can obtain all the values of $\epsilon$ for which
fictitious roots of $\det B(\epsilon)$ may appear. To check whether these
roots are true eigenvalues, one then needs to construct $B(\epsilon)$
by including the power-law solutions in the Ansatz. 

We further note that, while the Ansatz is not continuous at such values of 
$\epsilon$, the fact that the bulk solution space is the kernel of the
linear operator $P_B(H_N+W-\epsilon)$ implies that it must 
change {\em smoothly} with $\epsilon$.  
A way to improve numerical accuracy would be to construct an orthonormal basis
(e.g., via Gram-Schmidt orthogonalization)
of ${\cal M}_{1,N}(\epsilon)$
at each $\epsilon$, and use this basis to construct a modified boundary
matrix $\tilde{B}(\epsilon)$. In practice, one may {\em directly} compute the
new determinant by using 
\[
\det\tilde{B}(\epsilon) = \frac{\det B(\epsilon)}{\sqrt{\det \mathcal{G} (\epsilon)}},
\]
where $\mathcal{G}\equiv \langle \Psi_{\mathcal{B}} | \Psi_{\mathcal{B}}\rangle$ 
is the Gramian matrix \cite{grammatrix} of the basis of bulk solutions obtained in steps
(\ref{roots}) to (\ref{Kmatrix}) of the algorithm, with entries 
\(
\mathcal{G}_{s s'} \equiv  \langle \psi_s|\psi_{s'}\rangle,$ $s,s'=1,\dots,2Rd.
\)
In fact, it can be checked that the bulk solutions
\[
\Big \{|\phi_s\rangle \equiv  
\sum_{s'=1}^{2Rd} {\big[\mathcal{G}^{-1/2}\big]}_{s's}|\psi_{s'}\rangle,\quad s=1,\dots,2Rd \Big \}
\]
form an orthonormal basis of the bulk solution space $ {\mathcal M}_{1,N}$.
The calculation of the entries of the Gramian is straightforward thanks
to the analytic result 
\begin{align*}
\langle z,1 | z',1 \rangle = \left\{
\begin{array}{lcl}\frac{z^*z'-(z^*z')^{N+1}}{1-z^*z'}&  \;\text{if} & z' \ne 1/z^*\\
N & \;\text{if} & z'=1/z^*
\end{array}
\right..
\end{align*}

In regard to the time and space complexity of the algorithm, 
the required resources depend entirely on those needed to compute the boundary matrix.
For generic $\epsilon$, regardless of the invertibility of $h_R$, 
the size of $B(\epsilon)$ is $2Rd\times 2Rd$, independently of $N$. 
Calculation of each of its entries is also simple from the point of 
view of complexity, thanks to the fact that  $H=H_N+W$ is symmetrical \cite{RemarkSymm, JPA}.  
Accordingly, both the number of steps and the memory space used by this 
algorithm do not scale with the system size $N$, making this approach 
computationally more efficient than conventional methods of diagonalization
of generic Hermitian matrices \cite{Demmel07}.

\subsection{Algebraic diagonalization}
\label{easybethe}

The scan-in-energy algorithm can be further 
developed into an algorithm that yields an analytic solution (often closed-form), 
in the same sense as the Bethe Ansatz method does for a different class of 
(interacting) quantum integrable systems. 
The idea is to obtain, for generic values of $\epsilon$, an analytic expression for 
$B(\epsilon)$, since its determinant will then provide a 
condition for $\epsilon$ to be an eigenvalue, and the corresponding eigenvectors
can be obtained from its kernel. As mentioned, 
for generic $\epsilon$, the extended bulk solutions
do not include any power-law solutions. This property 
can be exploited to derive an analytic expression for $B(\epsilon)$ in 
such a generic setting. The values of $\epsilon$ for which
power-law solutions appear, or the analytic expression fails for other reasons,
can be dealt with on a case-by-case basis. 

By the Abel-Ruffini theorem, a  completely closed-form 
solution {\em by radicals} in terms of $\epsilon$
can be achieved if the degree in $z$ of the characteristic polynomial of the 
reduced bulk Hamiltonian is at most 
four. If this is not the case, the roots $\{z_\ell\}$ do not possess an algebraic 
expression in terms of $\epsilon$ and entries of $H$. The workaround 
is then to consider $\{z_\ell\}$ as free variables, with the constraint that each of them
satisfy the characteristic equation of $H(z)$. 
With these tools in hand, the following procedure can be used to find an analytical
solution for generic values of $\epsilon$:

\begin{enumerate}

\item 
Construct the polynomial $P(\epsilon,z)$ in Eq.\,\eqref{charpoly},
which is a bivariate polynomial in $\epsilon$ and $z$.
Determine $s_0$ using 
\( s_0 = 2Rd - \text{deg}(P(\epsilon,z)),\)
where $\text{deg}(.)$ denotes the degree of the polynomial in $z$.

\item \label{steptwo} Assuming that $\epsilon$ and $z$ satisfy 
$P(\epsilon,z)=0$, find an expression for the eigenvector 
$|u(\epsilon,z)\rangle$ of $H(z)$ with eigenvalue $\epsilon$. 

\item Consider variables $\{z_\ell,\ \ell=1,\dots,2Rd-2s_0\}$, each
satisfying $P_\epsilon(z_\ell)=0$.
Each of these corresponds to a bulk solution $|z_\ell,1\rangle|u(\epsilon,z_\ell)\rangle$. 

\item If $h_R$ is not invertible, construct matrices 
$K^-(\epsilon)$ and $K^+(\epsilon)= [K^-(\epsilon)^\dag]$ [Eq.\,\eqref{kmatrix}].

\item  \label{stepfive} Find bases for their kernels, each of which contains $s_0$ vectors. Let these 
be $\{|u_s^-(\epsilon)\rangle,\ s=1,\dots,s_0\}$ and 
$\{|u_s^+(\epsilon)\rangle,\ s=1,\dots,s_0\}$. These correspond to 
finite-support solutions of the bulk equation.

\item Construct the boundary matrix $B(\epsilon) \equiv B(\epsilon,\{z_\ell\})$ 
[Eq.\,\eqref{Boundary matrix}]. 

\item \label{eigchar} The condition for $\epsilon$ being an eigenvalue of $H$ is 
$\det B(\epsilon,\{z_\ell\})=0$.
Therefore, a complete characterization of eigenvalues is
\begin{equation*}
\qquad \{P(\epsilon,z_\ell)=0,\quad \ell=1,\dots,n\},\quad \det B(\epsilon,\{z_\ell\})=0.
\end{equation*}

\item If $\text{deg}(P(\epsilon,z))\leq 4$,
substitute for each $z_\ell$ the closed-form expression of the corresponding root $z_\ell(\epsilon)$. 
The eigenvalue condition in step\,(\ref{eigchar}) simplifies to
a single equation, $\det B(\epsilon,\{z_\ell(\epsilon)\})=0.$

\item \label{finalstep} For every eigenvalue $\epsilon$, the kernel vector 
$\bm{\alpha}(\epsilon,\{z_\ell\})$ of $B(\epsilon,\{z_\ell\})$ 
provides the corresponding eigenvector of $H$.
\end{enumerate}

In steps (\ref{steptwo}), (\ref{stepfive}) and (\ref{finalstep}), we need to
obtain an analytic expression for the basis of the kernel of a square 
symbolic matrix of fixed kernel dimension in terms of its entries. 
This can be done in many different ways, and often is possible by inspection. 
One possible way was described in Sec.\,\ref{flatbandid} 
in connection to evaluating Ker($H(z)-\epsilon\mathds{1}_{d})$
for singular values of $\epsilon$.
The above analysis does not hold when $\epsilon$ satisfies any
of the following conditions:

\begin{enumerate}[(i)]
\item $\det (H(z)-\epsilon\mathds{1})=0$ has one or more double roots. 
This  is equivalent to $D(P(\epsilon,z))=0$, as discussed in Sec.\,\ref{sec:scan}.
This is a polynomial equation in terms of $\epsilon$, the roots of which yield all 
required values of $\epsilon$.

\item The coefficient $p_{s_0}(\epsilon)$ of $z^{s_0}$ in $P(\epsilon,z)$ vanishes,
or equivalently, $\epsilon$ is a root of $p_{s_0}(\epsilon)=0$.

\item Each entry of $|u(\epsilon,z)\rangle$ vanishes. Such points are identified by solving
simultaneously the equations 
$\langle m|u(\epsilon,z)\rangle = 0,\ m=1,\dots,d$
and $P(\epsilon,z)=0$, 
Since a necessary and sufficient condition for these polynomials (in $z$)
to have a common root is that their resultant vanishes
\cite{Gelfand08}, we find the relevant values of $\epsilon$ by equating the pairwise resultants to zero.

\item $\{|u_s^-(\epsilon)\rangle,\ s=1,\dots,s_0\}$ or $\{|u_s^+(\epsilon)\rangle,\ s=1,\dots,s_0\}$
are linearly dependent. To find such values of $\epsilon$, one may form the corresponding
Gramian matrix and equate its determinant to zero.

\end{enumerate}
For all the values of $\epsilon$ thus identified, $B(\epsilon)$
is calculated by following steps\,(\ref{roots})-(\ref{last}) in the scan-in-energy algorithm.   
To summarize, this algebraic procedure achieves diagonalization in analytic form: 
the upshot is a system of {\em polynomial equations}, whose simultaneous roots are the eigenvalues, 
and an analytic expression for the eigenvectors, with {\em parametric dependence} on the eigenvalue.

\section{Illustrative examples}
\label{sec:examples}

This section contains three paradigmatic examples illustrating 
the use of our generalized Bloch theorem, along with the 
resulting algebraic procedure of diagonalization.

\subsection{The impurity model  revisited}
\label{sec:revisited}

Let us first reconsider the impurity model of Sec.\,\ref{motivating}.
The single-particle Hamiltonian 
is the corner-modified, banded block-Toeplitz matrix \(H=H_N+W\), with 
\begin{eqnarray*}
\hspace{-.1cm}
H_N=-t(T+T^\dagger),\quad \mbox{and}\quad  W= w P_\partial.
\end{eqnarray*}
The boundary consists of two sites, so that 
\(P_\partial= |1\rangle\langle 1|+|N\rangle\langle N|\), 
for any \(N>2\).  Likewise, 
$R=1=d$. 
The first step in diagonalizing \(H\) is solving the bulk 
equation. Since the reduced bulk Hamiltonian \(H(z)=-t(z+z^{-1})\),
\begin{eqnarray}
P(\epsilon,z) = z \,( H(z)-\epsilon) =-t (z^2+\frac{\epsilon}{t}z+1).
\label{roots2}
\end{eqnarray}
Thus, every value of \(\epsilon\) is regular and yields two 
(\(=\) the number of boundary degrees of freedom) solutions 
of the bulk equation. If \(\epsilon\neq \pm 2t\), the solutions 
are \(|z_\ell,1\rangle\), with 
\[ z_\ell=-\frac{\epsilon}{2t}+(-1)^\ell\sqrt{\frac{\epsilon^2}{4t^2}-1}, \quad \ell=1,2, \]
\noindent 
with $z_1 z_2=1$ and $\epsilon=-t(z_1+z_2)$.
The special values \(\epsilon=\pm 2t\) for which $H_N$
yields only one of the two bulk solution have an interpretation as the 
edges of the energy band. If \(\epsilon=2t\), then $H(z)$ yields only 
\(|z_1=-1, 1\rangle\), whereas 
if \(\epsilon=-2t\), it yields only 
\(|z_1=1,1\rangle\). In order to obtain the missing bulk solution in each case, one 
must consider the effective Hamiltonian [Eq.\,(\ref{genhb})] 
\begin{eqnarray*}
H_2(z)
=-t
\begin{bmatrix}
z+z^{-1}&  1-z^{-2}   \\
0       &  z+z^{-1}
\end{bmatrix}.
\end{eqnarray*}
One may check that \(H_2(z_1)-\epsilon\mathds{1} \equiv 0\)
if \(\epsilon=\pm 2t,\ z_1=\mp 1\). 
Thus, the two linearly independent solutions of the bulk equation
at these energies are \(|z_1=1,v\rangle,\) $v=1,2$, 
if \(\epsilon=-2t\), and \(|z_1=-1,v \rangle,\) $v=1,2$, if \(\epsilon=2t\).

For the purpose of solving the boundary equation, and hence
the full diagonalization problem, it is convenient to organize the 
solutions of the bulk equation as
\begin{eqnarray*}
|\epsilon\rangle=\left\{
\begin{array}{lcl}
\alpha_1|z_1,1\rangle+\alpha_2|z_2,1\rangle&\ \ \mbox{if}& \epsilon\neq \pm 2t\\
\alpha_1|z_1=-1,1\rangle+\alpha_2| z_1=-1,2\rangle&\ \ \mbox{if}& \epsilon = 2t\\
\alpha_1|z_1=1,1\rangle+\alpha_2| z_1=1,2\rangle&\ \ \mbox{if}& \epsilon=-2t
\end{array}
\right. .
\end{eqnarray*}
For comparison with Sec.\,\ref{motivating}, one should think of
\(z_1=e^{ik}\) and \(z_2=e^{-ik}\). Because the Ansatz is naturally broken 
into three pieces, so is the boundary matrix. For instance, when \(\epsilon \neq \pm 2t\),  
direct calculation yields 
\begin{equation*}
B(\epsilon)=
\begin{bmatrix}
-tz_1^2+(w-\epsilon)z_1       & -t z_2^2+(w-\epsilon) z_2       \\
-tz_1^{N-1}+(w-\epsilon)z_1^N & -t z_2^{N-1}+(w-\epsilon)z_2^N
\end{bmatrix}.
\end{equation*}
However, from Eq.\,(\ref{roots2}) it follows that  
\begin{eqnarray}
\label{recall}
-t(z_\ell+z_\ell^{-1})-\epsilon=0,\quad  \ell=1,2. 
\end{eqnarray}
This allows a simpler form to be obtained, by effectively changing the  
argument of the boundary matrix from $\epsilon$ to $z_\ell$ (or $k$). The complete final 
expression reads:
\begin{widetext}
\begin{align}
\label{bmbmatrix}
B(\epsilon)=
\left\{
\begin{array}{lcl}
\begin{bmatrix}
t+w z_1       & t+w z_2       \\
(z_1t+w)z_1^N & (z_2t+w)z_2^N
\end{bmatrix}
&\ \ \mbox{if}& \epsilon\neq \pm 2t\\
\\
\begin{bmatrix}
t-w            & w\\
(-1)^{N-1}(t-w)& \;(-1)^{N}(N(t-w)+t)
\end{bmatrix}
&\ \ \mbox{if}& \epsilon = 2t\\
\\
\begin{bmatrix}
w+t & w\\
w+t & (w+t)N+t
\end{bmatrix}
&\ \ \mbox{if}& \epsilon=-2t
\end{array}
\right. .
\end{align}
\end{widetext}
Notice that if $\epsilon$ approaches $\pm 2t$, the two distinct roots collide at  
$z_1=z_2=\mp 1$, and the boundary matrix becomes, trivially, a rank-one matrix, signaling 
the discontinuous behavior anticipated in Sec. \ref{sec:scan}. 
Furthermore, it follows from Eq.\,(\ref{shortcut}) that 
the power-law solution at $\epsilon=\pm2t$
may be written as $\partial_z(|z_1,1\rangle) = |z_1,2\rangle$.
The entries of the second column of the corresponding boundary matrices
satisfy $\langle b|H_\epsilon |z_1,2\rangle = \partial_{z_2}\langle b|H_\epsilon 
|z_2,1\rangle|_{z_2=z_1}$, where $z_1$ is the double root. 
Thus, the entries in the second column
of the boundary matrix for $\epsilon=\pm2t$ can be obtained by differentiating
with respect to $z_2$ the second column of the boundary matrix for 
other (generic) values of $\epsilon$, an observation we will use in other examples as well (see e.g. Sec. 
\ref{ccf}). 
We now analyze separately different regimes (see also Fig.\,\ref{fig:impurity} for illustration).

\subsubsection{Vanishing impurity potential}

If \(w=0\), then \(B(\epsilon=2t)\) and \(B(\epsilon=-2t)\) have a trivial 
kernel; the exotic 
states \(|\epsilon =\pm 2t\rangle\) cannot possibly arise as physical 
eigenvectors.
For other energies, we find that the kernel
of the boundary matrix 
\[
B(\epsilon)=t
\begin{bmatrix}
1         & 1\\
z_1^{N+1} & z_2^{N+1}
\end{bmatrix}\quad (w=0), \]
is nontrivial only if \(z_1^{N+1}=z_2^{N+1}\), in which
case we can take \(\alpha_1=1\) and \(\alpha_2=-1\).
From Eq.\,(\ref{roots2}), it also follows that \(z_1z_2=1\). 
Hence, there are \(2N+2\) solutions,  
\begin{eqnarray*}
z_1=z_2^{-1}=e^{i\frac{\pi q}{N+1}},\quad q=-N-1,-N,\dots, N.
\end{eqnarray*}
Of the associated \(2N+2\) (un-normalized) Ansatz vectors
\begin{eqnarray*}
|\epsilon_q\rangle&=& |z_1,1\rangle-|z_2,1\rangle = 
2i\sum_{j=1}^N \sin\Big(\frac{\pi q }{N+1} j \Big)|j\rangle,
\end{eqnarray*}
two vanish identically (\(q=-N-1\) and \(q=0\)). For \(q=\pm 1,\dots,\pm N\),
it is immediate to check that 
\(|\epsilon_{-q}\rangle=-|\epsilon_{q}\rangle\). This means
that the Ansatz yields exactly \(N\) linearly independent energy 
eigenvectors, of energy 
\[
\epsilon_q=-t(z_1+z_2)=-2t\cos\Big(\frac{\pi q}{N+1}\Big),\quad q=1,\dots,N.
\]
This is precisely the result of Sec.\,\ref{motivating}, where 
the solutions were labelled in terms of allowed quantum numbers
\(k=\pi q/(N+1),\ q=1,\dots,N\).

According to our general theory, the  eigenspaces  of $H$ 
are in one-to-one correspondence with the zeroes
of \(\det B(\epsilon)\). For this system then, there should be
at most \(N\) zeroes. The reason we find  \(2N+2\) zeroes 
is due to the above-mentioned (quadratic) change of argument 
in the boundary matrix from \(\epsilon\) to \(k\). Such a change of variables is
advantageous for analytic work, and the associated redundancy 
is always rectified at the level of the Ansatz.

\subsubsection{Power-law solutions}

What would it take for $|\epsilon=\pm 2t\rangle $ to become eigenvectors? 
The kernel of \(B(\epsilon=2t)\) is nontrivial only if 
\begin{eqnarray*}
w=t\quad\mbox{or}\quad 
w=t\, \frac{N+1}{N-1}.
\end{eqnarray*}
These two values coincide up to corrections of order \(1/N\),
but remember that our analysis is exact for any \(N>2\). 
Similarly, the kernel of \(B(\epsilon=-2t)\) is nontrivial only if
\begin{eqnarray*} 
\label{vN}
w=-t\quad\mbox{or}\quad 
w=-t\, \frac{N+1}{N-1}.
\end{eqnarray*}
Only one of these conditions can be met:
for fixed $w$, either \(|\epsilon=2t\rangle\) is
an energy eigenstate or \(|\epsilon=-2t\rangle\) is, but not 
both. Let us look more closely at the state at the bottom 
of the energy band. As we just noticed, this state
will be a valid eigenstate for either of
the two values of \(w\). Let us pick \(w \equiv w_N = -t ({N+1})/({N-1})\),
since it yields the most interesting ground state.
Then,
\begin{eqnarray*}
B(\epsilon=-2t)= 
\begin{bmatrix}
w_N+t & w_N\\
w_N+t & w_N
\end{bmatrix},
\end{eqnarray*}
so that one can set \(\alpha_1=1/(w_N+t)\), \(\alpha_2=-1/w_N\), and 
\begin{eqnarray*}
|\epsilon=-2t\rangle=
\sum_{j=1}^N
\Big(\frac{1}{w_N+t}-\frac{j}{w_N}\Big)|j\rangle.
\end{eqnarray*}
Notice that $\langle j|\epsilon=-2t\rangle=-\langle N-j+1|\epsilon=-2t\rangle$; that is, 
the power-law eigenvector of the impurity problem is an eigenstate of inversion 
symmetry.

\begin{figure}[htb]
\begin{center}
\includegraphics[width=7.8cm]{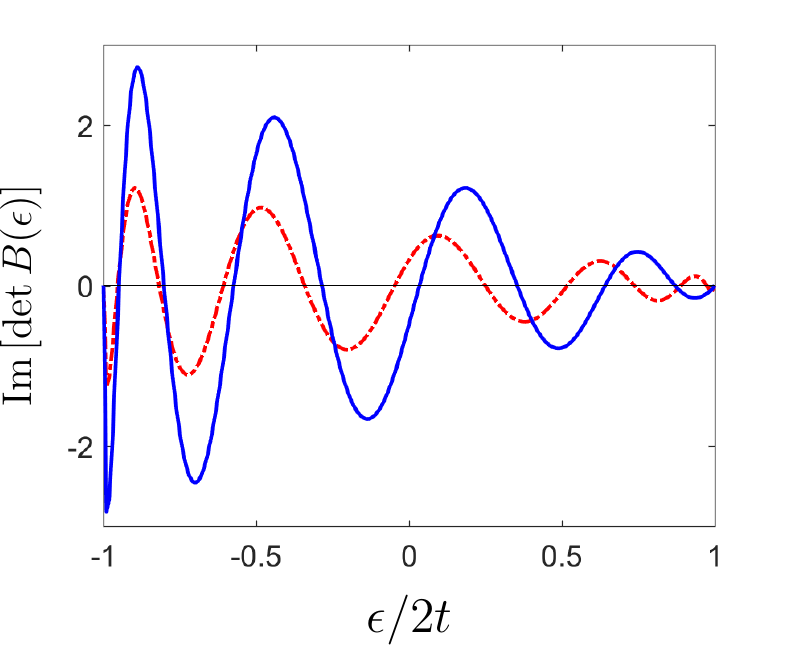}
\end{center}
\vspace{-4mm}
\caption{(Color online) 
Imaginary part of $\det B(\epsilon)$ for \(N=10\) as a function of the 
dimensionless parameter $\epsilon/2t$.
Here, \(B(\epsilon)\) is numerically evaluated from the top expression 
in Eq.\,(\ref{bmbmatrix}), $\epsilon\neq \pm 2t$.
Its real part vanishes identically in this range of energies. The impurity 
potential is $w=0.7>|w_N|$ for the solid 
blue curve, and $w=0.3<|w_N|$ 
for the dashed red curve. In the regime $w<w_N$ ($w>w_N$), the system 
hosts zero (two) edge modes, which is reflected in the number of zeroes
 ($N$ and $N-2$)
of the respective curves, in the energy range $-1<\epsilon<1$. 
In both cases, the crossings through zero at $\epsilon=\pm 1$ do not 
have associated eigenstates of the Hamiltonian. The origin of such 
fictitious zeroes was discussed in Sec.\,\ref{sec:scan}.}
\label{fig:impurity}
\end{figure}

\subsubsection{Strong impurity potential}

Lastly, consider the regime where \( t\ll |w|\), for large \(N\). 
Then, the values \(\epsilon=\pm 2t\) are excluded from the physical 
spectrum, and the eigenstates of the system can be determined from 
\(\det B(\epsilon)=0\). We expect bound states of energy \(w\) to
leading order and well-localized at the edges, so that \(0<|z_1|<1<|z_2|\),
say, with \(z_1\) (\(z_2\)) associated to the left (right) edge. It 
is convenient to take advantage of this feature and modify the original 
Ansatz to 
\begin{eqnarray*}
|\epsilon\rangle=\alpha_1|z_1,1\rangle+\alpha_2z_2^{-N}|z_2,1\rangle,
\end{eqnarray*}
so that \(|z_1,1\rangle\) (\(z_2^{-N}|z_2,1\rangle\)) peaks at the
left (right) edge, respectively. The boundary matrix becomes 
\begin{eqnarray*}
\tilde{B}(\epsilon) &=& 
\!\!\begin{bmatrix}
t+ z_1 w       & \!(t+w z_2)z_2^{-N}   \\
(z_1t+w)z_1^N & \!z_2t+w
\end{bmatrix} \\
& \approx &
\begin{bmatrix}
t+ w z_1       & \!\!\!0 \\
0 & \!\!\!z_2 t+w
\end{bmatrix},
\end{eqnarray*}
since \(|z_1|^N\approx 0\approx |z_2|^{-N}\). Keeping
in mind that \(z_1z_2=1\), we see that the kernel of $\tilde{B}(\epsilon)$ 
is two-dimensional for
\begin{eqnarray*}
z_1=-\frac{t}{w}=z_2^{-1}, \quad \epsilon_{b}=-t(z_1+z_2)=w-\frac{t^2}{w^2},
\end{eqnarray*}
and otherwise trivial. 
The corresponding energy eigenstates can be chosen to be
\begin{eqnarray*}
|\epsilon_b,1\rangle=\sum_{j=1}^N\Big(-\frac{t}{w}\Big)^j|j\rangle,\quad
|\epsilon_b,2\rangle=
\sum_{j=1}^N\Big(-\frac{w}{t}\Big)^{j-N}|j\rangle.
\end{eqnarray*}
Notice that \(|\epsilon_b,2\rangle\) is the mirror 
image of \(|\epsilon_b,1\rangle\), up to normalization. 
The large-\(N\) approach to boundary modes exemplified by
the preceding calculation can be made systematic, as we will
further explain in Sec.\,\ref{sec:genindicator}.

The remaining \((N-2)\) eigenstates consist of standing waves.
They can be computed from the original boundary matrix, approximated
for \(t \ll |w|\) as
\[ B(\epsilon\neq\epsilon_b)  \approx  w
\begin{bmatrix}
 z_1   &  z_2       \\
z_1^N  & z_2^N 
\end{bmatrix} .
\]
This boundary matrix has a nontrivial kernel only if
\[ z_1=z_2^{-1}=e^{i\frac{\pi s}{N-1}},\quad s=0,\dots, 2(N-1)-1, \]
in which case one may choose \(\alpha_1=z_2,\ \alpha_2=-z_1\).
Then, 
\[
|\epsilon_s\rangle=\sum_{j=1}^N(z_1^{j-1}-z_2^{j-1})|j\rangle=
2i\sum_{j=2}^{N-1}\sin\Big(\frac{\pi s(j-1)}{N-1}\Big)|j\rangle .
\]
Moreover, 
\( |\epsilon_{s}\rangle=-|\epsilon_{N-1+s}\rangle,$  $s=1,\dots,N-2.
\)
Hence, as needed, we have obtained \((N-2)\) linearly independent 
eigenvectors of energy \(\epsilon_s=-2t\cos[\pi s/(N-1)] \).

The above discussion is further illustrated in Fig.\,\ref{fig:impurity},
where the determinant of the exact boundary matrix is displayed as
a function of energy.

\subsection{Engineering perfectly localized \\ zero-energy modes: A periodic Anderson model}
\label{topocomb}

\begin{figure}
\includegraphics[width=8cm]{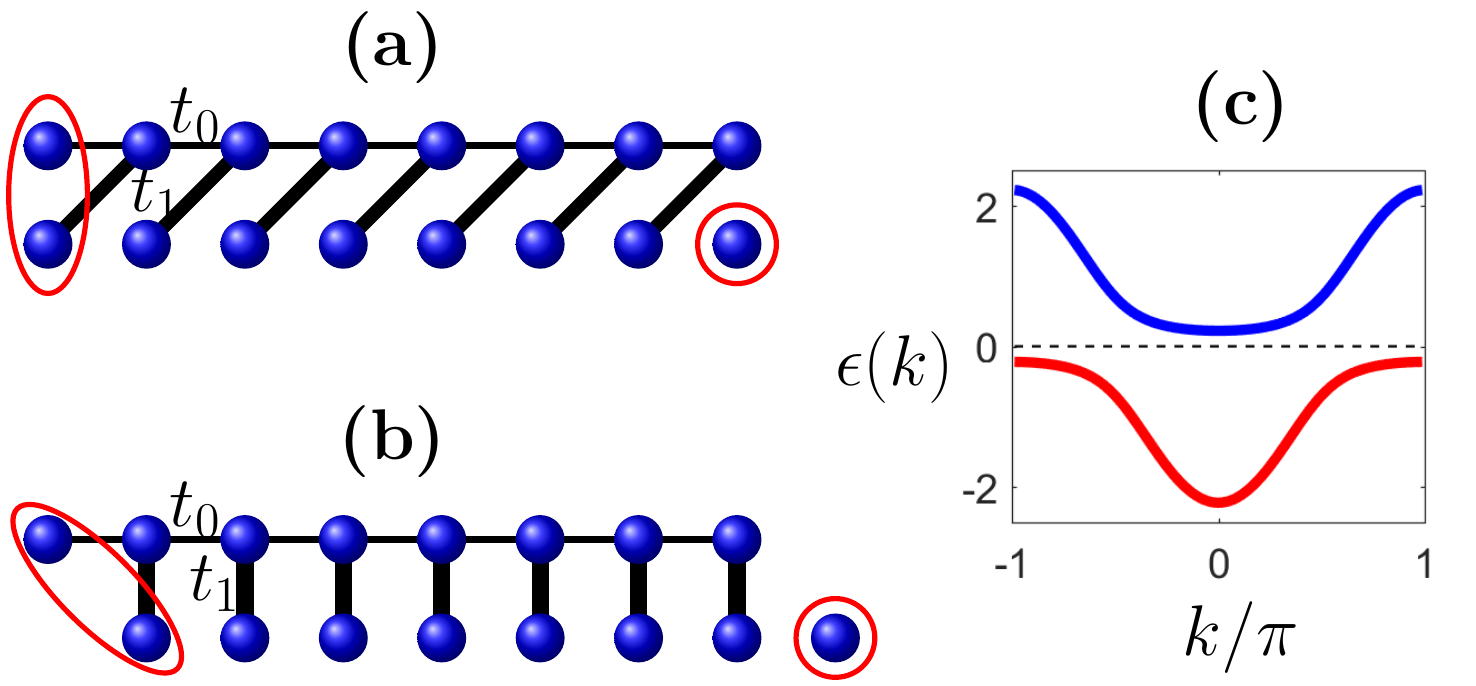}
\vspace*{-0mm}
\caption{(Color online) Two variants of the 
topological comb model. In (a), thin (thick) black lines indicate intra-ladder (diagonal) 
hopping with strength $t_{0}$ ($t_1$).
Red ovals or circles show the support of the zero energy edge modes.
In (b), upon shifting the lower chain by one site to the right, 
$t_{1}$ can be interpreted as direct inter-ladder hopping strength.
(c) Band structure for the parameter regime $t_1/t_0=0.7$. The (black) 
dashed line represents zero energy, which lies in the band gap.}
\label{comb}
\end{figure}

Having illustrated the algebraic diagonalization method on a simple
impurity model, we illustrate next its usefulness toward 
{\em Hamiltonian engineering}. In this section, we 
will design from basic principles a ``comb" model, see Fig.\,\ref{comb},
with the peculiar property of exhibiting a perfectly localized 
mode at zero energy while all other modes are dispersive. The zero 
mode is distributed over two sites on the same end of the comb, with 
weights determined by a ratio of hopping amplitudes. 
 
The starting point is the single-particle Hamiltonian 
\[ H = H_N=T\otimes h_{1}+T^{\dagger}\otimes h_{1}^{\dagger}.
\]
In order to have perfectly localized eigenvectors at zero energy, 
the bulk equation must bear emergent solutions.
Therefore, we assume that $h_1$ is non-invertible.
Let $|u^{-}\rangle$ be in the kernel of $h_{1}^{\dagger}$. 
Since $T$ annihilates $|j=1\rangle$, 
\begin{eqnarray*}
H(|j=1\rangle|u^{-}\rangle) & = & (T\otimes h_{1}+T^{\dagger}\otimes h_{1}^{\dagger})(|j=1\rangle|u^{-}\rangle)\\
 & = & T|j=1\rangle h_{1}|u^{-}\rangle+T^{\dagger}|j=1\rangle h_{1}^{\dagger}|u^{-}\rangle\\
 & = & 0.
\end{eqnarray*}
Similarly, if $|u^{+}\rangle$ is in the kernel of $h_{1}$, then
$|j=N\rangle|u^{+}\rangle$ is also in the kernel of $H$. Therefore,
$|j=1\rangle|u^{-}\rangle$ and $|j=N\rangle|u^{+}\rangle$ are perfectly
localized zero energy modes.

A concrete example may be obtained by choosing 
\[
h_{1}=-\begin{bmatrix}t_{0} & 0\\
t_{1} & 0
\end{bmatrix}\quad \text{and} \quad
h_{1}^\dagger=-\begin{bmatrix}t_{0} & t_1\\
0 & 0
\end{bmatrix},
\]
whose kernel is spanned by
\[
|u^{+}\rangle=\begin{bmatrix}0 \\ 1 \end{bmatrix}  \quad \text{and}\quad
|u^{-}\rangle=\begin{bmatrix}-t_{1}\\ t_{0} \end{bmatrix}, 
\]
respectively. This example 
corresponds to a many-body Hamiltonian of two coupled 
fermionic chains, as illustrated in Fig.\,\ref{comb}:  
\begin{eqnarray}
\label{combH}
\widehat{H} = 
-\sum_{j=1}^{N-1}(t_{0}c^\dagger_{j}c_{j+1}+
t_{1}c_{j+1}^\dagger f_j + \text{h.c.}),
\end{eqnarray}
where $c_{j}$ and $f_{j}$ denote the $j$th fermions in the upper 
and lower chain,
$t_{0}$ denotes intra-ladder hopping in one of the chains, and 
$t_{1}$ is the diagonal hopping strength between the two
chains of the ladder, respectively. 
Physically, this ``topological comb model'' is closely related to
the one-dimensional periodic Anderson 
model in its non-interacting (spinless) limit, see Ref.\,[\onlinecite{Dagotto}].

\subsubsection{Zero-energy modes}
\label{sec:comb1}

The perfectly localized zero-energy modes in this 
case are $|j=1\rangle|u^-\rangle$ and $|j=N\rangle|u^+\rangle$,
that translate, after normalization, into the fermionic operators
\begin{eqnarray}
\label{combmode}
\eta_{1}^\dagger=\frac{1}{\sqrt{t_{0}^{2}+t_{1}^{2}}}(t_{1}c_{1}^\dagger-t_{0}f_{1}^\dagger),\quad\eta_{2}^\dagger
=f_{N}^\dagger .
\end{eqnarray}
The operator $\eta_{2}^\dagger$ trivially describes
a zero-energy mode, since it corresponds to the last fermion on the 
lower chain, that is decoupled from the rest. However, $\eta_{1}^\dagger$ corresponds to a 
non-trivial zero energy mode, localized over the first sites of the two chains. 
For large values of  $|t_0/t_1|$,
$\eta_{1}^\dagger$ is localized mostly on the $f$-chain, whereas 
for small values it is localized mostly on the $c$-chain.

Remarkably, such a non-trivial zero-energy mode is {\em robust} against arbitrary 
fluctuations in hopping strengths, despite the absence of a protecting chiral symmetry. 
Imagine that in Eq.\,\eqref{combH} the hopping strengths $t_{0,j}$ and $t_{1, j}$ are 
position-dependent. Then, $\widehat{H}$ may be written as 
\begin{eqnarray*}
\widehat{H} = -(t_{0,1} 
c^\dagger_{1}c_{2} +t_{1,1} c^\dagger_{2}f_{1} + \text{h.c.}) 
+ \widehat{G},
\end{eqnarray*}
where $\widehat{G}$ does not contain terms involving $c_{1}$ and $f_{1}$, 
so that $[\widehat{G},c_{1}]=0=[\widehat{G},f_{1}]$.
Then it is easy to verify that the expression for the zero-energy mode is 
obtained from $\eta_1^\dagger$ in Eq.\,\eqref{combmode} after substituting 
$t_0\mapsto t_{0,1}$ and $t_1\mapsto t_{1,1}$.
We conclude that the zero-energy edge mode is protected by an 
``emergent symmetry'', that has a non-trivial action only on the sites 
corresponding to $j=1$.  Likewise, assume for concreteness that 
$t_0=\pm t_1$, and consider the inter-chain perturbation
described by
\begin{eqnarray*}
\widehat{H}_1 \equiv  \mu\sum_{j=1}^{N}(c^\dagger_{j}\pm 
f^\dagger_{j})(c_{j}\pm f_{j}),\quad \mu\in\mathds{R} .
\end{eqnarray*} 
In this case, the corresponding single-particle Hamiltonian becomes \(
H=\mathds{1}_N\otimes h_0+ T\otimes h_{1}+T^{\dagger}\otimes h_{1}^{\dagger} \)
with 
\begin{eqnarray*}
h_0=\mu \begin{bmatrix}1 & \pm1 \\ \pm1 & 1 \end{bmatrix}.
\end{eqnarray*}
Nevertheless, the zero-energy mode corresponding to $|1\rangle |u^-\rangle$ is still an emergent 
solution for $\epsilon=0$, and can be verified to satisfy the boundary 
equation as well.  The topological nature of this zero-energy mode is confirmed by 
its non-trivial Berry phase \cite{bernevig} at half-filling. 
Under periodic BCs, the Hamiltonian in momentum space  is
\[
H_k = -\begin{bmatrix}2t_0\cos k & t_1 e^{-ik} \\ t_1 e^{ik} & 0 \end{bmatrix},
\]
leading to the following eigenvectors for the two bands:
\[ |u_{m k}\rangle = \begin{bmatrix} -t_0\cos k +(-1)^m \sqrt{t_0^2\cos^2 k + t_1^2}\\ -t_1 e^{ik} 
\end{bmatrix},\quad m=1,2.\]
Direct calculation shows that the Berry phase 
has the non-trivial value $\pi$ $(\text{mod}\, 2 \pi)$, as long as $t_1\ne0$.

\subsubsection{Complete closed-form solution}
\label{ccf}

We now obtain a complete closed-form solution of the eigenvalue 
problem corresponding to Eq.\,\eqref{combH} (open BCs). 
The reduced bulk Hamiltonian is
\begin{equation*}
H(z)=-\begin{bmatrix}t_0(z+z^{-1}) & t_1 z^{-1} \\ t_1 z & 0 \end{bmatrix},
\end{equation*}
with the associated polynomial ($R=1, d=2$)
\begin{equation}
\label{combz}
P(\epsilon,z)=z^2 [\epsilon^2+\epsilon t_0(z+z^{-1})-t_1^2] .
\end{equation}  
The model has two energy bands with a gap containing \(\epsilon=0\), 
and no chiral symmetry. Because $H$ is real, this enforces
the symmetry \(z\leftrightarrow z^{-1}\) of the non-zero roots of $P(\epsilon,z)$ that 
satisfy $z_1 z_2=1$. For generic $\epsilon\ne 0$, 
there are two distinct non-zero roots and, therefore, two extended bulk solutions. 
The eigenvector of $H(z)$ may be generically 
expressed as
\begin{equation*}
|u(\epsilon,z)\rangle=
\begin{bmatrix} 
\epsilon \\ 
-t_1 z 
\end{bmatrix}. 
\end{equation*} 
Using Eq.\,(\ref{s0}), the number of  emergent bulk solutions is $2Rd-2=2=2s_0$, one localized 
on each edge.
As $K^{-}(\epsilon)=h_1^\dagger$ and $K^{+}(\epsilon)=h_1$, 
such solutions are found from their kernels, 
spanned by $|u^-\rangle$ and $|u^+\rangle$, independently of $\epsilon$. 
The boundary matrix
\begin{eqnarray*}
B(\epsilon)=
\begin{bmatrix}
t_0\epsilon-t_1^2z_1	& t_0\epsilon-t_1^2z_2	&	0 & \epsilon t_1	 \\
0	&	0	&		0 &	-\epsilon t_0 		\\
z_1^{N+1}t_0\epsilon		&		z_2^{N+1}t_0\epsilon		&	0		&		0\\
z_1^{N+1}t_1\epsilon		&		z_2^{N+1}t_1\epsilon	&		-\epsilon &		0	
\end{bmatrix}, 
\end{eqnarray*}
whose kernel is nontrivial only if
\begin{equation*}
\epsilon t_{0}(z_1^{N+1}-z_2^{N+1})-t_{1}^{2}z_1z_2(z_1^{N}-z_2^{N})=0.
\end{equation*}
In this case, since $z_1z_2=1$, we may reduce this system to one variable by substituting 
$z_2=z_1^{-1}$, which then yields the polynomial equation
\begin{equation}
\label{bdcomb}
\epsilon t_{0}z_1^{2N+2}-t_{1}^{2}z_1^{2N+1}+t_{1}^{2}z_1-\epsilon t_{0}=0.
\end{equation}
The algebraic system of equations \eqref{combz} and \eqref{bdcomb}
determine the ``dispersing'' extended-support bulk modes of the system. When these equations are 
both satisfied, the kernel of the boundary matrix is spanned by
\[\bm{\alpha} = \frac{i}{2}\begin{bmatrix}z_1^{-(N+1)} & -z_1^{N+1} & 0 & 0\end{bmatrix}^{\rm T} , \]
and the corresponding eigenvectors of $H$ are given by
\[
|\epsilon\rangle = \frac{iz_1^{-(N+1)}}{2}|z_1,1\rangle\begin{bmatrix} \epsilon \\ -t_1 
z_1 \end{bmatrix} - \frac{iz_1^{N+1}}{2}|z_1^{-1},1\rangle\begin{bmatrix} \epsilon \\ -t_1 z_1^{-1} 
\end{bmatrix}, 
\]
which, upon substituting $z_1=e^{ik}$, can be recast as \cite{remarkAlternative}
\begin{equation}
\label{combeig}
|\epsilon\rangle = \sum_{j=1}^{N}|j\rangle
\begin{bmatrix} \epsilon\sin k(N+1-j) \\ -t_1 \sin k(N-j)\end{bmatrix}.
\end{equation}

To check whether $|\epsilon\rangle$ in Eq.\,\eqref{combeig} indeed satisfies the eigenvalue equation,
notice that
\begin{widetext}
\[
\langle j|H-\epsilon\mathds{1} | \epsilon\rangle =\left\{
\begin{array}{lcl}
-\epsilon\langle 1|\epsilon\rangle + h_1\langle 2|\epsilon\rangle   & \text{if} & j=1\\
h_1^\dagger\langle j-1|\epsilon\rangle
+h_1\langle j+1|\epsilon\rangle  -\epsilon\langle j|\epsilon\rangle & \text{if} & 2\le j \le N-1\\
h_1^\dagger\langle N-1|\epsilon\rangle  -\epsilon\langle N|\epsilon\rangle & \text{if} & j=N
\end{array}\right..
\]
\end{widetext}
Using the expression for $|\epsilon\rangle$, $\langle N|H-\epsilon\mathds{1} | \epsilon\rangle$
vanishes trivially, while, for $j=1$, 
\[
\langle 1|H-\epsilon\mathds{1} | \epsilon\rangle = -
\begin{bmatrix}
\epsilon t_0 \sin k(N-1) +\epsilon^2 \sin kN \\ 0
\end{bmatrix},
\]
which is seen to vanish from the relation
\begin{multline*}
\epsilon t_0 \sin k(N-1) +\epsilon^2 \sin kN = 
\\
\sin kN [\epsilon^2-t_1^2 + 2\epsilon t_0 \cos k]
 +[-\epsilon t_0 \sin k(N+1) + t_1^2 \sin kN].
\end{multline*}
The first term on the right hand-side is equal to $P(\epsilon,e^{ik})=0$,
whereas the second term vanishes due to Eq.\,\eqref{bdcomb}.
Finally, for $2\le j \le N-1$, we get
\[
\langle j|H-\epsilon\mathds{1} | \epsilon\rangle = -
\begin{bmatrix}
\sin k(N+1-j)[\epsilon^2-t_1^2+2\epsilon t_0 \cos k] \\ 0
\end{bmatrix}, 
\]
which equals zero, completing the argument.

Next, we find the values of $\epsilon$ for which Eq.\,\eqref{combz}
has a double root. The discriminant of $P(\epsilon,z)$ is
$D(P(\epsilon,z)) = (\epsilon^2-t_1^2)^2-4\epsilon^2t_0^2$, 
and vanishes for $\epsilon=-t_0\pm\sqrt{t_0^2+t_1^2}$ 
and $\epsilon=t_0\pm\sqrt{t_0^2+t_1^2}$, for which the corresponding double
roots are $z_1=+1$ and $z_1=-1$, respectively. In these cases, the bulk equation
may have power-law solutions. While one could construct the reduced bulk Hamiltonian $H_2(z)$
to identify these solutions, another quick way to proceed is suggested by 
Eq.\,\eqref{shortcut}, as already remarked in Sec. \ref{sec:revisited}. A power-law solution 
may now be written as 
\begin{eqnarray*}
\partial_{z_1}(|z_1,1\rangle|u(\epsilon,z_1)\rangle) \!= \! |z_1,2\rangle|u(\epsilon,z_1)\rangle
\!+ \!|z_1,1\rangle\partial_{z_1}|u(\epsilon,z_1)\rangle,
\end{eqnarray*}
where $z_1$ is the double root corresponding to $\epsilon$. 
The first column of the new boundary matrix remains the same as the original one, while 
its second column is determined from  
the derivative of the second column of the original boundary matrix with respect to $z_2$,
computed at $z_2=z_1$. 
For $\epsilon=-t_0\pm\sqrt{t_0^2+t_1^2}$, we have $z_1=1$ and
\[
B(\epsilon)  = 
\begin{bmatrix}
t_0\epsilon-t_1^2	& -t_1^2	& 0& \epsilon t_1 \\
0	&	0	&	0 & 	-\epsilon t_0 		\\
t_0\epsilon		&		(N+1)t_0\epsilon		&	0		&		0\\
t_1\epsilon		&		(N+1)t_1\epsilon	&		-\epsilon & 0
\end{bmatrix}.
\]
Some algebra reveals that $\det B(\epsilon) \ne0$, so that these values of $\epsilon$ do not
appear in the spectrum of $H$ for any values of parameters $t_0,t_1$. Similar analysis
for $\epsilon=t_0\pm\sqrt{t_0^2+t_1^2}$ yields the same conclusion. 
Therefore, there are no power-law solutions compatible with open BCs.

We now derive the perfectly localized zero energy modes described in Sec.\,\ref{sec:comb1}. 
Notice that for $\epsilon=0$, the only possible roots of $P(\epsilon,z)$ are $z_0=0$, and from its degree
it follows that there are $s_0=2$ emergent solutions on each edge. In this case,
\begin{eqnarray*}
K^{-}(0) =\begin{bmatrix} h_1^\dagger		& 0	\\ 0 	&	h_1^\dagger\end{bmatrix},
\end{eqnarray*}
with its kernel spanned by 
\begin{eqnarray*}
|u_1^-\rangle = \begin{bmatrix} |u^-\rangle  \quad 0 \end{bmatrix}^{\rm T} \quad \text{and} 	\quad 
|u_2^-\rangle = \begin{bmatrix} 0 \quad |u^-\rangle  \end{bmatrix}^{\rm T}.
\end{eqnarray*}
Similarly, the kernel of $K^{+}(0)$ is spanned by 
\begin{eqnarray*}
|u_1^+\rangle = \begin{bmatrix} |u^+\rangle \quad 0 \end{bmatrix}^{\rm T} \quad \text{and} 	\quad 
|u_2^+\rangle = \begin{bmatrix} 0 \quad |u^+\rangle  \end{bmatrix}^{\rm T}.
\end{eqnarray*}
Thus, the Ansatz for $\epsilon=0$ consists of all four perfectly 
localized solutions (see Eqs. \eqref{basisj-} and \eqref{basisj+}). The boundary matrix in this case is
\begin{eqnarray*}
B(\epsilon=0)= 
\begin{bmatrix}
0	&	0	&	0 &	t_1t_0\\
0	&	0	&	0 &	t_1^2\\
-t_1	&	0 &	0	&	0	\\
0	&	0 &	0	&	0	
\end{bmatrix},  
\end{eqnarray*}
which has a two-dimensional kernel, spanned by 
\[
\bm{\alpha}_1 = \begin{bmatrix}0 & 0 & 1 & 0\end{bmatrix}^{\rm T},\quad 
\bm{\alpha}_2 = \begin{bmatrix}0 & 1 & 0 & 0\end{bmatrix}^{\rm T}.
\]
The corresponding two zero-energy edge modes are then 
\[
|\epsilon=0, \bm{\alpha}_1 \rangle = |1\rangle|u^-\rangle,\quad
|\epsilon=0,\bm{\alpha}_2 \rangle =  |N\rangle|u^+\rangle ,
\] 
consistent with the results of Sec.\,\ref{sec:comb1}.
The eigenvector $|\epsilon=0,\bm{\alpha}_1\rangle$
has support only on the first site of the two band chain. 
Since $|N\rangle|u^+\rangle = |N\rangle[0 \quad1]^{\rm T}$,
the eigenvector $|\epsilon=0,\bm{\alpha}_2\rangle$ represents 
the decoupled degree of freedom at the right end of the chain,
as shown in Fig.\,\ref{comb} (a) and (b).

\subsection{The Majorana Chain}
\label{sec:Kitaev}

Kitaev's Majorana chain\cite{kitaev01} is a prototypical model of 
$p$-wave topological superconductivity \cite{alicea12, beenakker13}. 
In terms of spinless fermions, the relevant many-body Hamiltonian in the 
absence of disorder and under open BCs reads
\[ 
\widehat{H}_K=-\sum_{j=1}^{N}\mu\,c_{j}^{\dagger}c^{\;}_{j}-
\sum_{j=1}^{N-1}\left(t \, c_{j}^{\dagger}c^{\;}_{j+1}-\Delta
\, c_{j}^{\dagger}c_{j+1}^{\dagger}+ \text{h.c.} \right),
\]
where $\mu,t,\Delta \in {\mathbb R}$ denote the chemical potential, hopping amplitude,
and pairing strengths, respectively.  This Hamiltonian, expressed in spin language 
via a Jordan-Wigner transformation, describes the well-known anisotropic XY spin chain,  
which has a long history in quantum magnetism, including analysis of boundary effects 
for both open BCs and periodic \cite{liebschultzmattis,Mikeska,pfeuty,XYremark}. 

Expressed in the form of Eq.\,\eqref{HBBT}, the corresponding single-particle 
Hamiltonian is  
\begin{eqnarray}
\label{kitaevmat}
H_N=\mathds{1}_N\otimes h_0+(T\otimes h_1+ T^\dagger \otimes h_1^\dagger), \nonumber \\
\hspace*{-3mm}h_0=\begin{bmatrix}
-\mu & 0\\ 
0 & \mu 
\end{bmatrix}, \quad
h_1=
\begin{bmatrix}-t & \Delta\\
 -\Delta & t
\end{bmatrix}.
\end{eqnarray}
Thus, \(R=1\), \(d=2d_\text{int}=2\),  
and $h_R=h_1$ (hence the model) is invertible in the generic 
parameter regime $|t| \ne |\Delta|$, for arbitrary $\mu$.  We have already 
characterized in detail both the invertible regime \cite{abc} 
and the non-invertible regime \cite{JPA} for generic, regular energy values. 
While, given the importance of the model, we will summarize 
some of these results in what follows, our emphasis here will be on 
(i) addressing singular energy values, in particular, by directly 
computing compactly-supported eigenstates of flat-band 
eigenvectors directly in real space; 
(ii) uncovering the existence of zero-energy Majorana modes with a power-law 
prefactor, emerging in an invertible but non-generic parameter regime 
recently discussed in the context of transfer-matrix analysis \cite{Hegde16}.

\subsubsection{The parameter regime $|t|=|\Delta|$, $\mu\ne0$}

We briefly recall some key steps and results presented in Sec.\,5.2 of 
Ref.\,[\onlinecite{JPA}].
For concreteness, we assume $t=\Delta$, but a similar analysis 
may be repeated for the case $t=-\Delta$. The reduced bulk Hamiltonian
in this case is
\[ H(z) = \begin{bmatrix}
-\mu-t(z+z^{-1}) & t(z-z^{-1})\\
-t(z-z^{-1}) & \mu+t(z+z^{-1})
\end{bmatrix},\]
with associated polynomial 
\begin{equation}
\label{charkitaev1}
P(\epsilon,z)=-z^2    [ 2\mu t(z+z^{-1}) + (\mu^2 + 4t^2-\epsilon^2)].
\end{equation}
As in the topological comb example, for generic values of $\epsilon$ the above 
has two distinct non-zero roots $z_1$ and $z_2$, which implies a two-dimensional
space of extended bulk solutions and one emergent solution on each edge.     
Let the two extended solutions be labeled by 
$z_1$ and $z_2=z_1^{-1}$, with $|z_1|\le 1$. Then, we get 
\begin{eqnarray*}
|u(\epsilon,z_\ell)\rangle = 
\begin{bmatrix} 
t(z_\ell-z_\ell^{-1}) \\ 
\epsilon + \mu +t(z_\ell+z_\ell^{-1})
\end{bmatrix},\quad \ell=1,2.
\end{eqnarray*}
The two emergent solutions are obtained from the one-dimensional kernels of the matrices
$K^-(\epsilon) = h_1^\dagger$ and $K^+(\epsilon) = h_1$, which are
spanned by 
\[
|u_1^-\rangle = \begin{bmatrix}1 \\ -1\end{bmatrix}\quad \text{and} \quad
|u_1^+\rangle = \begin{bmatrix}1 \\ 1\end{bmatrix},
\]
respectively.
Following Eq.\,\eqref{Boundary matrix}, the boundary matrix is 
\begin{widetext}
\[
B(\epsilon) = 
\begin{bmatrix}
2t^2z_1 +t(\epsilon+\mu) & 2t^2z_1^{-1} +t(\epsilon+\mu) & 0 & -\mu-\epsilon\\
-2t^2z_1 +t(\epsilon+\mu) & -2t^2z_1^{-1} +t(\epsilon+\mu) & 0 & -\mu+\epsilon\\
z_1^{N+1}[-2t^2z_1^{-1}-t(\epsilon-\mu)] & z_1^{-(N+1)}[-2t^2z_1-t(\epsilon-\mu)]
& -\mu-\epsilon & 0\\
z_1^{N+1}[-2t^2z_1^{-1}-t(\epsilon-\mu)] & z_1^{-(N+1)}[-2t^2z_1-t(\epsilon-\mu)]
& \mu-\epsilon & 0\\
\end{bmatrix}.
\]
\end{widetext}
Our analysis in Ref.\,[\onlinecite{JPA}] shows that open BCs do not allow
any contributions from the emergent solutions in the energy eigenstates, 
which are linear combinations of the two extended solutions. 
The condition for $\epsilon$ to be an energy eigenvalue is $\det B(\epsilon)=0$, 
which simplifies to
\begin{eqnarray}
\label{kitaevcondition1}
2t z_1+\epsilon+\mu=\pm \, z_1^{(N+1)}(2t z_1^{-1}+\epsilon+\mu).
\end{eqnarray}
Explicitly, as long as $\epsilon \notin \mathcal{S}\equiv \{\mu\pm2t,-\mu\pm2t\}$, the 
corresponding eigenstate is 
\[ 
|\epsilon\rangle = |z_1,1\rangle|u(\epsilon,z_1)\rangle \mp z_1^{N+1}|z_1^{-1},1\rangle|u(\epsilon,z_1^{-1})\rangle. \]

The above equation is particularly interesting for zero energy,
since it dictates the {\em necessary and sufficient} conditions for the existence of
Majorana modes. For $\epsilon=0$, the root $z_1$ takes values
\[
z_1 = \left\{\begin{array}{lcl}  -\mu/2t & \;\text{if} & |\mu|<2|t| \\
-2t/\mu & \;\text{if} & |\mu|>2|t| \end{array}\right..
\]    
In the large-$N$ limit, the factor $z_1^{N+1}$ in the right hand-side of 
Eq.\,\eqref{kitaevcondition1} vanishes thanks to our choice of 
$|z_1|<1$. However, the left hand-side vanishes
{\em only} in the topologically non-trivial regime characterized by $|\mu|<2|t|$,
giving rise to a localized Majorana excitation. 
The unnormalized Majorana wavefunction in this limit is characterized by an exact 
exponential decay (see also Fig.\,\ref{fig:kitaev}), namely, 
\[
|\epsilon=0\rangle =\Big(\frac{4t^2-\mu^2}{2\mu}\Big) \sum_{j=1}^{\infty}z_1^j|j\rangle\begin{bmatrix}1 \\ -1\end{bmatrix}.
\]

For the analysis of the non-generic energy values in $\mathcal{S}$, we return to the finite 
system size $N$. For such $\epsilon$, $P(\epsilon,z)$ has double roots
at $z_1=1$ and $z_1=-1$, so that the bulk equation has one power-law solution in each case \cite{JPA}. 
These solutions are compatible with the BCs 
for {\em certain} points in the parameter space, determined by the condition $2tN +\mu (N+1) = 0$. 
Explicitly, the eigenstates corresponding to eigenvalues $\epsilon=\pm(\mu+2t)$ are then 
\begin{eqnarray*}
&|\epsilon=\mu+2t\rangle = 
\sum_{j=1}^N|j\rangle
\begin{bmatrix}
1\\
-1+\frac{2 j}{N+1}
\end{bmatrix}
,\\
&|\epsilon=- \mu-2t\rangle = 
\sum_{j=1}^N|j\rangle
\begin{bmatrix}
-1+\frac{2 j}{N+1}\\
1
\end{bmatrix} .
\end{eqnarray*}

\subsubsection{The parameter regime $|t|=|\Delta|,\ \mu=0$} 

This regime, sometimes affectionately called the ``sweet spot,'' 
is remarkable. Since the analytic continuation of the Bloch 
Hamiltonian is
\[
H(z)=t
\begin{bmatrix}
-(z+z^{-1})& z-z^{-1}\\
-(z-z^{-1})& z+z^{-1}
\end{bmatrix}, 
\]
 one finds that
\(
\det(H(z)-\epsilon\mathds{1}_2)=\epsilon^2-4t^2.
\)
Thus, the energies \(\epsilon=\pm 2t\) realize
a flat band and its charge conjugate. From the point of view
of the generalized Bloch theorem, these two energies are singular.  
According to Sec.\,\ref{flatbandid}, they 
necessarily belong to the physical spectrum of the Kitaev 
chain {\it regardless of BCs}, each yielding $\mathcal{O}(N)$
corresponding bulk-localized eigenvectors. 

In order to construct such eigenvectors, note that  
for $\epsilon=\pm2t$, the adjugate of $H(z)-\epsilon\mathds{1}_d$ is the matrix
\[
\text{adj}(H(z)\mp 2t\mathds{1}_d) = t\begin{bmatrix}
z+z^{-1}\mp 2 & -z+z^{-1} \\ z-z^{-1} & -z-z^{-1}\mp 2
\end{bmatrix},
\]
which immediately provides  two kernel vectors 
\begin{eqnarray*}
|v_{1,\pm}(z)\rangle = \begin{bmatrix}
1+z^{-2}\pm 2  z^{-1} \\ 1-z^{-2} \end{bmatrix},\\
|v_{2,\pm}(z)\rangle = \begin{bmatrix}
-1+z^{-2} \\ -1-z^{-2}\pm 2z^{-1}\end{bmatrix}.
\end{eqnarray*}
In this  case, we see that
the kernel vectors contain polynomials in $z^{-1}$ of degree $2<\delta_0=(d-1)2Rd =4$ (recall 
Eq.\,(\ref{delta0})).
For a suitable range of lattice coordinates \(j\)s, the compactly-supported sequences 
\begin{eqnarray*}
\Psi_{j1,\pm} = |j\rangle\begin{bmatrix} 1 \\ 1\end{bmatrix}
\pm 2 |j+1\rangle\begin{bmatrix} 1 \\ 0\end{bmatrix}
+|j+2\rangle\begin{bmatrix} 1 \\ -1\end{bmatrix},\\
\Psi_{j2,\pm} = -|j\rangle\begin{bmatrix} 1 \\ 1\end{bmatrix}
\pm 2 |j+1\rangle\begin{bmatrix} 0 \\ 1\end{bmatrix}
+|j+2\rangle\begin{bmatrix} 1 \\ -1\end{bmatrix}, \\
\end{eqnarray*}
yield non-zero solutions
\(|\Psi_{j \mu,\pm}\rangle=\bm{P}_{1,N} \Psi_{j \mu,\pm},\) $\mu=1,2$, 
of the bulk equation. However, it is not 
{\em a priori} clear how many of these are linearly independent.
For example, it is immediate to check that 
\[
\Psi_{j1,\pm}+\Psi_{j2,\pm} = \mp(\Psi_{j+1,2,\pm}-\Psi_{j+1,1,\pm}).
\]
In this case, a {basis} of compactly-supported solutions can 
be chosen from the states    
\begin{eqnarray*}
|\tilde{\Psi}_{0}\rangle &=& |1\rangle\begin{bmatrix} -1 \\ 1\end{bmatrix}\hspace*{2.4cm}  \text{if  } j=0, \\
|\tilde{\Psi}_{j,\pm}\rangle &=&|j\rangle\begin{bmatrix} 1 \\ 1\end{bmatrix}\pm
|j+1\rangle\begin{bmatrix} 1 \\ -1\end{bmatrix}\quad  \text{if  }  1\le j \le N-1, \\
|\tilde{\Psi}_{N}\rangle &=&|N\rangle\begin{bmatrix} 1 \\ 1\end{bmatrix}\hspace*{2.55cm}   \text{if  }  j=N, \\
\end{eqnarray*}
Out of these $N+1$ states, the ones corresponding to $j=1,\dots,N-1$ 
can be immediately checked to be eigenstates of
energy \( \epsilon \pm 2t\) \cite{fn}.
In contrast, $|\tilde{\Psi}_{0}\rangle$ and $|\tilde{\Psi}_{N}\rangle$ 
are {\em not} eigenstates: they do not satisfy the 
boundary equation trivially like other states localized in the bulk. 
We have thus found $2N-2$ eigenstates 
of the Hamiltonian, $N-1$ for each band $\epsilon = \pm 2t$. 

\begin{figure}[t]
\begin{flushright}
\includegraphics[width=8cm]{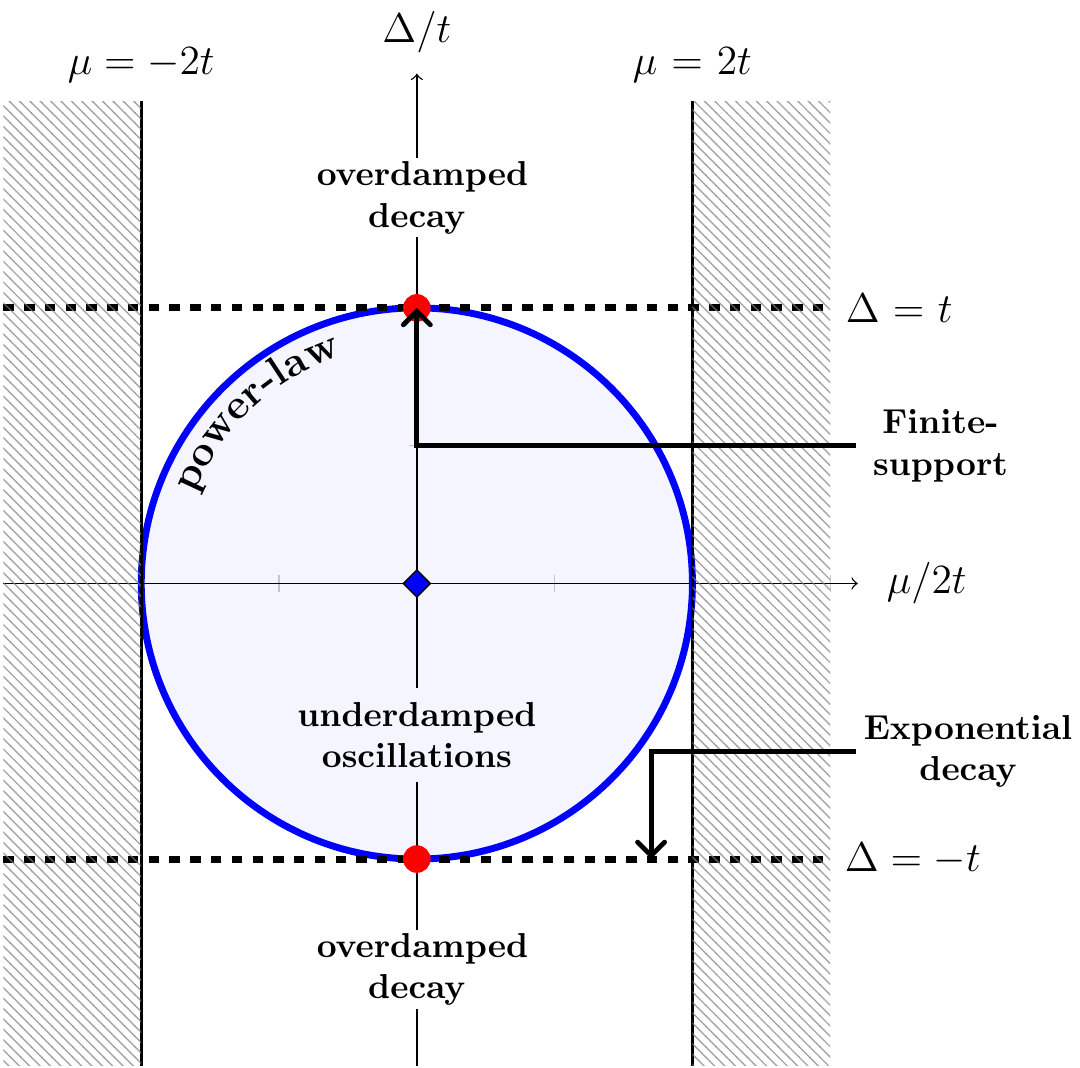}
\end{flushright}
\vspace*{-2mm}
\caption{(Color online) 
Spatial behavior of Majorana wavefunctions for
various parameter regimes of the Kitaev chain
under open BCs in the large-$N$ limit. 
The origin (blue diamond), $\mu=0,  \Delta =0$, corresponds to a metal 
at half filling. The region shaded in
black pattern is the trivial regime, which does not host 
Majoranas, and is separated from the non-trivial phase
by solid black lines indicating the critical points.
The interior of the circle of oscillations [Eq.\,\eqref{CircOsc}]
(shaded in light blue) hosts Majoranas whose wavefunction decays with 
oscillations, whereas the region outside show a behavior similar to 
overdamped decay of a
classical harmonic oscillator. On the circle, the wavefunction
decays exponentially with a power-law prefactor.
The ``sweet spots'' (red dots) host perfectly localized Majorana 
modes on the edge. 
\label{fig:kitaev}}
\end{figure}

The two missing eigenstates appear at $\epsilon=0$, which is a regular 
value of energy and so it is controlled by the generalized Bloch theorem. 
For \(\epsilon=0\), there are four emergent solutions (two on 
each edge), out of which only 
\[
|\psi^-\rangle = |1\rangle\begin{bmatrix} 1 \\ -1\end{bmatrix} = -|\tilde{\Psi}_{0}\rangle 
\;\;\mbox{and}\;\;
|\psi^+\rangle = |N\rangle\begin{bmatrix} 1 \\ 1\end{bmatrix} = |\tilde{\Psi}_{N}\rangle
\]  
are compatible with the BCs. Since these solutions are perfectly localized
on the two edges, they exist for any $N>2$ (see also Fig.\,\ref{fig:kitaev}).
Interestingly, the above states also appeared as solutions of the bulk equation at the singular 
energies \(\epsilon=\pm 2t\), and failed to satisfy the BCs at those values of energy. 
We do not know whether this fact is 
just a coincidence or has some deeper significance.

\vspace*{3mm}

\subsubsection{Majorana wavefunction oscillations in the regime $t\ne \Delta$}
 
Recently, it was shown \cite{Hegde16} that, 
inside the so-called ``circle of oscillations'', namely, the parameter regime
\begin{eqnarray}
\label{CircOsc}
\Big( \frac{\mu}{2t} \Big)^2 + \Big( \frac{\Delta}{t} \Big)^2 =1 ,
\end{eqnarray}
the Majorana wavefunction {\em oscillates while decaying} in space.
Such oscillations in Majorana wavefunction are not observed outside this circle.
This observation has consequences on the fermionic parity of the
ground state \cite{degottardi13}.
Because of duality, spin excitations in the XY chain 
show a similar behavior in the corresponding parameter regime \cite{XYremark} 
\( B_z^2=t^2-\Delta^2 = J_x J_y.\)
We now analyze this phenomenon by leveraging the analysis of Sec.\,\ref{sec:theory}.
For simplicity, we address directly the large-$N$ limit.

Clearly, whether a wavefunction
oscillates in space depends on the nature of the extended bulk solutions
that contribute to the wavefunction. In particular, let $|\psi\rangle = |z,1\rangle|u\rangle$
be one such bulk solution. For a wavefunction to be decaying 
asymptotically, we must have $|z|<1$. Further, if $z \in {\mathbb R}$,  
then $|\psi_j\rangle = z|\psi_{j-1}\rangle$ implies that the part of the
wavefunction associated to this bulk solution simply decays exponentially 
without any oscillations. On the other hand, if $z \equiv |z|e^{i\phi}$ with non-zero
phase, then a linear combination of vectors
\[ |z,1\rangle + |z^*,1\rangle = \sum_{j=1}^{N}2|z|^j \cos( \phi j )|j\rangle , 
\]
{\em can} show oscillatory behavior while decaying. This is precisely the phenomenon 
observed in this case. When $t\ne \Delta$,
the reduced bulk Hamiltonian is
\[
H(z) = \begin{bmatrix}
-\mu-t(z+z^{-1}) & \Delta(z-z^{-1})\\
-\Delta(z-z^{-1}) & \mu+t(z+z^{-1})
\end{bmatrix},
\]
with associated characteristic equation
\begin{equation}
\label{charkitaev2}
(z+z^{-1})^2(t^2-\Delta^2) + (z+z^{-1})(2\mu t) + (\mu^2 + 4\Delta^2-\epsilon^2)=0.
\end{equation}
For $\epsilon=0$, the above admits four distinct roots in general, out of which two
lie inside the unit circle and contribute to the Majorana mode on the left edge.
Whether any of these two roots is complex decides if the 
Majorana wavefunction oscillates for those parameter values. 
Notice that the characteristic equation is quadratic in the variable $\omega=z+z^{-1}$.
We get the two values of $\omega$ to be
\[
\omega_\pm = \frac{-\mu t\pm\Delta\sqrt{\mu^2-4(t^2-\Delta^2)}}{(t^2-\Delta^2)}.
\]
Likewise, notice that for $\mu^2<4(t^2-\Delta^2)$, we get both $\omega_+$ and $\omega_-$ to 
be complex, which necessarily means that both $z_1,z_2$ inside the unit 
circle are also necessarily complex.
Further, the symmetry of Eq.\,\eqref{charkitaev2} forces that $z_2=z_1^*$.
This leads to the oscillatory behavior of the Majorana wavefunction in the regime
\(\mu^2<4(t^2-\Delta^2)\), that is, {\em inside} the circle defined by Eq.\,\eqref{CircOsc}.
Thus, the spatial behavior of Majorana excitations in this regime is
formally similar to the solution of an underdamped classical harmonic
oscillator (see Fig.\,\ref{fig:kitaev}).
{\em Outside} the circle, the roots $\omega_\pm$
are real. With some algebra, it can be shown that $|\omega_\pm|>2$ in this regime,
which also means that both $z_1,z_2$ are real roots. This is  why 
oscillations are not observed in this parameter regime, in agreement
with the results of Ref.\,[\onlinecite{Hegde16}]. The Majorana
wavefunction in this case resembles qualitatively the solution of
a overdamped harmonic oscillator.

The situation when the parameters lie precisely {\it on} the circle is particularly interesting.
In this case, we find that $\omega_+= \omega_- \equiv \omega_0 = -4 t/\mu$.
Let us assume $t/\Delta>0$ for simplicity. It then follows 
that $z_1 =z_2  = -2(t-\Delta)/\mu$, which  
rightly indicates appearance of a power-law solution. Let us 
specifically analyze the case of open BCs on one end (for $N\gg 1$ as stated).
One of the two decaying bulk solutions
is $|\psi_{1,1}\rangle = |z_1,1\rangle|u(z_1)\rangle$, where
\[  |u(z)\rangle = \begin{bmatrix} 
\Delta(z-z^{-1}) \\ \mu + t(z+z^{-1}) \end{bmatrix}.
\]
The other bulk solution is obtained from
\begin{eqnarray*}
&|\psi_{1,2}\rangle &= \partial_{z_1}|\psi_{11}\rangle\\
&&\hspace*{-0.5cm}=z_1^{-1}|z_1,1\rangle\begin{bmatrix}
\Delta(z_1+z_1^{-1}) \\  t(z_1-z_1^{-1}) \end{bmatrix}
+|z_1,2\rangle\begin{bmatrix}
\Delta(z_1-z_1^{-1}) \\ \mu + t(z_1+z_1^{-1}) \end{bmatrix}.
\end{eqnarray*}
The relevant boundary matrix, 
\[
B(\epsilon=0) \equiv  \begin{bmatrix}
B_{11}(z_1) & B_{12}(z_1)\\
B_{21}(z_1) & B_{22}(z_1)
\end{bmatrix},
\]
may be computed by relating its second column to the partial derivative of 
the first column at $z=z_1$ as also done previously.  Explicitly: 
\begin{eqnarray*}
&&\begin{bmatrix}B_{11}(z_1) \\B_{21}(z_1)\end{bmatrix} = 
\begin{bmatrix}(2t z_1+\mu)\Delta \\-\mu t -z_1(t^2+\Delta^2)-z_1^{-1}(t^2-\Delta^2)\end{bmatrix}, \\
&&\begin{bmatrix}B_{12}(z_1) \\B_{22}(z_1)\end{bmatrix}= 
\begin{bmatrix}2t\Delta \\ -(t^2+\Delta^2) +z_1^{-2}(t^2-\Delta^2)
\end{bmatrix},
\end{eqnarray*}
where we also used Eq.\,\eqref{charkitaev2} for simplification.
Some algebra reveals that $B(0)$ has a one-dimensional kernel, spanned by the vector
\[\bm{\alpha} = \begin{bmatrix}-\mu t & 2\Delta(t-\Delta)\end{bmatrix}^{\rm T}.\]
This leads to the power-law Majorana wavefunction
\begin{eqnarray}
&|\epsilon=0\rangle &= -\mu t \,  |\psi_{1,1}\rangle + 
2\Delta(t-\Delta) \,  |\psi_{1,2}\rangle \nonumber\\
&&=\frac{8\Delta^2(t-\Delta)}{\mu}
\sum_{j=1}^{\infty}    j z_1^{j-1} |j\rangle 
\begin{bmatrix}
1 \\ -1
\end{bmatrix},
\end{eqnarray}
which decays exponentially with a linear prefactor (see Fig.\,\ref{fig:kitaev}).
In principle, the existence of such exotic Majorana modes could be probed 
in proposed Kitaev-chain realizations based on linear quantum dot arrays 
\cite{Fulga13}, which are expected to afford tunable control on all parameters.

\section{An indicator of the bulk-boundary correspondence}
\label{sec:indicator}

As stated in the Introduction, a main motivation behind the 
development of the generalized Bloch theorem is to elucidate 
the bulk-boundary correspondence. In this section, we start presenting an
indicator of bulk-boundary correspondence based on the results from 
Sec.\,\ref{sec:theory}, generalizing the original definition in Ref.\,[\onlinecite{abc}]. 
The indicator is built out of the boundary matrix and, therefore, encodes information 
from the bulk {\it and} the BCs. 
We will then consider an application of the indicator to study 
the Josephson response of an $s$-wave two-band topological superconductor 
\cite{swavePRL,swavePRB}. 
Interestingly, and to the best of our knowledge, this system provides the first example of 
an unconventional (fractional) Josephson effect {\em not} accompanied by a 
fermionic parity switch. We explain the physical reasons behind such a result. 

\subsection{Derivation of the indicator}
\label{sec:genindicator}

For a system of size $N$, the existence of localized modes at energy 
$\epsilon$ reflects into a non-trivial kernel of the corresponding
boundary matrix, which we now denote by $B_N(\epsilon)$ in order
to emphasize the dependence on $N$ and $\epsilon$.
As we increase $N$ without changing the BCs, 
the energy $\epsilon$ of the bound modes 
(that is, modes that remain asymptotically normalizable) attains a 
limiting value. For instance, in topologically non-trivial, 
particle-hole or chiral- symmetric systems under hard-wall BCs, the 
mid-gap bound modes attain zero energy in the large-$N$ limit. 
This convergence of bound modes and their energies is nicely
captured by a modified version of the boundary matrix in the limit $N \gg 1$, 
which we now construct.

\begin{figure}
\includegraphics[width=8cm]{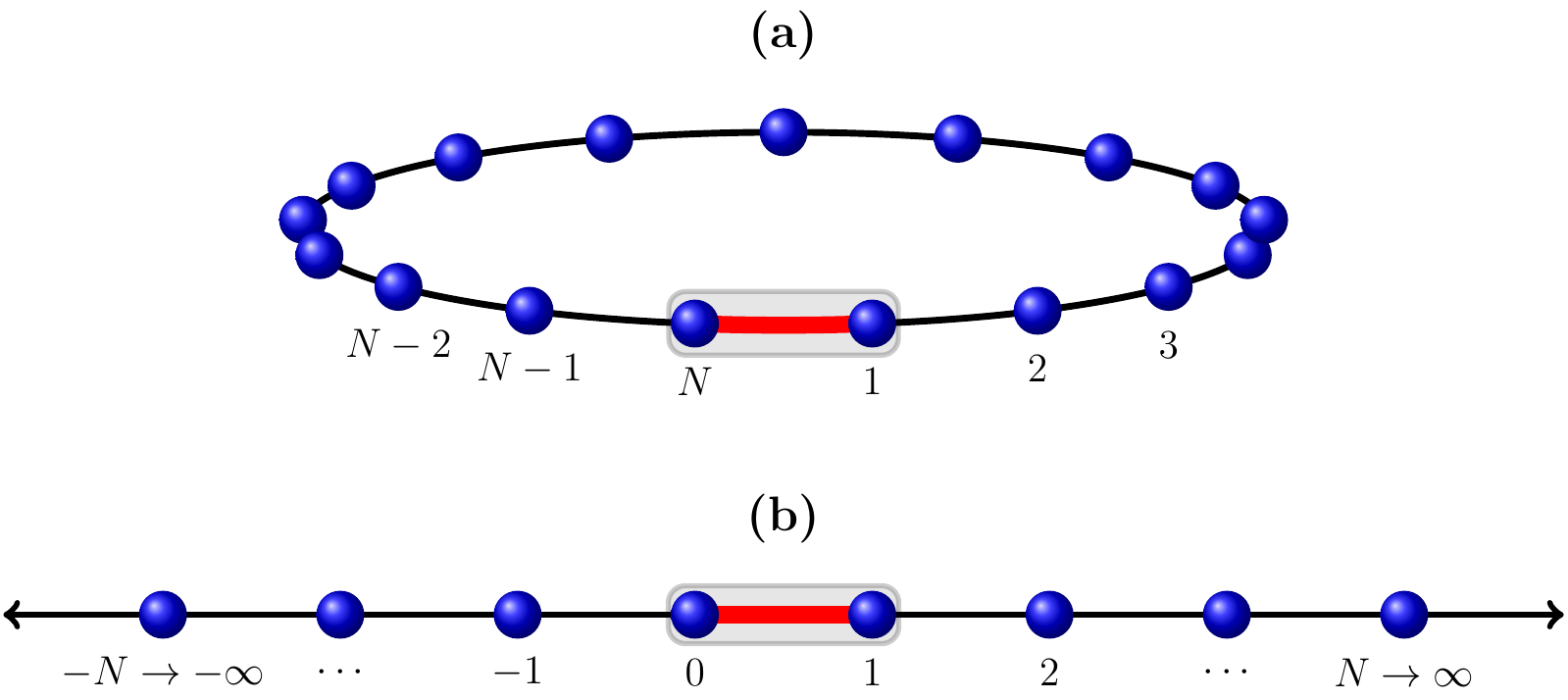}
\caption{(Color online) 
Ring (a) vs bridge (b) configurations of a chain Hamiltonian, $d=1=R$.
The solid (black) lines denote nearest-neighbor bulk 
hopping, whereas the thick (red) line indicates hopping
between the left ($j=N$) and the right ($j=1$) boundary 
(shaded gray rectangle). The bound states of (a) converge 
to the ones of (b) in the large-$N$ limit.
\label{fig:ringtobridge}}
\end{figure}

Consider a system of $N$ sites in a ring topology, as 
shown in Fig.\,\ref{fig:ringtobridge}(a), so as to allow non-zero contribution from 
the matrix $w_{bb'}$ in the BCs described by $W$ 
(see Eq.\,\eqref{Wmatrix}). Let us assume 
that the system hosts one or more bound modes near the junction formed 
by the two ends, which converge in the large-$N$ limit to energy 
$\epsilon$. The resulting modes are the bound modes
of a bridge configuration that extends to infinity on both sides, 
and where the boundary region is shown in Fig.\,\ref{fig:ringtobridge}(b).
For each $N$, we may express the bound eigenstate as in Eq.\,\eqref{ansatz}. 
Such bound states have contributions {\em only} from those bulk
solutions that are normalizable for $N\gg 1$.
The extended-support solutions corresponding to 
$|z_\ell |=1$ are not normalizable, and therefore 
must drop out from the Ansatz. 
Further, while the amplitude of those 
corresponding to $|z_\ell|>1$ blows up near $j=N$, they {\it remain}  
normalizable in the limit. This becomes apparent once
we rescale such solutions by 
$z_\ell^{-N}$. 
These rescaled solutions almost vanish at $j=1$ for large $N$.
Based on these considerations, we propose a modified Ansatz for finite $N$,
\begin{multline}
\label{ansatzind}
|\epsilon,\bm{\alpha}\rangle_N \equiv 
\sum_{|z_\ell|<1}\sum_{s=1}^{s_\ell}\alpha_{\ell s}|\psi_{\ell s}\rangle
+\sum_{s=1}^{s_-}\alpha^{-}_{s}|\psi^{-}_{s}\rangle + \\ 
\sum_{|z_\ell|>1}\sum_{s=1}^{s_\ell}\alpha_{\ell s}{z_\ell}^{-N}|\psi_{\ell s}\rangle 
+\sum_{s=1}^{s_+}\alpha^{+}_{s}|\psi^{+}_{s}\rangle  .
\end{multline}
expressed in terms of up most $2Rd$ amplitudes. 

The above Ansatz may be used to compute a corresponding boundary matrix 
${B}_N(\epsilon)$ in the same way as described in Sec.\,\ref{sec:boundeq}. Note that 
${B}_N(\epsilon)$ may not capture the bound modes appearing at finite $N$ since,
by construction, it does
not incorporate contributions from extended support solutions
corresponding to $|z_\ell|=1$. However, ${B}_\infty(\epsilon) \equiv \lim_{N\rightarrow \infty} {B}_N(\epsilon)$
is now well-defined, and describes accurately the presence and exact form of bound modes in the limit.
The condition for a non-trivial kernel becomes 
$\det [{B}_N^\dagger(\epsilon){B}_N(\epsilon)]=0$. 
Based on this condition, we define the quantity
\begin{equation}
\label{indicator}
\mathcal{D}_{\epsilon}\equiv\log\{\det[{B}_\infty(\epsilon)^\dagger{B}_\infty(\epsilon)]\}, 
\end{equation}
as an {\em indicator of bulk-boundary correspondence}. 
This captures precisely the interplay between the bulk properties and the BCs
that may lead to the emergence of bound modes, in the sense that, 
as we parametrically change either or both of the reduced bulk Hamiltonian and the BCs, 
$\mathcal{D}_\epsilon$ shows a singularity at (and only at) the 
parameter value for which the system hosts bound modes at energy $\epsilon$.  
Unlike most other topological indicators that are derived from bulk 
properties (i.e., in a torus topology), our indicator is constructed from a boundary matrix, 
that incorporates the relevant properties of the bulk. In cases where the 
bound modes are protected by a symmetry, this allows for the indicator to 
be computed for arbitrary BCs that respect the symmetry, paving the way to 
characterizing the robustness of the bound modes 
against classes of boundary perturbations.

An interesting situation is that of $w_{bb'}=0$, in which case the large-$N$
limit consists of two disjoint semi-infinite chains.  Then 
${B}_\infty(\epsilon)$ is block diagonal,  
\begin{eqnarray*}
{B}_\infty (\epsilon) = \begin{bmatrix} B_\infty^- (\epsilon) & 0 \\ 0 & B_\infty^+ (\epsilon)
\end{bmatrix},
\end{eqnarray*}  
where $B_\infty^-$ ($B_\infty^+$) may be interpreted as the 
boundary matrix of a semi-infinite chain, describing the edge 
modes at the left (right) edge, respectively.

While the indicator ${\cal D}_\epsilon$ of Eq.\,\eqref{indicator} signals the presence
of bound states, it does not convey information about the degeneracy 
of that energy level, which is nevertheless contained in the boundary matrix.
Therefore, it is often useful to also study the behavior of the {\em degeneracy indicator}
as a function of $\epsilon$:
\[
\mathcal{K}_{\epsilon}\equiv  \dim {\rm{Ker}} [{B}_\infty(\epsilon)]. \] 
In practice, the dimension of the kernel is
obtained by counting the number of zero singular values
of ${B}_\infty(\epsilon)$.

{\em Remark.---}
With reference to the discussion in Sec.\,\ref{sec:scan}, recall that
in numerical computations, ${B}_\infty(\epsilon)$
signals fictitious roots whenever the bulk equation has a power-law solution. In such cases, we 
once again remedy the issue by resorting to the Gramian. Then the corrected value of
the indicator is given by
\[ \mathcal{D}_{\epsilon}
= \log\left\{\frac{\det[{B}_\infty(\epsilon)^\dagger{B}_\infty(\epsilon)]}
{\det \mathcal{G} (\epsilon)}\right\}.
\]
Thus, the correct degeneracy of the energy is obtained by counting
zero (within numerical accuracy) singular values of 
the matrix $\tilde{B}_\infty(\epsilon) = B_\infty(\epsilon)\mathcal{G} (\epsilon)^{-1/2}$.

\subsection{Application: An $s$-wave topological superconducting wire}
\label{1Dswavetoponductor}

The usefulness of the proposed indicator of bulk-boundary correspondence was 
demonstrated in the context of characterizing the Josephson response of a two-band 
time-reversal invariant $s$-wave topological superconducting wire in Ref.\,[\onlinecite{abc}]. 
While the calculations reported there employed a simplified Ansatz, including only 
extended-support solutions of the bulk equation, 
we now validate the analysis by using the complete Ansatz given in Eqs. (\ref{ansatz}) 
and (\ref{ansatzind}), and further analyze and interpret our results in terms of fermionic parity switches.

The relevant $s$-wave, spin-singlet, two-band superconductor model
\cite{swavePRL,swavePRB} derives its topological nature from the interplay between 
a Dimmock-type intra-band spin-orbit coupling and inter-band hybridization terms.
Due to the spin degree of freedom in each of the two relevant orbitals, say, 
$c$ and $d$, the Nambu basis corresponding to an atom at position $j$
consists of 8 fermionic operators, that we write as the vector 
$$\hat{\Psi}_{j}^{\dagger}=\begin{bmatrix}c_{j,\uparrow}^{\dagger} & c_{j,
\downarrow}^{\dagger} & d_{j,\uparrow}^{\dagger} & d_{j,\downarrow}^{\dagger} 
& c_{j,\uparrow} & c_{j,\downarrow} & d_{j,\uparrow} & d_{j,\downarrow}\end{bmatrix}.$$
In this basis, the single-particle Hamiltonian under open BCs is given by 
\begin{eqnarray*} \label{intmat}
H_N&=&\mathds{1}_N\otimes h_0+(T\otimes h_1+ T^\dagger \otimes h_1^\dagger),\\
h_{0}&=&
\begin{bmatrix}
-\mu & u_{cd} & -i\Delta\sigma_y & 0\\
u_{cd} & -\mu & 0 & i\Delta\sigma_y\\
i\Delta\sigma_y & 0 & \mu & -u_{cd} \\
0 & -i\Delta\sigma_y & -u_{cd} & \mu
\end{bmatrix}\\
&=&-\mu\tau_{z}+u_{cd}\tau_{z}\nu_{x}
+\Delta\tau_{y}\nu_{z}\sigma_{y},\nonumber \\
\nonumber\\
h_1&=&
\begin{bmatrix}
i\lambda\sigma_x & -t & 0 & 0\\
-t & -i\lambda\sigma_x & 0 & 0\\
0 & 0 & i\lambda\sigma_x & t\\
0 & 0 & t &-i\lambda\sigma_x
\end{bmatrix}\\
&=&-t\tau_{z}\nu_{x}+i\lambda\nu_{z}\sigma_{x},
\end{eqnarray*}
where the real parameters $\mu,u_{cd},t,\lambda,\Delta$ denote the chemical potential, 
the interband hybridization, hopping, spin-orbit coupling and
pairing potential strengths, respectively, and $\tau_\alpha, \nu_\alpha,\sigma_\alpha$, 
$\alpha=\{x,y,z\}$, are Pauli matrices in Nambu, orbital and spin spaces. 

The topological properties of the above Hamiltonian were analyzed in Ref.\,[\onlinecite{swavePRB}].
The BdG Hamiltonian is time-reversal invariant, which places it in the symmetry class DIII. 
The topological phases may thus be distinguished by a ${\mathbb Z}_2$-invariant, 
given by the parity of the sum of the Berry phases for the two occupied negative bands in 
one of the Kramers' sectors {\em only} \cite{swavePRB}.  For open BCs and for non-vanishing 
pairing, the system in its trivial phases was found to host zero or two pairs of Majoranas 
on each edge, in contrast to the topologically non-trivial phase supporting one pair of 
Majoranas per edge. Similar to the two-dimensional version of the model, 
one may see that the existence of such Majorana modes is protected by a non-trivial chiral 
symmetry, of the form $\tau_y \sigma_z$.
The single-particle Hamiltonian $H_N$ for open BCs can be exactly 
diagonalized as described in Sec. \ref{sec:algo}. 
In the large $N$-limit, the boundary matrix $B_\infty(\epsilon=0)$ calculated 
by using the Ansatz in Eq.\,(\ref{ansatzind}) yields degeneracy ${\cal K}_0= 0,4,8$ 
in the no-pair, one-pair, and two-pair phases, respectively, verifying the bulk-boundary 
correspondence previously established through numerical diagonalization.

\subsubsection{Josephson response} 

\begin{figure}
\includegraphics[width=8cm]{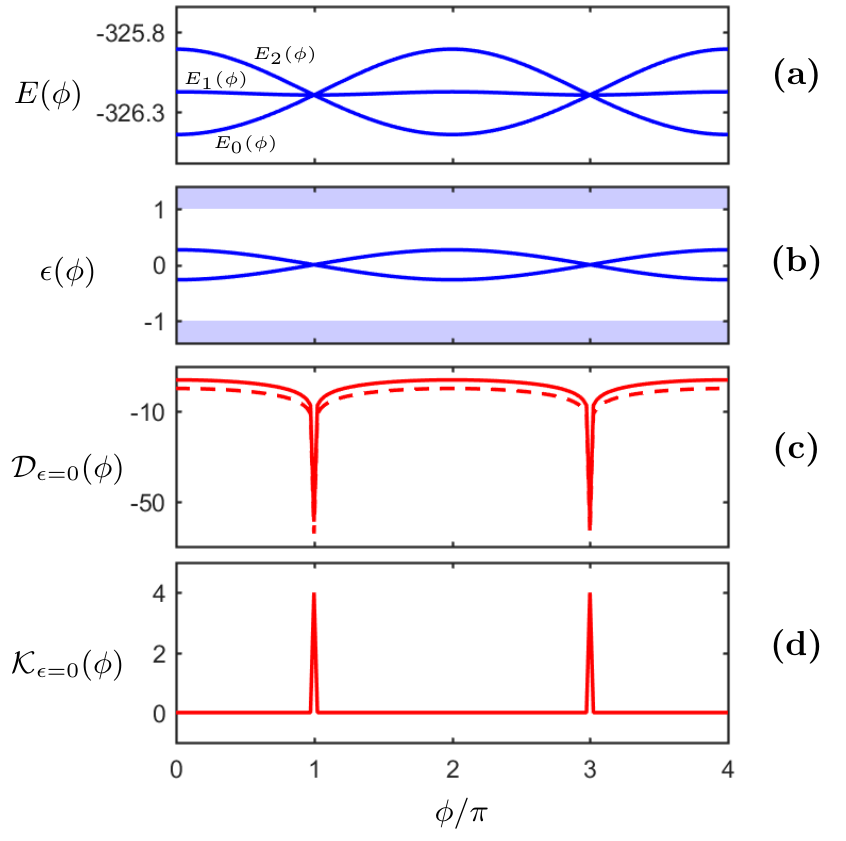}
\vspace*{-3mm}
\caption{(Color online)  
(a) Low-lying many-body energy eigenvalues in the Josephson
ring configuration, as a function of flux $\phi$. The energy level $E_1(\phi)$
is doubly degenerate.
(b) Energy of the bound mode and its anti-particle excitation.
The shaded (blue) area denotes the continuum of
energy states in the bulk.
(c) Comparison of the indicator defined 
in Ref.\,[\onlinecite{abc}] (dashed red line) and 
the generalized indicator of Eq.\,(\ref{indicator}) (solid red line)
in the topologically non-trivial phase. 
(d) Degeneracy of the zero-energy level inferred from the
dimension of the kernel of $B_\infty(\epsilon=0,\phi)$.
The parameters are $w=0.2$,
$\mu=0$, $u_{cd}=t=\lambda=1$,
$\Delta = 2$, $N=60$ in (a) and (b).
\label{fig:swave}}
\end{figure}

In the Josephson ring configuration considered in Ref.\,[\onlinecite{abc}],  
the first and last sites
of the open chain are coupled by the same 
hopping and spin-orbit terms as in the rest of the chain, only weaker by a factor 
of $1/w$. A flux $\phi$ is introduced between the two
ends via this weak link. In the large-$N$ limit, this link acts as a junction, 
with the corresponding tunneling term in the many-body Hamiltonian being given by 
$$\widehat{H}_{T}(\phi)=\hat{\Psi}_{N}^{\dagger}(wh_{1}U_\phi)\hat{\Psi}_{1}
+ \,\text{h.c.}, \, U_\phi=\begin{bmatrix}e^{i\phi/2}\mathds{1}_{4} & \!\!\!0\\
0 & \!\!\!e^{-i\phi/2}\mathds{1}_{4}
\end{bmatrix}\!.$$ 
The total Hamiltonian is then 
$\widehat{H}(\phi) = \widehat{H}_N + \widehat{H}_{T}(\phi)$.
It was demonstrated \cite{abc} that the Hamiltonian 
displays fractional Josephson effect in the topologically non-trivial phase, 
as inferred from its $4\pi$-periodic many-body ground state energy [Fig.\,\ref{fig:swave}(a)], 
with the phenomenon being observed {\em only} if the open-chain Hamiltonian 
correspondingly hosts an odd number of Majorana pairs per edge.
The physics behind the $4\pi$-periodicity was explained in terms of the crossing of 
a positive and a negative single-particle energy level happening at precisely zero energy 
as a function of flux $\phi$. 

The singular behavior resulting at flux values $\phi=\pi,3\pi$ from the indicator 
${\cal D}_{\epsilon=0}(\phi)$ computed using both the simplified Ansatz as in 
Ref.\,[\onlinecite{abc}] and the complete Ansatz of Eq.\,(\ref{ansatzind}) is shown 
in Fig.\,\ref{fig:swave}(c). The qualitative features are clearly unchanged,  
indicating that in the large-$N$ limit the bound modes formed near the junction are linear 
combinations {\em only} of extended-support solutions, with no contributions from emergent ones.
As seen in Fig.\,\ref{fig:swave}(d), at both $\phi=\pi $ and $\phi=3\pi$ the junction hosts a total of four Majoranas.

\subsubsection{Parity switch and decoupling transformation}

Despite the $4\pi$-periodic Josephson response witnessed in the topologically non-trivial phase, 
it turns out that the ground state fermionic parity {\it remains unchanged
for all flux values}. In the non-trivial regime of interest, we may focus on the three low-lying energy levels. 
Specifically, for values of $\phi < \pi$, let $|\Phi(\phi)\rangle$
denote the many-body ground state, with energy $E_0(\phi)$, as in 
Fig.\,\ref{fig:swave}(a).  As we will show, 
there are two degenerate quasi-particle excitations, say, 
$\eta_1(\phi),\eta_2(\phi)$, with small positive energy $\epsilon_0(\phi)$.
This results in a {\em two-fold degenerate first excited many-body state}, with energy 
$E_1(\phi)=E_0(\phi)+\epsilon_0(\phi)$, and a corresponding 
eigenspace is spanned by 
$\{ \eta_1^\dagger(\phi)|\Phi(\phi)\rangle,\eta_2^\dagger(\phi)|\Phi(\phi)\rangle\}$. 
The second excited state, 
$\eta_1^\dagger(\phi)\eta_2^\dagger(\phi)|\Phi(\phi)\rangle$, is not degenerate and has 
energy $E_2(\phi)=E_0(\phi)+2\epsilon_0(\phi)$. Note that this state
has the same (even) fermionic parity as the ground state. At $\phi=\pi$, the 
quasi-particle excitation has exactly zero energy,  $\epsilon_0(\pi)=0$,
causing all three energy levels to become degenerate. As 
$\phi$ crosses $\pi$, $\epsilon_0(\phi)$ becomes negative. Therefore, for 
$\pi<\phi<3\pi$, we find that $E_2(\phi)<E_1(\phi)<E_0(\phi)$. The continuation 
of the state $\eta_1^\dagger(\phi)\eta_2^\dagger(\phi)|\Phi(\phi)\rangle$ with energy $E_2(\phi)$ 
thus becomes the new ground state, whereas
the continuation of the original ground state $|\Phi(\phi)\rangle$ now attains the maximum
energy among these three levels. Since the new ground state has the same parity as the original one,
the system shows no parity switch, with a similar analysis holding for the crossover at $\phi=3\pi$.
We conclude that the absence of a fermionic parity switch originates from 
the twofold degeneracy of the single-particle energy levels.

While the system under open BCs is time-reversal invariant, away from $\phi=0,2\pi$  
this symmetry is broken by the tunneling term $\widehat{H}_{T}(\phi)$. 
Therefore, Kramer's theorem is not responsible in general for the 
degeneracy in the single-particle levels. Instead, we now explain the physical origin of 
this degeneracy in terms of a ``decoupling transformation'' in real space, thanks to which 
the system in the Josephson bridge configuration is mapped into 
{\em two decoupled systems} in the same configuration, each with half the number
of internal degrees of freedom as the original one.  Although each of these smaller systems 
does undergo a parity switch, the total parity being the sum of individual parities remains
unchanged. 

Observe that the Hamiltonian $\widehat{H}(\phi)$ is invariant under the unitary symmetries
$\hat{S}_1$ and $\hat{S}_2$, defined by the action 
\begin{eqnarray*}
&&\hat{S}_1:\; c_\uparrow (d_\uparrow) \mapsto d_\uparrow (c_\uparrow),\quad
c_\downarrow (d_\downarrow) \mapsto -d_\downarrow (-c_\downarrow),\\
&&\hat{S}_2: \; c_\uparrow (d_\uparrow) \mapsto ic_\downarrow (id_\downarrow),\quad
c_\downarrow (d_\downarrow) \mapsto ic_\uparrow (id_\uparrow).
\end{eqnarray*}
We can use the eigenbasis of $\hat{S}_1$ to decouple $\widehat{H}(\phi)$
into two independent Hamiltonians. Consider, for each site $j=1,\ldots, N$, the canonical transformation 
\begin{eqnarray}
\label{dec}
a_{j\sigma} \equiv \frac{c_{j\sigma} + d_{j\sigma}}{\sqrt{2}},\quad 
b_{j\sigma} \equiv \frac{c_{j\sigma} - d_{j\sigma}}{\sqrt{2}},\quad 
\sigma= \uparrow,\downarrow.\quad
\end{eqnarray}
and let $\hat{U}_1$ be the unitary change of basis defined by $\hat{U}_1: \hat{\Psi}_{j}^{\dagger} \mapsto
[\,\hat{\Psi}_{+,j}^{\dagger} \; \hat{\Psi}_{-,j}^{\dagger}]$, where 
\begin{eqnarray*} 
\hat{\Psi}_{+,j}^{\dagger}&\equiv &[\,a_{j,\uparrow}^{\dagger} \quad\; b_{j,
\downarrow}^{\dagger} \quad\;  a_{j,\uparrow} \quad\; b_{j,\downarrow}\,],\nonumber\\
\hat{\Psi}_{-,j}^{\dagger}&\equiv &[\,a_{j,\downarrow}^{\dagger} \ -b_{j,
\uparrow}^{\dagger} \quad  a_{j,\downarrow} \ -b_{j,\uparrow}\,]. 
\end{eqnarray*} 
By letting $\hat{\Psi}_{\pm}^{\dagger} \equiv [\hat{\Psi}_{\pm,1}^{\dagger} \; \dots \;\hat{\Psi}_{\pm,N}^{\dagger}]$,
the action of $\hat{U}_1$ then decouples $\widehat{H}(\phi)$ according to 
\begin{eqnarray*}
&\widehat{H}(\phi) \equiv \widehat{H}_+(\phi) + \widehat{H}_-(\phi)= \hat{\Psi}_{+}^{\dagger}H_+(\phi)\hat{\Psi}_{+}+ 
\hat{\Psi}_{-}^{\dagger}H_-(\phi)\hat{\Psi}_{-},
\end{eqnarray*} 
where $\widehat{H}_\pm(\phi)$ describes two smaller systems, each in a Josephson ring configuration, 
with hopping and pairing amplitudes given by 
\begin{eqnarray*}
\label{intmat1}
h_{\pm,0}& = &
\begin{bmatrix}
-\mu+u_{cd}\tilde{\sigma}_{z} & -i\Delta\tilde{\sigma}_{y}\\
i\Delta\tilde{\sigma}_{y} & \mu-u_{cd}\tilde{\sigma}_{z}
\end{bmatrix} 
\nonumber\\[2pt]
&= &\;
-\mu\tau_{z}+u_{cd}\tau_{z}\tilde{\sigma}_{z}+
\Delta\tau_{y}\tilde{\sigma}_{y},\nonumber \\
\nonumber\\[-6pt]
h_{\pm,1}&= &
\begin{bmatrix}
\pm i\lambda\tilde{\sigma}_{x} -t\tilde{\sigma}_{z} & 0\\
0 & \pm i\lambda\tilde{\sigma}_{x} + t\tilde{\sigma}_{z}
\end{bmatrix}\\[2pt]
&=&\;\pm i\lambda \tilde{\sigma}_{x}-t\tau_{z}\tilde{\sigma}_{z},
\end{eqnarray*}     
with $\tilde{\sigma}_\alpha$ denoting Pauli matrices in the 
modified spin basis.
The decoupling transformation in Eq.\,(\ref{dec}) is close in spirit to the one already 
employed under periodic BCs \cite{swavePRL,swavePRB}. Indeed, it is worth remarking that 
$\hat{\Psi}_{+,j}$ and $\hat{\Psi}_{-,j}$ are still time-reversals of 
each other, in the sense that $\mathcal{T}\hat{\Psi}_{+,j}^{\dagger}\mathcal{T}^{-1}=
\hat{\Psi}_{-,j}^{\dagger}$, with ${\cal T}$ being the anti-unitary time-reversal operator for 
the system. Because of the tunneling term, however, 
the two decoupled (commuting) Hamiltonians $\widehat{H}_\pm(\phi)$
are related by $\mathcal{T} \widehat{H}_{+}(\phi)\mathcal{T}^{-1} = \widehat{H}_{-}(4\pi-\phi)$.

It now remains to show that $\widehat{H}_\pm(\phi)$ 
have identical single-particle energy spectrum, and therefore lead to the desired
degeneracy in the energy levels of $\widehat{H}(\phi)$.
This follows by examining the symmetries of the single-particle BdG Hamiltonian $H(\phi)$. 
Corresponding to $\hat{S}_1$, 
$H(\phi)$ has a unitary symmetry 
$S_1=\mathds{1}_N \otimes \nu_x\sigma_z$, and thus gets block-diagonalized 
into two blocks, $H_\pm(\phi)$, upon the action of $U_1$.
Similarly, corresponding to $\hat{S}_2$, $H(\phi)$ has another unitary symmetry 
$S_2 = i\mathds{1}_N\otimes \tau_z \sigma_x$.
Further, $S_1$ and $S_2$ satisfy the anti-commutation relation $\{S_1,S_2\}=0$,
which is responsible for the doubly degenerate eigenvalue spectrum \cite{fn6}.
In fact, one can also verify directly that $\widehat{H}_+(\phi)$ and $\widehat{H}_-(\phi)$
satisfy  $\hat{S}_2 \widehat{H}_+(\phi) \hat{S}_2^\dagger = \widehat{H}_-(\phi)$.
This explains the origin of the double degeneracy of each single-particle energy level,
and hence of the absence of fermionic parity switch.

\section{Transfer matrix in the light of the generalized Bloch theorem}
\label{sec:TM}

Starting with the work in Refs. [\onlinecite{HatsugaiPRL}]-[\onlinecite{HatsugaiPRB}], the 
transfer matrix has remained the tool of choice for analytical 
investigations of the bulk-boundary correspondence 
\cite{DelplaceZakPhase,MongEdge,MaoAnalytic,DwivediBB} including, as
mentioned, recent studies of Majorana wavefunctions in both clean and 
disordered Kitaev wires \cite{Hegde16}.
In this section, we revisit the transfer matrix approach to band-structure 
determination in the light of our generalized Bloch theorem. In particular, we show how, in 
situations where the transfer matrix fails to be diagonalizable, our analysis makes it possible to give physical 
meaning to the generalized eigenvectors by relating them to the power-law solutions 
discussed in Sec. \ref{sec:bulkeq}.

\subsection{Basics of the standard transfer matrix method}

While our conclusions apply more generally to arbitrary finite-range clean models, for concreteness 
we refer in our discussion to the simplest setting 
where both approaches are applicable, namely, a one-dimensional chain with 
nearest-neighbor hopping.  We further focus on {\em open (hard-wall) BCs}, as most commonly 
employed in transfer-matrix studies. The relevant single particle-Hamiltonian $H_N$
is then a tridiagonal block-Toeplitz matrix, with entries $h_1^\dagger,h_0$ and $h_1$
along the three diagonals. Generically, $h_{1}$ is assumed to be invertible. 
The starting point of the method entails obtaining the recurrence relation between 
eigenvector components. Specifically, if 
$|\epsilon\rangle=\sum_{j=1}^{N}|j\rangle|\psi_j\rangle$ is an eigenvector
of $H$ with energy eigenvalue $\epsilon$ relative to the usual Hilbert-space factorization 
${\cal H}={\cal H}_L \otimes {\cal H}_I$, the components $|\psi_j\rangle$ satisfy the recurrence relation 
\begin{equation}
\label{bulkcomp1}
h_1^\dagger |\psi_{j-1}\rangle + (h_0-\epsilon\mathds{1})|\psi_j\rangle 
+ h_1|\psi_{j+1}\rangle = 0,\quad 2\le j\le N-1.
\end{equation}
In terms of the $2d \times 2d$ {\it transfer matrix}
\begin{equation}
\label{TM}
t(\epsilon) \equiv 
\begin{bmatrix}0 & \mathds{1}_d \\ -h_{1}^{-1}h_{1}^{\dagger} & -h_{1}^{-1}(h_{0}-\epsilon\mathds{1})\end{bmatrix},
\end{equation}
the above recurrence relation may be reformulated as
\begin{eqnarray}
\label{TM1}
\bm{P}_{j,j+1}|\epsilon\rangle = t(\epsilon)\bm{P}_{j-1,j}|\epsilon\rangle,
\quad 2\le j\le N-1,
\end{eqnarray}
where we have written 
$\bm{P}_{j,j+1}|\epsilon\rangle \equiv \begin{bmatrix}|\psi_{j}\rangle &
|\psi_{j+1}\rangle \end{bmatrix}^{\rm T}$. 
Thus, 
\begin{eqnarray}
\label{TMsteps}
\bm{P}_{j+1,j+2}|\epsilon\rangle = t(\epsilon)^j\bm{P}_{1,2}|\epsilon\rangle,
\quad 0\le j \le N-2, 
\end{eqnarray}
which can be leveraged for obtaining the complete set of eigenvectors 
of $H_N$.  We can define $|\psi_{0}\rangle,|\psi_{N+1}\rangle$
by using the relations
\[
\bm{P}_{1,2}|\epsilon\rangle = t(\epsilon)\bm{P}_{0,1}|\epsilon\rangle,\quad
\bm{P}_{N,N+1}|\epsilon\rangle = t(\epsilon)\bm{P}_{N-1,N}|\epsilon\rangle, \]
so that $\bm{P}_{N,N+1}|\epsilon\rangle = T(\epsilon)\bm{P}_{0,1}|\epsilon\rangle$
in terms of the matrix $T(\epsilon) \equiv {t(\epsilon)}^{N}$. Hard-wall BCs enforce
$|\psi_{0}\rangle=0= |\psi_{N+1}\rangle$.
Substituting these boundary values leads to
\begin{equation*}
\label{TMbound}
\begin{bmatrix} |\psi_{N}\rangle \\ 0 \end{bmatrix}=\begin{bmatrix}
T_{11}(\epsilon) & T_{12}(\epsilon) \\ 
T_{21}(\epsilon) & T_{22}(\epsilon)
\end{bmatrix}\begin{bmatrix} 0 \\ |\psi_{1}\rangle \end{bmatrix},
\end{equation*}
which has a non-trivial solution if and only if 
\begin{eqnarray}
\label{TMboundmat}
\det \,T_{22}(\epsilon)=0.
\end{eqnarray}    
Therefore, all values of $\epsilon$ that obey the above condition are 
eigenvalues of $H_N$. For each eigenvalue, the corresponding $|\psi_1\rangle$
is obtained as the kernel of $T_{22}(\epsilon)$.
In practice, $T(\epsilon)$ is calculated by first diagonalizing 
$t(\epsilon)$ by a similarity transformation, and then 
exponentiating the eigenvalues along its diagonal \cite{Lee}.

As can be appreciated from this example,
the standard version of the transfer matrix method relies on 
invertibility of certain matrices, although ``inversion-free'' \cite{Boykin, Biczo}
or partially inversion-free \cite{DwivediBB} modifications have also been 
suggested. In the standard case, the only prerequisite 
for constructing $t(\epsilon)$ at each step is the banded structure of the single-particle Hamiltonian 
and, most importantly, the resulting matrix $T(\epsilon)$ is assumed to be diagonalizable.

\subsection{Connections to the generalized Bloch theorem}
\label{sec:transbulk}

In order to relate the above analysis to the generalized Bloch formalism, the key observation is to 
note that the set of equations in Eq.\,\eqref{bulkcomp1} constitute the complete bulk equation, 
as described in Sec.\,\ref{sec:BBseparation}.
Consequently, Eq.\,\eqref{TM1} is satisfied
by any bulk solution $|\psi\rangle \in \mathcal{M}_{1,N}$, where $\mathcal{M}_{1,N}$ 
denotes the bulk solution space as usual. It is insightful to recast Eq.\,\eqref{TMsteps}
in the form
\[
t(\epsilon)^j\bm{P}_{1,2}|\psi\rangle = \bm{P}_{1,2} \, (T)^j|\psi\rangle, \quad 0\le j \le N-2,
\]
suggesting that the action of the transfer matrix in the bulk solution space 
is closely related to the one of the left shift $T$.
When restricted to $\mathcal{M}_{1,N}$, the above yields the following operator identity: 
\begin{eqnarray}
\label{connection}
(t(\epsilon)-z\mathds{1}_{d})^j\bm{P}_{1,2}\Big|_{\mathcal{M}_{1,N}} = 
\bm{P}_{1,2}(T-z\mathds{1}_{N})^j\Big|_{\mathcal{M}_{1,N}}\!,\quad
\end{eqnarray}
with $ z\in\mathds{C}$.
This relation may be used to establish a direct connection between the basis of the bulk solution space
described in the generalized Bloch theorem, and the Jordan structure of the transfer matrix. 
In the absence
of power-law solutions, each bulk solution $|\psi_{\ell s}\rangle$ is annihilated by 
$\bm{P}_{1,2}(T-z_\ell\mathds{1}_{N})=\bm{P}_{1,2}[P_B(T-z_\ell\mathds{1}_{N})]$.
In such cases, Eq.\,\eqref{connection} reads 
\[
(t(\epsilon)-z_\ell\mathds{1}_{d})\bm{P}_{1,2}|\psi_{\ell s}\rangle = 
\bm{P}_{1,2}(T-z_\ell\mathds{1}_{N})|\psi_{\ell s}\rangle=0,
\]
implying that $\bm{P}_{1,2}|\psi_{\ell s}\rangle$
is an eigenvector of $t(\epsilon)$ with eigenvalue $z_\ell$. 
Naturally, a Bloch wave-like bulk solution corresponds to an
eigenvalue on the unit circle, whereas an exponential
solution corresponds to one inside or outside the unit circle, in agreement with
the literature \cite{Lee}.

While, as remarked, the transfer matrix is typically assumed to be diagonalizable, we now 
show that generalized eigenvectors of $t(\epsilon)$ {\em are}
physically meaningful, and in fact related to the power-law solutions of the bulk
equation. Let $\epsilon$ be a value of energy for which power-law solutions are present. 
We can then generalize our earlier calculation for the eigenvectors of the
transfer matrix by noting 
that each $|\psi_{\ell s}\rangle$ is annihilated by 
$\bm{P}_{1,2}(T-z_\ell\mathds{1}_N)^{s_\ell}$, where $s_\ell$ is the multiplicity of the root $z_\ell$
as usual. Then, a similar calculation reveals that $\bm{P}_{1,2}|\psi_{\ell s}\rangle$ 
is a generalized eigenvector of $t(\epsilon)$, satisfying
\[
(t(\epsilon)-z_\ell\mathds{1}_{d})^{s_\ell}\bm{P}_{1,2}|\psi_{\ell s}\rangle = 0.
\]
Thus, {\em generalized eigenvectors of the transfer matrix are projections of 
solutions with a power-law prefactor}. In some non-generic scenarios, they indeed contribute
to the energy eigenstates, as we discussed \cite{TMremark}.

This analysis is vividly exemplified by the parameter regime corresponding to the circle of oscillations
in the Majorana chain, Eq.\,\eqref{CircOsc}, which we found to be associated
to a zero-energy power-law Majorana wavefunction. Accordingly, we expect 
the corresponding transfer matrix to possess generalized eigenvectors 
of rank two, failing to be diagonalizable. Let us verify this
explicitly. Except for the points $\mu=0, \Delta/t=\pm1$ in this regime, 
the matrix $h_1$ in Eq.\,\eqref{kitaevmat} is invertible.
The transfer matrix is then 
\[
t(\epsilon=0) \!=\!
 \frac{1}{\mu^2}\!\!\begin{bmatrix}
0 & 0 & \mu^2 & 0\\
0 & 0 & 0 & \mu^2\\
-4(t^2+\Delta^2) & -8t\Delta & 
-4t\mu & -4\Delta\mu\\
-8t\Delta &-4(t^2+\Delta^2) &  
-4\Delta\mu & -4t\mu \\
\end{bmatrix}\!\!,
\]
where $\mu,t$ and $\Delta$ satisfy Eq.\,(\ref{CircOsc}).
It can be checked that $t(\epsilon=0)$ has only two eigenvalues, namely, 
\( z_\ell = {-2(t + (-1)^\ell \Delta)}/{\mu},\) \(\ell=1,2,\)
each of algebraic multiplicity two, and that both of
these eigenvalues have only one eigenvector, given by 
\[ \bm{P}_{1,2}|z_\ell,1\rangle|u_\ell\rangle = \begin{bmatrix}
z_\ell \\ (-1)^\ell z_\ell\\ 
z_\ell^2 \\
(-1)^\ell z_\ell^2
\end{bmatrix}, \]
hence geometric multiplicity equal to one.  
Both $z_1,z_2$ are then defective, making 
$t(\epsilon=0)$ not diagonalizable. In fact, $t(\epsilon=0)$ 
has one generalized eigenvector of rank two corresponding to each eigenvalue,
given by 
\[ \bm{P}_{1,2}|z_\ell,2\rangle|u_\ell\rangle=
\begin{bmatrix}
1 \\ (-1)^\ell \\ (2z_\ell) \\ (-1)^\ell (2z_\ell)\\
\end{bmatrix}. \]

Returning to the general case, a number of additional remarks are worth making,
in regard to points of contact and differences between the transfer matrix approach 
and our generalized Bloch theorem. First, the eigenstate Ansatz obtained from the 
analytic continuation of the Bloch Hamiltonian provides a {\em global} characterization 
of energy eigenvectors (and generalized eigenvectors), as opposed to the local
characterization afforded within the transfer-matrix approach, whereby each 
eigenvector is reconstructed ``iteratively'' for any given eigenvalue. Further to that,  
the generalized Bloch theorem unveils the role of non-unitary representations of 
translational symmetry for finite systems.  Perhaps most importantly, the two 
methods differ in the way BCs are handled.  Clearly, in both approaches 
it is necessary to match BCs in order to obtain the physical energy spectrum. 
While open BCs are most commonly used in transfer-matrix calculations, 
the method has also been applied to relaxed surfaces \cite{Lee} and 
generalized periodic BCs \cite{Molinari}, all of which belong
to the class of BCs considered in this paper. In this sense, 
it is tempting to compare Eq.\,\eqref{TMboundmat} with the condition on 
the determinant of the boundary matrix, $\det B(\epsilon)=0$. However, 
the class of BCs to which the transfer matrix 
approach can be successfully applied is not {\em a priori} clear, thus 
whether such a condition can be established for as general a class of 
BCs as our theorem covers has not been investigated to the best of 
our knowledge.

From a numerical standpoint, the 
computational complexity of the standard transfer matrix method 
for clean systems (when applicable) is independent of the system size $N$, 
as is the case of our scan-in-energy algorithm in Sec.\,\ref{sec:scan}.
In those cases where inversion of certain matrices is a difficulty and 
inversion-free approaches are used \cite{Boykin, Biczo}, the latter also have 
a comparable computational complexity to our method. Interestingly, 
all approaches so far that are truly inversion-free rely at some point or 
another on the solution of a non-linear eigenvalue problem \cite{TMremark}.
Thanks to the fact that, as noted, the construction of $t(\epsilon)$ in 
the generic case relies only on the banded structure of $H_N$, bulk disorder 
can be handled efficiently within transfer-matrix approaches, albeit for a limited 
class of BCs. For general BCs as we consider, it is thus natural to 
combine the transfer matrix approach with the bulk-boundary 
separation we have introduced, in order to still find solutions
efficiently: the transfer matrix can be employed to find
all possible solutions of the bulk equation in the presence of 
bulk disorder, and the latter can then be used as input for the boundary
matrix, that provides a condition for energy eigenstates.

\section{Discussion and outlook}
\label{sec:end}

We have formulated a generalization of Bloch's theorem applicable to 
clean systems of independent fermions on a lattice, subject to 
BCs that are arbitrary -- other than respecting the finite-range nature 
of the overall Hamiltonian.  This generalization, which leverages
a reformulation of the problem in terms of corner-modified block-Toeplitz matrices, 
affords exact, analytical expressions for all the energy eigenvalues and eigenstates of 
the system -- which consistently recovers the ones derived from the standard Bloch's 
theorem for periodic BCs.  As a key component to this theorem, one obtains an 
exact {\em structural Ansatz}, close in spirit to the Bethe Ansatz, for all (regular) energy 
eigenstates in {\em dispersive} bands. This Ansatz is easy to construct since it depends
only on the energy eigenvalue and the bulk properties of the Hamiltonian. 
The individual components of this Ansatz reflect translation invariance
in a way we have made precise and are, as such, determined by the analytic continuation 
of the Bloch Hamiltonian, as shown. 

Based on the generalized Bloch theorem, we have provided both a numerical and an algebraic 
diagonalization algorithm for the class of quadratic Hamiltonians under consideration. 
For generic energy values, the former is computationally more efficient than existing ones 
in that its complexity is independent upon the system size; the latter is especially well-suited for 
symbolic computation or pen-and-paper solutions, as we explicitly demonstrated by solving in 
closed form a number of tight-binding Hamiltonians of interest, under various BCs. With an eye 
toward applications in synthetic quantum matter, we have also used the generalized Bloch 
theorem to engineer a quasi one-dimensional Hamiltonian that support a perfectly localized, 
robust zero-energy mode, notwithstanding the lack of chiral and charge-conjugation protecting 
symmetries.

Remarkably, our generalized Bloch theorem predicts the existence, under specific 
(non-generic) conditions, of edge states that decay exponentially in space 
with a {\em power-law prefactor}. Such exotic states were previously believed to arise 
only in systems with long-range couplings. In our framework, their origin may be 
traced back to the description of the system's eigenstates in terms of {\em non-unitary} 
representations of translation symmetry ``outside Hilbert space'' -- again 
capturing the fact that such a symmetry is only mildly broken by the BCs, in a 
precise sense.  Notably, we have shown how the emergence of zero-energy 
Majorana modes with a linear prefactor is possible in the paradigmatic Kitaev chain 
by proper Hamiltonian tuning {\em on} the so-called ``circle of oscillations''. 
Their ``critical'' spatial behavior separates the theoretically observed Majorana 
wavefunction oscillations inside such a circle from the simple exponential decay outside.

Our generalized Bloch theorem makes no prediction about the (singular) energy values 
which correspond to {\em dispersionless}, or flat, bands of eigenstates.  
We have nonetheless provided a prescription for identifying such energy values
without diagonalizing the full Hamiltonian, and showed how 
such energy values necessarily enter the physical energy spectrum
irrespective of the BCs.  In such singular cases, we have further provided 
a procedure to effectively obtain a (possibly overcomplete) basis of {\em perfectly 
localized states} using an analytic continuation of the Bloch Hamiltonian,
and explicitly illustrated such a procedure in the Kitaev's Majorana chain 
Hamiltonian at its sweet spot.

Building on our proposal in Ref.\,[\onlinecite{abc}], we have rigorously derived and further 
explored a proposed {\em boundary indicator} for the bulk-boundary correspondence.
This indicator leverages the other key component to our generalized Bloch theorem, the 
{\em boundary matrix}, and is unique in the sense that, unlike most other indicators in the 
literature, it combines information from both the bulk and the boundary. The utility of this 
indicator is seen from our analysis of the $4\pi$-periodic Josephson effect in a model of a 
$s$-wave topological superconductor.  In the process, we show how, remarkably, the 
$4\pi$-periodicity that distinguishes a topologically nontrivial response is {\it not} 
accompanied by a fermionic parity switch in this system. We have provided a physical 
explanation of this behavior by exhibiting a decoupling transformation, which maps the 
relevant Hamiltonian to two uncoupled ``virtual'' wires -- each undergoing a parity switch.

Finally, for systems where no bulk disorder is present, and subject to BCs for which the 
well-known transfer matrix approach is also applicable, we have shown how the generalized 
Bloch theorem may be used to obtain a physical interpretation of the transfer matrix's 
{\em generalized eigenvectors}, in terms of bulk solutions with a power-law prefactor.  
An explicit example is seen, again, in the semi-infinite Kitaev's chain with open BCs, precisely 
in the same circle-of-oscillations parameter regime that hosts power-law zero-energy Majorana 
modes. While, in this way, our method may be seen to provide yet another inversion-free 
alternative to the standard transfer-matrix approach, the connections we have identified 
in this work naturally point to further possibilities for fruitfully combining the two approaches.
In particular, since the bulk-boundary separation we proposed remains useful in the presence of 
bulk disorder, one may envision a hybrid approach for solving disordered systems subject 
to arbitrary BCs, by employing transfer-matrix techniques to handle the resulting bulk equation.

The tools we have developed here may serve as the starting point for a number of 
additional studies and applications. As mentioned, in the companion paper \cite{PRB2}, 
we will provide a formulation of the bulk-solution Ansatz and the generalized Bloch theorem 
further accounting for the role played by the transverse momentum ($\k_\perp$) in 
higher-dimensional systems with non-trivial boundaries -- as opposed to the single 
$\k_\perp$-analysis presented here. We will show that topological power-law modes 
discussed in this paper are not just a feature of one-dimensional systems, 
and indeed are present in higher dimensions too. 
Beside exploring the interplay between $\k_\perp$, 
the boundary matrix, and the edge states in a number of paradigmatic 
model Hamiltonians, we will also demonstrate how the treatment of one-dimensional 
homogeneous systems can be effectively extended to those of interfaces. 
From a computational standpoint, we expect that the diagonalization algorithms 
emerging from our approach will be useful for large-scale electronic calculations   
in both one- and higher- dimensions, possibly in conjunction with perturbative 
approaches for incorporating interactions. 

Towards a deeper understanding of bulk-boundary correspondence in topological 
insulators and superconductors, our approach can be instrumental in studying 
robustness against boundary perturbations.
It is natural to start by asking how certain symmetries of the system influence the 
nature of the proposed indicator, or the boundary matrix from which the indicator 
itself is derived. This can possibly lead to identifying a symmetry principle 
which dictates the bulk-boundary correspondence, as well as an interpretation 
at the basic dynamical-system level in terms of stability theory. Likewise, the 
framework we have developed may also serve as a concrete starting point for  
rigorously deriving an effective boundary theory for lattice systems.

Lastly, while we have focused on fermions in this paper, the general foundation of 
our method laid out in Ref.\,[\onlinecite{JPA}] is equally valid for bosons and immediately 
applicable to non-Hermitian effective Hamiltonians with non-trivial boundaries, as 
often arising in semi-classical models of open quantum systems in various contexts 
\cite{nonhermitianrev,MandalEP,tayebi16,Chong}. We plan to explore the corresponding 
generalized Bloch theorems in forthcoming publications, and to ultimately provide
extensions to Markovian open quantum systems described by quadratic Lindblad 
master equations.

\section*{Acknowledgements}
We gratefully acknowledge useful discussions with Smitha Vishveshwara.
Work at Dartmouth was supported in part by the US NSF through Grant No. 
PHY-1620541 and the Constance and Walter Burke Special Projects Fund 
in Quantum Information Science.

\appendix

\section{Further discussion on arbitrary BCs}
\label{finitediff}

Section\,\ref{sec:setting} imposes two restrictions on the allowed form
of BCs, described by $\widehat{W}$. The first restricts 
the non-trivial action of $\widehat{W}$ to the boundary hyperplanes.
Since the corresponding single-particle operator $W$ satisfies the 
relation \(P_BW=0\), with $P_B$ being the bulk projector associated to $H_N$,
$W$ can be thought of as a corner-modification of the banded block-Toeplitz matrix \(H_N\).
The operators $H_N+W$ represent boundary value problems in such a way that a change 
of BCs is encoded in a change of \(W\). The 
intuition behind these ideas comes from finite-difference 
methods for solving differential equations. We briefly
illuminate this connection here. 

Consider for concreteness the Schr\"odinger boundary value problem  
\begin{align*}
\psi(0)=\psi(L)=0&,\\
\Big(-\frac{1}{2}\frac{d^2}{dx^2}-\epsilon\Big)\psi(x)=0&
\quad \mbox{for} \quad x\in(0,L),
\end{align*}
describing a particle in an infinite one-dimensional potential well. 
The discretization \(
x\mapsto x_j=j\Delta x
\), with \(j=0,1,\dots,N+1=L/\Delta x\), reduces this problem to the 
lattice boundary value problem
\begin{align}
\psi(x_0)=\psi(x_{N+1})=0, &\\ 
-\frac{1}{2}\psi(x_{j-1})+(1-\epsilon)\psi(x_j)-\frac{1}{2}\psi(x_{j+1})=0, &
\label{latlap}
\end{align}    
in terms of the centered second difference approximation to the Laplacian. 
This set of linear equations is equivalent to the eigenvalue
equation \((H_N-\epsilon\mathds{1}_N)|\psi\rangle=0\), with
\begin{align*}
H_N=-\frac{1}{2}(T+T^\dagger)+\mathds{1}_N
\quad \mbox{and}\quad
|\psi\rangle\equiv \sum_{j=1}^N|j\rangle\psi(x_j).
\end{align*}
By comparison, the more general BCs 
\begin{align*} 
\alpha_1\psi(0)+\beta_1\frac{d\psi}{dx}(0^+)=0,\quad 
\alpha_2\psi(L)+\beta_2\frac{d\psi}{dx}(L^-)=0   ,
\end{align*}    
lead to the lattice boundary value problem  
\begin{align}
\alpha_1\psi(x_0)+\beta_1\frac{\psi(x_1)-\psi(x_0)}{\Delta x}=0&, \label{a3}\\
\alpha_2\psi(x_{N+1})+\beta_2\frac{\psi(x_{N+1})-\psi(x_{N})}{\Delta x}=0& \label{a4},
\end{align}
together with Eq.\,\eqref{latlap}.
The system of linear equations in Eqs. (\ref{latlap})--(\ref{a4}) is equivalent to the eigenvalue
problem \((H_N+W-\epsilon\mathds{1}_N)|\psi\rangle=0\), with
\[
W=\frac{\beta_1}{2(\alpha_1\Delta x-\beta_1)}|1\rangle\langle 1|
-\frac{\beta_2}{2(\alpha_2\Delta x+\beta_2)}
|N\rangle\langle N| , 
\]
a corner modification of the lattice Laplacian \(H_N\). For the 
special case \(\alpha_1=\alpha_2,\ \beta_1=-\beta_2\), 
we have discussed the exact diagonalization of \(H_N+W\) 
in Sec.\,\ref{sec:revisited}.

\section{Algebras of shift operators}
\label{app:opalg}

\begin{figure}
\includegraphics[angle=0,width=0.8\columnwidth]{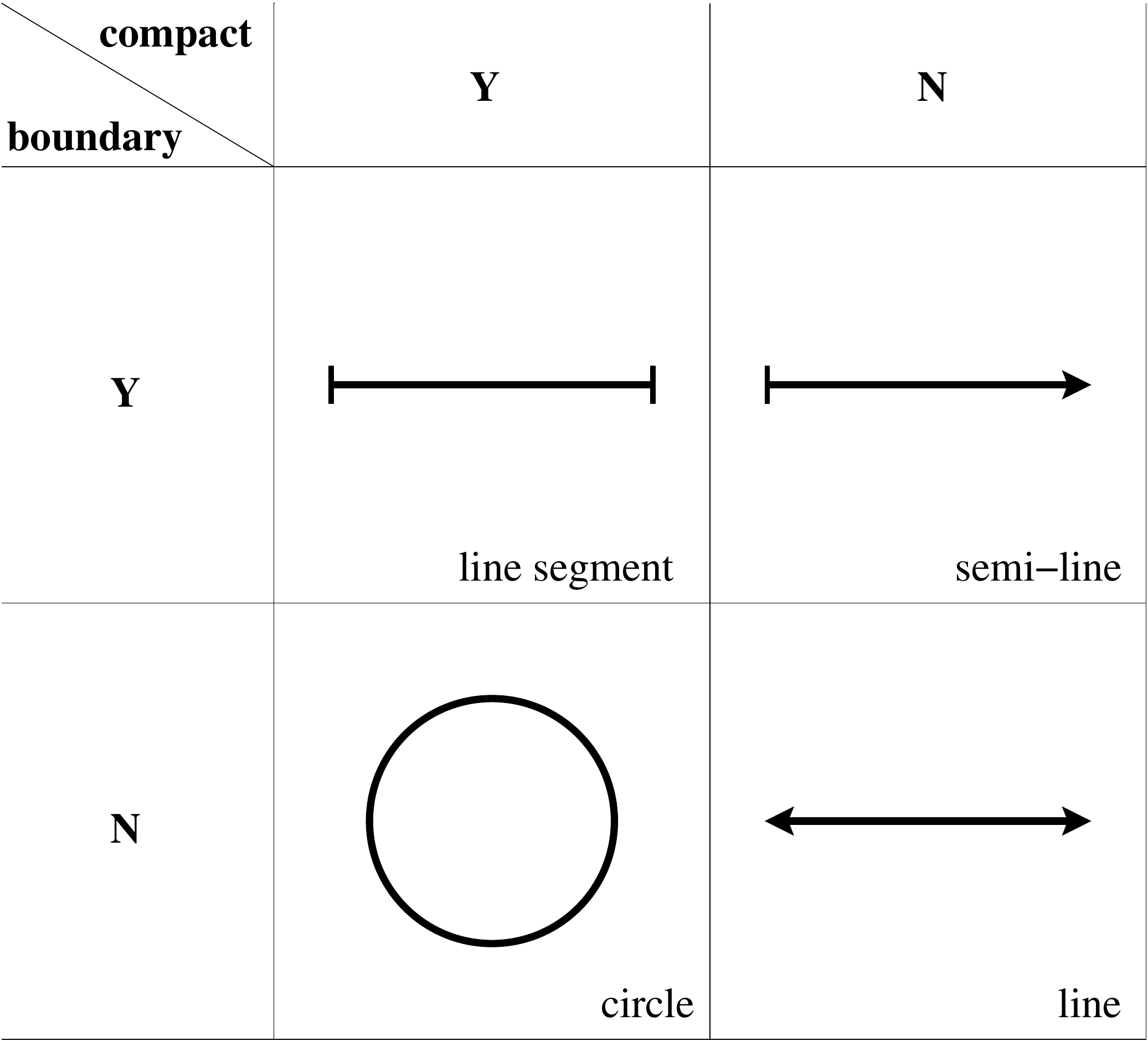}
\caption{Four topologically inequivalent one-dimensional
manifolds. The classification (Yes $=$ Y, No $ =$ N) encompasses 
compactness and whether the boundary is empty.}
\label{topolines}
\end{figure}

Consider the topologically inequivalent manifolds 
corresponding to the finite line segment, the circle (of finite or infinite radius), 
the semi-infinite line, and the infinite line, as illustrated in Fig.\,\ref{topolines}. Given a 
physical system whose state space has support on those manifolds, 
one can define distinct shift (or translation by a distance $a$) operators
acting on the physical states. Certainly, those shift operators encode topological 
information that depending on the circumstances may have physical consequences. 
In the following we 
will study the algebra of those shift operators.
The subtle difference between the various 
shift (or translation) operators is reflected in the fundamental discussions that led to 
the modern theory of macroscopic electric polarization in many-body systems
in terms of Berry phases \cite{macroP1,macroP2}, and the concomitant definition 
of the position operator in extended systems \cite{positionop}.

\vspace*{1mm}

{\it The finite line segment.---}
This section is based on Ref.\,[\onlinecite{fockpfs}], where the 
matrices we are about to consider appeared with a different physical 
meaning. Consider a line of finite length $L=N a$, written in terms 
of a characteristic length $a$, typically defined by a periodic 
potential or lattice. The left shift operator is given by
\( T=\sum_{j=1}^{N-1}|j\rangle\langle j+1| ,\) in terms of the orthonormal 
lattice states $|j\rangle$. The lattice state \(|1\rangle\) is annihilated by 
\(T\), \(T|1\rangle=0\), and \(|N\rangle\) is annihilated by
\(T^\dagger\), mirroring the fact that the  boundary of 
a line segment consists of two points. For states other 
than \(|1\rangle,|N\rangle\), \(T\) and \(T^\dagger\) act as 
ordinary translations, to the left or right respectively, i.e., 
$T |j \rangle = | j-1 \rangle$ and $T^\dagger |j \rangle = | j+1 \rangle$.

While \(T\) can be regarded as the generator of 
bulk translations, it is not a unitary transformation. Instead,
\[ T^s(T^\dagger)^s+(T^\dagger)^{N-s}T^{N-s}=\mathds{1}, \quad s=1,\dots,N-1,
\]
and notice also that \(T^N=0\). The commutator
\( [T,T^\dagger]=|1\rangle\langle 1|-|N\rangle\langle N| \)
captures the extent of translation-symmetry breaking introduced
by the BCs. The lattice-regularized position operator
\( X=\sum_{j=1}^Nj \ |j\rangle\langle j|  \)
satisfies the commutation relation
\begin{eqnarray}
\label{u1finite} [X,T]=-T.
\end{eqnarray}
While this is formally analogous to \([x,e^{i p/\hbar}]=-e^{i p/\hbar}\), care 
must be exercised with such analogy, precisely because of 
issues of definition of the domains of functions where operators 
act upon. 

\vspace*{1mm}

{\it The circle.---}
The other compact one-dimensional manifold is the circle. 
The standard (periodic) left shift operator in this case is given by
\begin{eqnarray*}
V=\sum_{j=1}^{N-1}|j\rangle\langle j+1|+|N\rangle\langle 1|=T+(T^{\dagger})^{N-1}.
\end{eqnarray*}
No lattice state \(|j\rangle\) is annihilated by either \(V\) or
\(V^\dagger\), because the circle is a manifold with no boundary.
One can further check that \(VV^\dagger =\mathds{1}=V^N . \)
The relation between periodic shifts and the position operator $X$
is better described in terms of $U\equiv e^{i\frac{2\pi}{N} X},$
since then we have the Heisenberg-Weyl relation
\begin{eqnarray}
\label{weylrel}
VU=e^{i\frac{2\pi}{N}}\,UV.
\end{eqnarray}
This Heisenberg-Weyl algebra is well-known in statistical
mechanics in connection to clock models \cite{pclock}, but its relevance to 
tight-binding models appears to have gone unnoticed. 
The two generators are related by the discrete Fourier transform 
$F$ as $F U F^\dagger = V^\dagger$ and $F V F^\dagger = U$, see for
example Ref.\,[\onlinecite{pclock}] for more details and references. 

By comparing Eq.\,\eqref{u1finite} to Eq.\,\eqref{weylrel}, one 
sees that the \(U(1)\) symmetry of the shift algebra associated to 
the line segment is broken to a \(\mathds{Z}_N\) symmetry for 
the circle. In practice, the full \(U(1)\) symmetry is recovered
by introducing twisted generalizations of the Heisenberg-Weyl algebra,  
$V_\phi U_\phi =e^{i\frac{2\pi}{N}}U_\phi V_\phi,$ $V_\phi^N=e^{i\phi}\mathds{1},$
with \(U_\phi\) and \(V_\phi\) unitary. Their meaning is clear in 
terms of tight-binding models. Twisted Heisenberg-Weyl algebras 
describe physical problems subject to generalized Born-von-Karman 
BCs, needed for example for defining topological 
invariants such as the Chern number. A representation of these algebra 
is given by
\begin{eqnarray*}
U_\phi=U,\ \ 
V_\phi=\sum_{j=1}^{N} e^{i\frac{\phi}{N}} |j\rangle\langle j+1| +
e^{i\frac{\phi}{N}} |N\rangle\langle 1| .
\end{eqnarray*}
In statistical mechanics, our twisted Heisenberg-Weyl algebras
are connected to chiral Potts models, but this connection
seems to be unknown in the literature. 

\vspace*{1mm}

{\it The semi-infinite line.---} The left and right unilateral shifts \(\bm{T}_-,\ \bm{T}_-^\star\)
were introduced in Sec.\,\ref{sec:emergent}. The commutator
\( [\bm{T}_-,\bm{T}_-^\dagger]=|1\rangle\langle 1| \)
captures in some sense the extent of translation symmetry breaking. The lattice position operator  
\(
X_-^{}=\sum_{j=1}^\infty j \, |j\rangle\langle j|
\)
satisfies the commutation relations  
\(
[X_-^{},\bm{T}_-^{}]=-\bm{T}_-^{},\quad
[X_-^{},\bm{T}_-^\star]=\bm{T}_-^\star.
\)
The relation \(\bm{T}_-^\star=\bm{T}_-^\dagger\) holds if the 
domain of these linear transformations is restricted to
the Hilbert space of square summable half-infinite sequences. 

\vspace*{1mm}

{\it The real line.---}
The shift operator is
\( \bm{T} \equiv \sum_{j\in\mathds{Z}}|j\rangle\langle j+1|,
\) and it is unitary when restricted to the Hilbert space of 
square-summable sequences, that is, \(\bm{T}^{-1}=\bm{T}^\dagger\).
We carefully refrained from restricting \(\bm{T}\) so in 
Sec.\,\ref{sec:extended}. With 
\(X\equiv\sum_{j \in\mathds{Z}} j \, |j\rangle\langle j|\)
(an unbounded Hermitian operator in Hilbert space), one can 
show that 
\( [X,\bm{T}]=-\bm{T},\quad [X,\bm{T}^{-1}]=\bm{T}^{-1}
\)
both in and out of Hilbert space.

\vspace*{1mm}

In summary, the shift operators associated to the finite 
and the semi-infinite line segment do {\em not} commute with their 
adjoints, reflecting the presence of boundary points for these topologies. 
In contrast, the shift operators defined on the circle and the line  
\(V\) and \(\bm{T}\) {\em do} commute with their adjoints (or inverses) 
and are unitary (or just invertible) -- which is why they can represent 
translation symmetry. As a consequence, \(V,V^\dagger\) can be diagonalized 
simultaneously, and the same goes for \(\bm{T},\bm{T}^\dagger\) \cite{footapp2}. 
Their eigenvalues lay on the unit circle due to unitarity. The key 
difference between these two types of translation symmetry stems from 
their interplay with lattice position operators. For all the shift 
algebras but the one associated to the circle, the position 
operators generate \(U(1)\) rotations of the shift operators. 
For the Heisenberg-Weyl algebra, this \(U(1)\) symmetry appears 
instead as a family of inequivalent unitary irreducible representations of the 
defining relation Eq.\,\eqref{weylrel}.

\section{Emergent solutions at regular energies}
\label{app:ext2eme}

This appendix provides further mathematical detail 
on the procedure for computing emergent bulk solutions outlined in 
Sec.\,\ref{sec:emergent}. Specifically, we pick up the discussion 
where we left it therein, right after the definition of the matrix polynomial 
\(K^-(\epsilon,\bm{T}_-)\)  in Eq.\,\eqref{kminus}.

\vspace*{1mm}

{\it Left-localized emergent bulk solutions.---} 
In analogy to the sequences $\Phi_{z,v}$ associated to ${\bm T}$ in Eq.\,(\ref{geneig}), 
let us define states
\begin{align}
\Upsilon^{-}_{z,1}\equiv &\ \sum_{j=0}^{\infty}z^j|j+1\rangle,\nonumber \\
\Upsilon^{-}_{z,v} \equiv &\
\frac{1}{(v-1)!}\frac{d^{v-1}}{dz^{v-1}}\Upsilon_{z,1}^{-},\quad v=2,3,\dots.
\label{higherus}
\end{align}
in such a way that 
\( \Upsilon^{-}_{0,v}=|j=v\rangle \) and, also,
\begin{align*}
\Upsilon_z^{-}|u\rangle\equiv\sum_{x=1}^v\Upsilon_{z,x}^{-}|u_x\rangle=
\begin{bmatrix}
\Upsilon_{z,1}^{-}&\dots &\Upsilon_{z,v}^{-}
\end{bmatrix}
\begin{bmatrix}
|u_1\rangle\\
\vdots\\
|u_v\rangle
\end{bmatrix}.
\end{align*}
It is then immediate to verify that
\[ K^-(\epsilon,\bm{T}_-)\Upsilon^-_{z,1}|u_1\rangle=\Upsilon^-_{z,1}K^-(\epsilon,z)|u_1\rangle.
\]
Moreover, using Eq.\,\eqref{higherus}, one also obtains 
the more general relation 
\begin{eqnarray}
\label{emleft}
\hspace*{-5mm}
K^-(\epsilon,\bm{T}_{-})\Upsilon_z^{-}|u\rangle\!=\!
\begin{bmatrix}
\Upsilon_{z,1}^{-}&\dots &\Upsilon_{z,v}^{-}
\end{bmatrix}
K^-_v(\epsilon,z)\!\!
\begin{bmatrix}
|u_1\rangle\\
\vdots\\
|u_{v}\rangle
\end{bmatrix}\!,\ 
\end{eqnarray}
in terms of the upper-triangular \(v\times v\) block matrix
\begin{eqnarray*}
[K^-_v(\epsilon,z)]_{xx'}=
\frac{1}{(x'-x)!}\frac{d^{x'-x} K^{-}(\epsilon,z)}{dz^{x'-x}},\ \ 1\leq x\leq x'\leq v.
\end{eqnarray*}
It will be crucial for later use to notice that \(K^-_v(\epsilon,z)\)
is a block-Toeplitz matrix. 

Both \(K^-_v(\epsilon,z)\) and \(H_v(z)\) are defined by the
same formula, recall Eq.\,\eqref{genhb}. The key difference
between the two is that \(K^-_v(\epsilon,z)\) is well-defined
also at \(z=0\). So suppose that \(z_0=0\) is a root of \(P(\epsilon,z)\) 
of multiplicity \(s_0>0\). Then, one can show using tools from 
Ref.\,[\onlinecite{JPA}], that there are precisely \(s_0\) independent
solutions of the equation 
\begin{eqnarray*}
K_{s_0}^-(\epsilon,z_0=0) |u^-_s\rangle=0,\quad s=1,\dots,s_0.
\end{eqnarray*}
The corresponding emergent bulk solutions are 
\begin{eqnarray*}
|\psi^-_s\rangle=\bm{P}_{1,N} \Upsilon^-_0|u^-_s\rangle=\sum_{j=1}^{s_0}
|j\rangle|u_{sj}^-\rangle.
\end{eqnarray*}
They are localized on the left edge  over the first 
\(s_0\) sites. For Hermitian Hamiltonians, \(s_0\leq dR\) necessarily. 

\vspace*{1mm}

{\it Right-localized emergent bulk solutions.---}
Left-localized emergent bulk solutions cannot appear alone; 
they can only appear in conjunction with a set of right-localized 
emergent bulk solutions. The reason is as follows. Consider the 
unitary, Hermitian operator
\[
U=U^\dag \equiv\sum_{j=1}^N|N-j+1\rangle\langle j|\otimes \mathds{1}_d , \quad
U^2=\mathds{1}_{dN}, 
\]
which implements a mirror transformation of the lattice, 
by acting trivially on internal states. The transformed Hamiltonian is the Hermitian block-Toeplitz 
matrix
\[
\widetilde{H}_N = UH_N U=\mathds{1}_N\otimes h_0+
\sum_{r=1}^R(T^r\otimes h_r^\dagger+T^{r\,\dagger}\otimes h_r) ,
\]
in which the hopping matrices have been exchanged as \(h_r\leftrightarrow h_r^\dagger\).
Therefore, the left-localized  emergent bulk solutions for $\widetilde{H}_N$ are dictated by the
matrix $\widetilde{K}^-(\epsilon)$ with entries 
$[\widetilde{K}^-(\epsilon)]_{ij}=[K^-(\epsilon)]_{ij}^\dagger$.
If \(|\widetilde{\psi}^-\rangle\) denotes a left-localized emergent 
solution for \(\widetilde{H}_N\), then   
\[
0=P_B(\widetilde{H}_N-\epsilon)|\widetilde{\psi}_s^-\rangle=
UP_B(H-\epsilon)U|\widetilde{\psi}_s^-\rangle,
\]
implying that the state 
\( U|\widetilde{\psi}_s^-\rangle=\sum_{j=1}^{s_0}|N-j+1\rangle|\tilde{u}^{-}_{sj}\rangle  \)
is an emergent bulk solution for \(H_N\), localized on the {\it right} 
edge. Similarly, the left-localized emergent bulk solutions of \(H_N\) 
are in one-to-one correspondence with the right-localized emergent 
solutions of \(\widetilde{H}_N\). This conclusion relies havily 
on the commutation relation \(P_BU=UP_B\), which is always necessarily
true for {\em closed} systems (Hermitian Hamiltonians), as we considered here.

But how can we compute the right-localized emergent bulk solutions
directly in terms of \(H_N\)? In Sec.\,\ref{sec:emergent}, we answered
this question with the help of the matrix \(K^+(\epsilon)\equiv K^-(\epsilon)^\dagger\). 
We will justify this answer here. Let 
\(|\widetilde{\psi}_s^-\rangle=\sum_{j=1}^{s_0}|j\rangle|\tilde{u}^-_{sj}\rangle$,
$s=1,\dots,s_0$, denote the left-localized  emergent solutions associated
to \(\widetilde{H}_N\), and let
\[
|\psi^+_s\rangle \equiv \sum_{j=1}^{s_0}|N-s_0+j\rangle|u^+_{sj}\rangle =
U|\widetilde{\psi}_s^-\rangle,\quad s=1,\dots,s_0, 
\] denote
the corresponding right-localized emergent solutions of \(H_N\),
so that \(|u^+_{sj}\rangle\equiv|\tilde{u}^-_{s,s_0-j+1}\rangle\).
Our goal is to show that the arrays 
\[
|u^+_s\rangle=
\begin{bmatrix}
|u^+_{s1}\rangle& \dots & |u^+_{ss_0}\rangle
\end{bmatrix}^T,\quad  s=1,\dots,s_0,
\]
are annihilated by \(K^+(\epsilon)\). 
Because $|u^+_s\rangle = \tilde{U}|\tilde{u}^-_{s}\rangle$, with 
$\tilde{U} = \tilde{U}^\dagger = \sum_{j=1}^{s_0}|j\rangle\langle s_0-j+1|$,
we conclude that $K^+(\epsilon)$ is related to $\widetilde{K}^-(\epsilon)$ via 
\( K^+(\epsilon) = \tilde{U}\widetilde{K}^-(\epsilon)\tilde{U}.
\)
This leads to the entries
\[
[K^+(\epsilon)]_{ij} = [\widetilde{K}^-(\epsilon)]_{s_0-i+1,s_0-j+1} 
=[K^-(\epsilon)]_{ji}^\dagger ,
\]
thanks to the fact that \(K^-(\epsilon)\) is a block-Toeplitz matrix. 
Hence,  $K^+(\epsilon)=[K^-(\epsilon)]^\dagger$, as desired.


\begin{thebibliography}{}
\bibitem{commentAM}
It is illuminating to review the footnote 6 in Ref. 
[\onlinecite{ashcroftmermin}] (p. 135), when the authors discuss
the proof of the standard Bloch's theorem.
\bibitem{footnote2}
For the vector space of semi-infinite systems extending to infinity in the positive direction 
(see Appendix \ref{app:ext2eme}), only the former relation, that is,  
$\bm{T}\mathcal{M}_{0,\infty}\subseteq\mathcal{M}_{0,\infty}$
holds true.
\bibitem{normalization}
Notice that our choice of normalization for ${\mathbf{\Phi}}_{z,v}$ differs 
from the one we used in Ref.\,[\onlinecite{JPA}] by the inclusion of a factorial 
factor. This allows for a unified formalism to be used for emergent solutions, 
as detailed in Appendix \ref{app:ext2eme}.
\bibitem{remarkAlternative}
Alternatively, we could have substituted the analytic expression for 
either of the roots $z_1(\epsilon)$ 
or $z_2(\epsilon)$ of Eq.\,\eqref{combz} in Eq.\,\eqref{bdcomb},  to obtain
a single equation in $\epsilon$, whose roots coincide with the eigenvalues of $H$.
\bibitem{XYremark} Explicitly, the Jordan-Wigner mapping yields the Hamiltonian 
\( \widehat{H}_{XY} \!= \!-\sum_{j=1}^{N}B_z(\sigma_z^j+1)-
\sum_{j=1}^{N-1}\big(J_x\sigma_x^{j}\sigma_x^{j+1}
+J_y\sigma_x^{j}\sigma_x^{j+1}\big),\)
where $\sigma_x^j,\sigma_y^j, \sigma_z^j $ are Pauli matrices for spin $j$, $B_z = \mu/2$ 
is the strength of the magnetic field along the $z$-direction, and $J_x=t-\Delta$, $J_y=t+\Delta$
are coupling strengths along $x$ and $y$, respectively.
\bibitem{fn}
Observe that the states corresponding to $j=1, \ldots, N-2$ are related to the 
basis states $\Psi_{j1,\pm}$ and $\Psi_{j2,\pm}$ as follows:
$\Psi_{j1,\pm} = \tilde{\Psi}_{j,\pm} \pm \tilde{\Psi}_{j+1,\pm},$ and 
$\Psi_{j2,\pm} = -\tilde{\Psi}_{j,\pm} \pm \tilde{\Psi}_{j+1,\pm}$.
\bibitem{fn6}
If $|\epsilon\rangle$ is an eigenstate of $H$ with energy $\epsilon$, satisfying 
$S_1|\epsilon\rangle=|\epsilon\rangle$, then $S_2|\epsilon\rangle$
is also an eigenstate with the same energy. Further, $S_2|\epsilon\rangle$
is orthogonal to $|\epsilon\rangle$, as the relation $\{S_1,S_2\}=0$ leads to 
$S_1(S_2|\epsilon\rangle) = -(S_2|\epsilon\rangle)$.
\bibitem{footapp2}
In contrast, \(\bm{T},\ \bm{\T}^{-1}\) are not diagonalizable in the space of all sequences, as  
explained in Sec.\,\ref{sec:extended}, but share a common Jordan basis.
\end{thebibliography}
\end{document}